
\input harvmac.tex
\hfuzz=10pt
\hoffset=-.1in
\voffset=-.2in

\vsize=7.5in
\hsize=5.6in
\tolerance 10000
\def\ll{\left\langle}
\def\rr{\right\rangle}
\def\svec#1{\skew{-2}\vec#1}
\def\pmb#1{\setbox0=\hbox{$#1$}%
\kern-.025em\copy0\kern-\wd0
\kern.05em\copy0\kern-\wd0
\kern-.025em\raise.0433em\box0 }
\def\IR{\relax{\rm I\kern-.18em R}}
\def\IZ{{\hbox{{\rm Z}\kern-.4em\hbox{\rm Z}}}}
\def\to{\rightarrow}
%
%
\lref\Azbel{M. Ya. Azbel', Zh.Eksp.Teor.Fiz. {\bf 46}, (1964) 929 [Sov.Phys.
JETP {\bf 19}, (1964) 634].}
\lref\Hofstadter{D. R. Hofstadter, Phys. Rev. {\bf B14} (1976) 2239.}
\lref\Wannier{G. H. Wannier, Phys. Status Solidi  {\bf B88} (1978) 757.}
\lref\PD{C. G. Callan and D. Freed, Nucl. Phys. {\bf B374} (1992) 543.}
\lref\CL{A. O. Caldeira and A. J. Leggett, Physica  {\bf 121A}(1983) 587;
Phys. Rev. Lett. {\bf 46} (1981) 211; Ann. of Phys. {\bf 149} (1983) 374.}
\lref\CT{C. G. Callan, A. G. Felce and D. E. Freed, ``Critical Theories of
the Dissipative Hofstadter Model'', PUPT-1292, CTP\#2073,
to appear in Nucl. Phys. {\bf B}.}
\lref\OSDQM{C. G. Callan, L. Thorlacius, Nucl. Phys. {\bf B329} (1990) 117.}
\lref\CLNY{C. G. Callan, C. Lovelace, C. R. Nappi, and S. A. Yost,
Nucl. Phys. {\bf B293} (1987) 83; Nucl. Phys. {\bf B308} (1988) 221.}
\lref\RIWI{C. G. Callan, L. Thorlacius, Nucl.  Phys. {\bf B319} (1989) 133.}
\lref\FZ{M. P. A. Fisher and W. Zwerger, Phys. Rev. {\bf B32} (1985) 6190.}
\lref\GHM{F. Guinea, V. Hakim and A. Muramatsu,
Phys. Rev. Lett.  {\bf 54} (1985) 263.}
\lref\Muir{T. Muir, {\it A Treatise on the Theory of Determinants},
Longmens, Green and Co., New York, 1933.}
\lref\Schmid{A.~Schmid, Phys. Rev. Lett. {\bf 51} (1983) 1506.}
\lref\DEF{D.~E.~Freed,
``Open String Reparametrization Ward Identities in the Presence of
Tachyon and Constant Gauge Fields", in preparation}
\lref\DH{J. J. Duistermaat, G. J. Heckman, Invent. Math. {\bf 69}
(1982)259.}
%
%
\nfig\figone{The graphs for the free energy of the fermionized theory, at
order $V_0^4$.}
\nfig\figtwo{The corrections to the fermionized graphs at order $V_0^4$.}
\nref\figthree{A magnetic field graph with 1PI subgraphs, $G_1, \ldots,
G_6$, joined with magnetic-field edges, $e_1, \ldots, e_7$.}
\nfig\figfour{The evaluation of the residue at $z_1 = e^{-a\epsilon}z_2$,
where edges $e_1$ and $e_2$ correspond to $1/(z_1-e^{-a\epsilon}z_2)$ and
$1/(z_1-e^{-b\epsilon}z_2)$, respectively.}
\nfig\figfive{The evaluation of the residue at $z_1 = e^{-a\epsilon}z_2$,
where edges $e_1$ and $e_1'$ both correspond to $1/(z_1-e^{-a\epsilon}z_2)$.}
\nfig\figsix{The evaluation of a $z_1=0$ residue, when all the derivatives
act on only on the edges joining $z_1$ and $z_2$.}

{\nopagenumbers

\baselineskip 12pt plus 1pt minus 1pt
{\hfill hep-th/9304006}
\smallskip
\centerline{\bf CONTACT TERMS AND DUALITY SYMMETRY IN }
\smallskip
\centerline{{\bf THE CRITICAL DISSIPATIVE HOFSTADTER MODEL}\footnote{*}{This
work is supported in part by National Science Foundation grant \#87-8447
and funds provided by the U. S. Department of Energy (D.O.E.) under contract
\#DE-AC02-76ER03069.}}
\vskip 24pt
\centerline{Denise E.~Freed}
\vskip 12pt
\centerline{\it Center for Theoretical Physics}
\centerline{\it Laboratory for Nuclear Science}
\centerline{\it and Department of Physics}
\centerline{\it Massachusetts Institute of Technology}
\centerline{\it Cambridge, Massachusetts\ \ 02139\ \ \ U.S.A.}
\vskip .7in
\centerline{\bf Abstract}
\smallskip
The dissipative Hofstadter model describes the quantum
mechanics of a charged particle in two dimensions subject to a periodic
potential, uniform magnetic field, and dissipative force. Its phase diagram
exhibits an SL(2,Z) duality symmetry and has an infinite number of critical
circles in the dissipation/magnetic field plane.  In addition, multi-critical
points on a particular critical circle correspond to non-trivial solutions
of open string theory.  The duality symmetry is expected to provide relations
between correlation functions at different multi-critical points.  Many of
these correlators are contact terms.  However we expect them to have
physical significance because under duality they transform into functions
that are non-zero for large separations of the operators.

Motivated by the search for exact, regulator independent
solutions for these contact terms, in this paper we derive many properties
and symmetries of the coordinate correlation functions at the special
multi-critical points.  In particular, we prove that the correlation
functions are homogeneous, piecewise-linear functions of the momenta, and
we prove a weaker version of the anticipated duality transformation.
Consequently, the possible forms of the correlation functions are limited to
lie in a finite dimensional linear space.  We treat the potential
perturbatively and these results are valid to all orders in
perturbation theory.
\vskip .8in
\centerline{Submitted to: {\it Nuclear Physics B\/}}
\vfill
\centerline{ Typeset in $\TeX$ by Roger L. Gilson}
\vskip -12pt
\noindent CTP\#2170\hfill March 1993
\eject}
\pageno=1
\baselineskip 16pt plus 2pt minus 1pt
\newsec{INTRODUCTION}
The dissipative Wannier--Azbel--Hofstadter model is a quantum-mechanical
description of a particle moving in two dimensions, subject to a doubly
periodic potential, a transverse magnetic field and a dissipative force.  In
the absence of friction, the energy spectrum, as a function of flux quanta per
unit cell, exhibits a remarkable fractal
structure \refs{\Azbel, \Hofstadter, \Wannier}.
As the friction/unit cell, $\alpha$, and flux/unit cell, $\beta$,
are varied, the system undergoes a
series of phase transitions between localized and delocalized behavior.  In
the limit as the friction goes to zero, the phase diagram becomes fractal, and
throughout the entire $\beta$-$\alpha$ plane it exhibits an $SL(2,Z)$
symmetry \PD.

In addition to having an unusual phase diagram, this system is also of
interest because it is related to several other models.  It is equivalent to
a generalized, neutral Coulomb gas restricted to one dimension.  If the gas is
allowed to have a total net charge, then the correlation functions of this
theory can be used to find the boundary state in open string
theory \OSDQM. As a
consequence, if the critical theories also satisfy a set of Ward identities
reflecting the reparametrization independence of the boundary state, then they
represent new solutions of open string theory in a non-trivial background of
tachyons and gauge fields \RIWI.  One can also directly use these
correlation functions to calculate the scattering of fields off a boundary
with background tachyon and gauge fields.

In earlier papers \PD\ \CT,
we have shown that there is a critical circle when
$\alpha/\left(\alpha^2+\beta^2\right) = 1$, and that the points on this circle
where $\beta/\alpha\in{\IZ}$ are multi-critical points.
  At the critical point $\alpha=1$,
$\beta=0$, the theory fermionizes and the large-time behavior of the system can
be solved exactly.  In this theory, all coordinate correlation functions (other
than two-point functions) are zero, except for contact terms.  However,
because the fermionization is valid only for large-time behavior, it does not
appear to give the correct form for the contact terms that satisfy the Ward
identities.
Because of the $SL(2,R)$ invariance of
the phase diagram, these contact terms are related by a duality transformation
to correlation functions with non-zero magnetic field.   In Ref.~\CT\ we
have shown that these correlation functions with $\beta=0$ are finite at large
times.  This result suggests that the contact terms are physical and that the
symmetries of the theory should determine them.  However, this line of
reasoning is not entirely reliable because the duality transformation has not
been proven with the regulator that satisfies the Ward identity.

In this paper, we address these issues and the evaluation of the contact terms
by giving a detailed study
of the theories on the critical circle $\alpha/\left(
\alpha^2+\beta^2\right)=1$.  Using methods motivated by the derivation of
fermionization and the duality symmetry, we
derive several ``large time'' properties or
symmetries that greatly restrict the possible form of the correlation
functions when $\beta/\alpha$ is an integer.  In addition, we prove that a weak
version of the duality transformation is true for the regulated theories.
These results are
valid to {\it all\/} orders in perturbation theory.
In a later paper we will show how these
properties, combined with the Ward identity,
can be used to obtain exact solutions for
some of the coordinate correlation functions while setting stringent
requirements on the others.

The outline of this paper is as follows. In Section~2,
we begin by briefly defining the
regulated dissipative Hofstadter model.
In the following section, after setting the cut-off to
zero in what appear to be non-divergent functions, we show how the partition
function of this model is equivalent to one for a gas of fermions, and we
derive the duality transformation of the correlation functions.  At this
point, readers interested only in the results may wish to skip to the end of
Section~7.  In
Section~4 we use an analog
of fermionization to find Feynman rules for the regulated graphs and ascertain
their relevant properties.  In the following section, we use these properties
of the graphs and simple properties of contour integration to find general
rules for integrating these types of graphs.  In Section~6, we restrict our
attention to the case when $\beta/\alpha\in{\IZ}$.  We apply the results of
the previous sections to the calculation of the degree of divergence of
the free energy, with the result that it has a pole divergence and no
logarithmic divergences as the ultraviolet cutoff is taken to 0.
This is essential for
showing that the Ward identity is satisfied and that the theories are at a
zero of the $\beta$-function.  Then we calculate the Fourier transform of the
correlation functions of $e^{\pm ix(t)}$ and $e^{\pm iy(t)}$ when there is no
periodic potential because these are necessary for calculating the correlation
functions of the dissipative Hofstadter model.  In Section~7, we finally use
all of these results to prove that, for $\beta/\alpha\in{\IZ}$,
all the correlation functions of the coordinates $\dot x(t)$ and
$\dot y(t)$ are piecewise linear and homogeneous in the Fourier variables.
This result restricts the possible forms of these correlation functions to
lie in a finite-dimensional linear space of functions.  We also
prove a weaker version of the duality transformation, and we present a
calculation to second order in the potential of all the correlation
functions of $\dot x(t)$ and $\dot y(t)$.  In the last section, we
summarize our main results and give a brief discussion of them.
\newsec{THE REGULATED DISSIPATIVE HOFSTADTER MODEL}
We begin with a brief outline of the dissipative Hofstadter model (DHM).  For
details the reader is referred to Refs.~\CL, \OSDQM, \PD\
and \CT.  In this model, a macroscopic
particle with mass, $M$, and coordinates,
${\svec X}(t)= \left(x(t),y(t)\right)$, is confined
to two dimensions and is subject to a doubly periodic potential, $V({\svec
X})$, and a uniform, transverse magnetic field, ${\svec B}$.
In addition, it is also subject
to dissipative forces caused by its interaction with its environment.  In
order to treat this model quantum mechanically, we describe the friction using
the Caldeira--Leggett model, in which the environment is modeled by a bath of
harmonic oscillators.  The particle interacts linearly with these
oscillators.  The frequencies of the oscillators and the interaction
strengths can be
chosen so that the classical equations of motion for ${\svec X}(t)$ reduce to
the usual phenomenological equations of motion for a particle subject to
friction, after the oscillator coordinates are eliminated via these equations
of motion.  Because the dependence on the oscillator coordinates in the
Lagrangian is quadratic, it is also possible to integrate them out from the
quantum mechanical path integral.  This results in a quantum, effective action
for the $X$ variables, which includes a non-local piece containing the effect
of dissipation. For arbitrary scalar and vector potentials, $V$ and
$A$, the action has the form
\eqn\Sdef{S[{\svec X}] = \int dt \left\{ {1\over 2} M \dot{\svec X}^2 +
V({\svec X}) +
i A_\mu ({\svec X}) \dot x^\mu + {\eta\over 4\pi} \int^\infty_{-\infty} dt'
{\left({\svec X}(t) - {\svec X}(t')\right)^2\over (t-t')^2}\right\}\ \ .}
Remarkably, the only dependence on the oscillator parameters that remains in
the action is through the $\eta$-term, where $\eta$ is the classical
coefficient of
friction.  Because this term is non-local, the path integral is effectively
that of a one-dimensional statistical system with long-range interactions.
Such systems, unlike one-dimensional {\it local\/} systems, do have phase
transitions.  In the dissipative quantum mechanics context, the phase
transitions are between different
regimes of long-time behavior of Green's functions.  When there is a
periodic potential, but no magnetic field, these transitions are discussed
in \Schmid, \FZ, and \GHM.  For the dissipative
Hofstadter model, and a similar discrete version of the model, we have
shown \PD\
that as the flux/unit cell, $\beta$, and friction/unit cell, $\alpha$, are
varied, the particle undergoes a series of localization--delocalization phase
transitions and, as the friction goes to zero the phase diagram in the
$\alpha$-$\beta$ plane becomes fractal.  For the discrete model, these results
are exact.

At these critical theories, the one-dimensional field theories describing
dissipative quantum mechanics correspond to solutions of open string
theory \OSDQM.  In the presence of open-string background fields,
interactions
between a string and the background take place at the boundary of the string,
and their effects can be represented by a boundary state $|B\rangle$.  Such a
boundary state can be calculated in terms of a functional integral over the
action $S$, where $M$ acts as a cutoff which must be taken to 0;
$\eta$ is $1/(2\pi\alpha')$, where $\alpha'$ is the string constant;
and $A_\mu({\svec X})$
and $V({\svec X})$ are the gauge fields and tachyon fields, respectively.  For
more details, we refer the reader to \OSDQM\ and \CLNY.

Specializing to the dissipative Hofstadter model, we see that the Euclidean
action is the sum of a quadratic piece and a more complicated potential term,
\eqn\last{S = S_q + S_V\ \ ,}
where
\eqn\last{{1\over \hbar} S_q
= {1\over\hbar} \int^{T/2}_{-T/2} dt \left\{ {M\over 2}
\dot{\svec X}^2 + {ieB\over 2c} (\dot xy - \dot yx) + {\eta\over 4\pi}
\int^\infty_{-\infty} dt' \left( {\left({\svec X}(t) -{\svec
X}(t')\right)^2\over(t-t')^2}\right)\right\},}
and
\eqn\last{S_V = \int^{T/2}_{-T/2} dt\, V(x,y)\ \ .}
For the periodic potential we take
\eqn\last{V(x,y)
= -V_0 \cos \left( {2\pi x(t)\over a}\right) - V_0 \cos \left( {2\pi
y(t) \over a}\right)\ \ .}
It is convenient to define the dimensionless parameters
\eqn\last{2\pi\alpha = {\eta a^2\over\hbar}\ \ ,\qquad 2\pi\beta
= {eB\over \hbar c} a^2,}
and to rescale $x$ and $y$ by $a/2\pi$ and $V_0$ by $\hbar$.  In
\PD\ and \CT\ we have shown that this model is
critical whenever $\alpha/(\alpha^2+\beta^2)=1$
and $\beta/\alpha\in{\IZ}$; and we expect these points to be multi-critical
points.  These critical theories will be the main focus of this paper.

Because the ordinary kinetic term ${1\over 2} M\dot{\svec X}^2$ is a
dimension-two operator, it is irrelevant and acts only as a regulator as far
as the large-time behavior is concerned.  Since we are studying behavior at
the critical theories, it should be legitimate to set $M=0$ and use some
other, more convenient regulator instead.  In fact, we actually will have to
be quite careful about our handling of the regulator.  This is because the
correlation functions we calculate will contain contact terms.  When we change
the regulator, these contact terms also change, so we must use a  consistent
treatment of the regulator throughout all our calculations.
In addition, if the contact terms are to have any
physical meaning they must be determined by large-distance properties of the
system
and by symmetries of the system.  One set of
symmetries we would like these theories to obey is the broken
reparametrization invariance Ward identities \RIWI.  These identities are
conditions which the one-dimensional field theory described by the action in
Eq.~\Sdef\ must satisfy if it is to be a solution of open string theory.
They follow
from the reparametrization invariance of the actual boundary state
$|B\rangle$.  In a forthcoming paper,\DEF\ we will show that whenever
$\alpha/\left(\alpha^2+\beta^2\right)=1$ and $\beta/\alpha\in{\IZ}$,
the dissipative Hofstadter model does indeed satisfy these Ward identities.
In the proof, we set the
mass term to zero and instead use a high-frequency cutoff in which we multiply
the bosonic propagator due to the quadratic action, $S_q$, by
$e^{-\epsilon|\omega|}$ where $\epsilon$ is a dimensionless cut-off and
$\omega$ is the frequency.  Because we hope to use the Ward identities as a
means of fixing the contact terms, and because we are also interested in the
solutions to open string theory which obey these identities, in this paper we
will use the regulator that we know satisfies these Ward identities.
However, most of our proofs in the following section will be general enough to
be applicable to
other regulators, including the fermionic regulator which
will be defined below.

In order to calculate the free energy and correlation
functions of the dissipative Hofstadter model, we will make use of the Coulomb
gas expansion, whose derivation is described in detail in \PD\ and
\CT.  To
obtain this expansion, we treat the cosine potential in $S_V$ as a
perturbation and calculate all expressions in terms of the propagator
due to $S_q$.  After Taylor expanding $\exp \int\left[ \cos
x(t) + \cos y(t)\right]dt$, and writing
\eqnn\cosx
\eqnn\cosy
$$\eqalignno{\cos x(t) &= {1\over 2} \left( \exp \left( \hat x \cdot {\svec
X}(t)\right) + \exp \left( - \hat x \cdot {\svec X}(t)\right)\right)
&\cosx \cr\noalign{\hbox{and}}
\cos y(t) &= {1\over 2}\left( \exp \left( \hat y \cdot {\svec X}(t)\right) +
\exp \left( - \hat y \cdot {\svec X}(t)\right)\right)\ \ ,&\cosy \cr}$$
we find that the partition function is given by
\eqn\Zdef{\eqalign{Z &= \exp \left( - {1\over\hbar}S\right) \cr
&= \sum^\infty_{n=0} \int dt_1 \ldots dt_n \left( {V_0\over 2}\right)^n {1\over
n!}
\sum_{{{\svec q} = \pm \hat x}\atop{\pm\hat y}} \ll \prod^n_{j=1} e^{i{\svec
q}_j\cdot {\svec X}(t_j)} \rr_0\ \ , \cr}}
The correlation functions in Eq.~\Zdef\ are evaluated with
respect to the regulated propagators
\eqn\Gdiag{\eqalign{G^{\mu\nu} (t_1-t_2) &=
{\alpha\over\alpha^2+\beta^2} \sum_{m\not=0} {1\over
|m|} e^{im(t_1-t_2)2\pi/T} e^{-\epsilon|m|} \cr
&=  {\alpha\over\alpha^2 + \beta^2} \ln \left[ {z_1z_2\over \left( z_1 -
e^{-\epsilon} z_2\right) \left( z_2 - e^{-\epsilon}z_1\right)}\right]
\cr}}
for $\mu=\nu$; and
\eqn\Goffd{\eqalign{G^{\mu\nu} (t_1-t_2) &= -\epsilon^{\mu\nu} {\beta\over
\alpha^2+\beta^2} \sum_{m\not=0} {1\over m} e^{im(t_1-t_2)2\pi/T}
e^{-\epsilon|m|} \cr
&= \epsilon^{\mu\nu} {\beta\over \alpha^2 + \beta^2}\ln \left[ {z_1 - z_2
e^\epsilon\over z_2 - z_1 e^\epsilon} \cdot {z_1\over z_2}\right] \cr}}
for $\mu\not=\nu$.  In these equations, we have defined $z_j = e^{2\pi
it_j/T}$.

In these expressions for $G^{\mu\nu}$, we have taken care of the
infrared divergences by putting the time coordinate, $t$, on a circle of
circumference $T$, and we have introduced an ultraviolet cutoff by
multiplying the Fourier-space propagator by $e^{-\epsilon|m|}$.  When we take
$\epsilon$ and $T$ to zero in the sums in Eqs.~\Gdiag\ and \Goffd,
we find the propagator has the simpler form
\eqn\Gsim{G^{\mu\nu} (t_i - t_j)
= {\alpha\over\alpha^2 + \beta^2} \ln (t_i - t_j)^2
\delta^{\mu\nu} - {i\over 2}\ {2\pi\beta\over \alpha^2 +
\beta^2}\hbox{sign}(t_i - t_j) \epsilon^{\mu\nu} \ \ .}

A subtlety to be born in mind is that for the dissipative quantum mechanics
system, we must integrate over the zero mode in Eq.~\Zdef.  This imposes the
neutral-charge requirement, $\Sigma {\svec q}_j=0$.  For calculating the
$\beta$-function and the string-theory boundary state path integral,
we omit the integration over the zero mode, and the $q_i$'s are unconstrained.
In this paper, we will concentrate only on the zero-charge sector.  However,
many of our methods and results can be extended to the charged sector.

For the connected correlation functions, similar calculations yield
\eqn\cordef{\ll \prod_i {\cal O}(t_i)\rr^{\rm con}
   = \sum^\infty_{n=0} \int dt_1\ldots dt_n
\left( {V_0\over 2}\right)^n {1\over n!} \sum_{{\svec q}=\pm\hat x,\hat y} \ll
\prod_i {\cal O}(t_i) \prod_{j=1}^n
e^{i{\svec q}_j \cdot {\svec X}(t_j)}\rr_0^{\rm con} \ \ ,}
with a similar charge conservation condition.

Equations \Zdef\ and \cordef\ for the partition function and correlation
functions in terms of the propagator $G^{\mu\nu}$ will be the starting points
for our proofs in the following sections.  Apart from replacing the
$M\dot{\svec X}^2$ term with another cutoff, we have made no approximations
in obtaining these expressions.
In the following sections it is assumed that we are using the renormalization
scheme of Ref.~\CT.  In this prescription,
we hold $V_R = {V_0 T\epsilon}$ finite as we
take $\epsilon \to 0$, and the divergence of the free energy can be removed by
a constant shift to the action which depends on
$V_R$ and $\beta/\alpha$.
\newsec{FERMIONIZATION AND DUALITY SYMMETRY}
In this section we describe the fermionization of the free energy and the
duality
properties of the critical dissipative Hofstadter model. The material in this
section directly builds on references \PD\ and \CT,
to which we refer the reader for
further information.  When the magnetic field is zero and $\alpha=1$,
we can actually fermionize the whole theory, and then, upon regulating
the fermionic propagators, solve for the free energy and all of the
correlation functions of this theory.  We find that all of the coordinate
correlation functions are contact terms, except for the two-point function.
(Similarly, in \GHM, the Hamiltonian for the tight-binding version of
the DHM is also shown to be equivalent to one for independent fermions and
therefore exactly solvable.)

In Ref.~\PD\ we show that there is a duality symmetry which relates the
model when $\beta/\alpha\in\IZ$ to the model at $\beta=0$.  From this
symmetry, we can derive transformation rules that give all the correlation
functions and the free energy at $\beta/\alpha\in\IZ$ in terms of those when
$\beta/\alpha=0$.  When $\beta/\alpha\not=0$, calculations in Ref.~\CT\
show that some of the correlation functions are finite at large times and
consequently
are not just contact terms.  This relation between the contact terms at zero
magnetic field and finite correlation functions at non-zero magnetic field
suggest that the contact terms are physical and not just a relic of the
regularization.   Furthermore, the connection to the solvable fermionized
theory suggests that the critical DHM with $\beta/\alpha\in\IZ$ and
$\alpha/\left( \alpha^2+\beta^2\right)=1$ is exactly
solvable.   There are two problems with these conclusions.  The first is that,
in general, fermionization and bosonization are correct for large distance
behavior, but not necessarily for contact terms.
As a result, the fermionized theory is
not necessarily the same as the original theory, in which the bosonic
propagators, $\ll X^\mu (t) X^\nu(0)\rr$
are regulated in a consistent way and for which the Ward
identities are satisfied. The second problem is that even though the proof of
the duality symmetry was exact for a discrete version of the DHM, in the
derivation of the symmetry for the continuous DHM some subtleties
about the regulator
were ignored.  Therefore, we do not know if we can rely on this symmetry to
express contact terms and finite terms as  functions of one another.

In this section, we will first show how the free energy fermionizes when
$\beta/\alpha\in\IZ$ and the
cutoff, $\epsilon$, is taken to zero.
Next, when the magnetic field is not regulated,
we will derive the duality transformation relating correlation functions
 at $\beta=0$ with those
at $\beta/\alpha\in\IZ$.  The remaining sections will be
devoted to showing to what extent these properties are still true when we
include the full bosonic regulator.
\goodbreak
\noindent{\bf Fermionization}
\medskip
\nobreak
Using the properties of Gaussian propagators, we can write the ${\cal
O}(V^n_0)$ term in expression \Zdef\ for the partition function as a sum over
${\svec q} = \pm\hat x$ and $\pm\hat y$ of
\eqn\Hndef{
H_n = {1\over n!} \left( {V_0\over 2} e^{-{1\over 2}\ll x^2(0)\rr}\right)^n
\int \prod^n_{i=1} dt_i \exp \left[ - \sum_{i<j} q^\mu_i q^\nu_j G^{\mu\nu}
\left(t_i-t_j\right)\right] \ \ .}
The object in the exponent can be viewed as the energy of a generalized
Coulomb gas.  In this gas, there are two species of particles, one
corresponding to the $x$-component of ${\svec q}$ and one corresponding to the
$y$-component of ${\svec q}$.  Each type of particle can have either a
negative charge or a positive charge.  For finite $T$, the particles live on a
circle, and for $T\to\infty$, they live on a line.  Ignoring the regulators,
according to Eq.~\Gsim, particles of the same species interact
logarithmically, and particles of differing species interact via a sign
function.  Lastly, the particles have fugacity given by $\left( {V_0\over 2}
e^{-{1\over 2}\ll x^2(0)\rr}\right)$, and the condition $\Sigma{\svec q}=0$
requires the gas to be neutral.  Because we are taking the exponential of this
energy function, the sign interaction becomes a phase.  This means that the
only effect of the magnetic (off-diagonal) interaction is that
whenever the positions, $t_1$ and $t_2$, of two particles of differing
species are interchanged,
the wave
function picks up the phase $e^{\pm i2\pi\beta/\left(\alpha^2+\beta^2\right)}$.
  On the
critical circle, we have $\beta/\left(\alpha^2+\beta^2\right)
= \beta/\alpha$, so whenever
$\beta/\alpha\in\IZ$, this phase equals 1.  We conclude that the magnetic
field ought to have no effect at all on the free energy or partition function
of these theories.  We shall show momentarily that this remains
true even when we
put the particles on a circle and regulate the diagonal part of $G^{\mu\nu}$.

According to the regulated expressions for $G^{\mu\nu}$, \Gdiag\ and \Goffd,
the expression, $H_n$, for the terms in the perturbation
series for the partition function has the form
\eqn\Hdef{
H_n = {1\over n!} \left[ {V_0T\over 2} \left( e^\epsilon + e^{-\epsilon} - 2
\right)^{1/2} e^{-\epsilon/2} \right]^n
\oint \prod^n_{\ell=1} {dz_\ell\over 2\pi i z_\ell}
Z_{n_x,n_y}}
where
\eqn\Znxny{
Z_{n_x,n_y} = \prod^n_{i<j=1} \left[ - {e^\epsilon z_iz_j\over \left( z_i -
e^\epsilon z_j\right)\left( z_i - e^{-\epsilon} z_j\right)}\right]^{\zeta_{ij}}
\left[ {z_i - z_j e^\epsilon\over z_j - z_i e^\epsilon}\cdot {z_i\over
z_j}\right]^{\eta_{ij}} \ \ .}
On the critical circle, $\alpha/(\alpha^2+\beta^2) = 1$, we have
\eqn\xieta{\zeta_{ij}
= - {\svec q}_i\cdot {\svec q}_j\qquad\hbox{and}\qquad\eta_{ij} = -
q^\mu_i q^\nu_j \epsilon^{\mu\nu} {\beta\over\alpha}\ \ .}
Also, we have defined $n_x=\sum^n_{i=1} q^x_i$ to be the number of
$x$-particles and $n_y = \sum^n_{i=1} q^y_j$ to be the number of $y$-particles,
so that $n = n_x+n_y$.  We can factor $Z_{n_x,n_y}$ into three parts.  The
first is due to only the $x$-particles interacting with each other.  The
second is due to only the $y$-particles interacting with each other, and the
third is due to the $x$- and $y$-particles interacting with each other via the
magnetic field interaction.  To do this, we will let $z_j$ for $j=1$ to $n_x$,
be the ``position'' of the $x$-particles and $w_j$, for $j=1$ to $n_y$, be the
position of the $y$-particles. Also, we will let $q_j = \pm 1$ and $p_j=\pm 1$
be the charges of the $x$- and $y$-particles, respectively.  Then the
integrand, $Z_{n_x,n_y}$, for the partition function is
\eqn\ZZC{Z_{n_x, n_y}
= Z_{n_x} (z) Z_{n_y} (w) C_{n_x,n_y} (z,w)\ \ ,}
where
\eqnn\Znxdef
\eqnn\Znydef
\eqnn\Cnxnydef
$$\eqalignno{ Z_{n_x} (z) &= \prod^{n_x}_{i<j=1} \left[ -{e^\epsilon z_i
z_j\over \left( z_i - e^\epsilon z_j\right)\left( z_i - e^{-\epsilon}
z_j\right)}\right]^{-q_i q_j}\ \ ;&\Znxdef\cr
Z_{n_y} (w) &= \prod^{n_y}_{i<j=1} \left[-{e^\epsilon w_iw_j\over \left( w_i -
e^\epsilon w_j\right)\left( w_i - e^{-\epsilon} w_j\right)}\right]^{-p_ip_j}
\ \ ;&\Znydef\cr\noalign{\hbox{and}}
C_{n_x,n_y} (z,w) &= \prod_{i,j} \left( {z_i - w_j e^\epsilon\over
w_j-e^\epsilon z_i}\ {z_i\over w_j}\right)^{-q_i p_j(\beta/\alpha)}\ \ .
&\Cnxnydef \cr}$$
Because we are considering only the neutral gas
case, we have $\sum^{n_x}_{i=1} q_i = \sum^{n_y}_{j=1} p_j=0$.

First, we note that $Z_{n_x}$ and $Z_{n_y}$ are the integrands for the
partition functions for the $x$- and $y$-particles in the absence of the
magnetic field.  All the information about the magnetic field is contained in
$C_{n_x,n_y}$.  If we set $\epsilon$ to zero in $C_{n_x,n_y}(z,w)$, we obtain
\eqn\Cze{C_{n_x,n_y} (z,w) = \prod_{i,j}
\left( e^{-i\pi\,{\rm sign}\left[(t_i-s_j)
\pi/T\right]} {z_i\over w_j}\right)^{-q_ip_j\beta/\alpha}\ \ ,}
where $z_i = e^{2\pi i t_i/T}$ and $w_j = e^{2\pi i s_j/T}$.  (In order to get
this equation we must actually be very careful about how we treat the branch
cuts that are introduced in Eq.~\Goffd\ when we set $\epsilon$ to zero.)  If
we use the fact that $\Sigma q_i = \Sigma p_j=0$, and that ${\rm sign}(x) =
\theta(x) - \theta(-x)$, we can write $C_{n_x,n_y}$ as
\eqn\Cphase{\eqalign{C_{n_x, n_y}(z,w) &= \exp \left\{ i\pi {\beta\over\alpha}
\sum_{i,j} q_i p_j\,{\rm sign}\left[ (t_i - s_j) {\pi\over T}\right]
\right\} \cr
&= \exp \left\{ i2\pi {\beta\over\alpha}\sum_i q_i\left[\sum_{s_j<t_i}
p_j\right] \right\} \ \ ,\cr}}
as long as $t_i\not=s_j$.  Because $q_i$
and $p_j$ are integers, whenever $\beta/\alpha\in\IZ$,
$C_{n_x,n_y}$ just reduces to 1.  Ignoring questions of what happens if
$t_i=s_j$, we can conclude that when $\beta/\alpha\in\IZ$ and we do not use an
ultraviolet cutoff for the magnetic field propagator, the perturbative part of
the partition function is the same as when there is no magnetic field.  In
that case, $Z_{n_x,n_y}$ is given by $Z_{n_x,n_y}=Z_{n_x}(z)Z_{n_y}(w)$.  This
is equivalent to the statement of the duality transformation of the free energy
when $\beta/\alpha$ is an integer.

Next, we will show that, for a neutral
gas, $Z_{n_x}$ (or $Z_{n_y}$) looks like a partition function for a gas of
fermions.  If we set $\epsilon$ to zero in Eq.~\Znxdef, we obtain
\eqn\Znxze{Z_{n_x} (z)
= \prod^{n_x}_{i<j=1} \left[ - {z_i z_j\over (z_i - z_j)^2}
\right]^{-q_iq_j}\ \ .}
Because the gas is neutral, $n_x$ must be even, so call $n_x=2N$.  Then we
will define $z_j = e^{2\pi i t_j/T}$ and $w_j = e^{2\pi i s_j/T}$, \
for $1\le j\le
N$, where now the positive charges lie at $t_j$ and the negative charges lie
at $s_j$.  With these definitions, $Z_{n_x}$ becomes
\eqn\Ztn{Z_{2N}\bigg|_{\epsilon\to 0}
= (-1)^N {\prod\limits^N_{i<j=1} (z_i - z_j)^2
 \prod\limits^N_{i<j=1}
(w_i - w_j)^2  \over \prod\limits^N_{i,j=1} (z_i-w_j)^2} \prod^N_{i=1} z_i
w_i\ \ .}

This expression can be ``simplified'' with the aid of the following identity:
\eqn\fermid{{\prod\limits_{i<j} (z_i-z_j) \prod\limits_{i<j} (w_i-w_j) \over
\prod\limits_{i,j} (z_i - w_j)} = \pm \det M (z,w) \ \ ,}
where $M$ is a matrix defined by
\eqn\Mdef{M_{ij} = {1\over z_i - w_j}\ \ .}
This identify can be proved by using partial fractions or properties of
determinants (see \Muir, section 353).
It is also generalizable to the case
when the number of $z$'s and $w$'s are unequal.  This
generalization can be used to analyze the charged sector of the theory.
Using identity \fermid, we obtain for $Z_{2N}$,
\eqn\last{Z_{2N}\bigg|_{\epsilon\to0} = (-1)^N
\left[ \det M (z,w)\right]^2
\prod^N_{i=1} z_i w_i\ \ .}
We can actually take the identity \fermid\ one step further and write
\eqn\eqnMM{\eqalign{
(-1)^N&{\prod\limits_{i<j} (z_i-z_j)^2 \prod\limits_{i<j} (w_i-w_j)^2 \over
\prod\limits_{i,j} (z_i - e^\epsilon w_j)
\prod\limits_{i,j}(z_i-e^{-\epsilon}w_j)}
\prod^N_{i=1} z_iw_i \cr
= (-1)^N&\det M (z,e^\epsilon w)
\det M (z,e^{-\epsilon}w) \prod^N_{i=1} z_iw_i .\cr}}
The left-hand side of this equation is equal to $Z_{2N}$ if we set all the
$\epsilon$'s in the numerator to 0, and keep all the $\epsilon$'s in the
denominator.  It corresponds to a Feynman diagram for a neutral gas where every
pair of like-charged particles is joined by a propagator $-(z_i-z_j)^2/z_iz_j$,
and every pair of oppositely charged particles is joined by the propagator $-
z_iw_j/(z_i-e^\epsilon w_j)(z_i-e^{-\epsilon}w_j)$.  The right-hand side of
the equation is a product of Slater determinants.  Each determinant of $M$
is a sum over diagrams in which each particle is joined to exactly one
particle of the opposite charge by the propagator $1/\left(z_i-
e^{-\epsilon}w_j\right)$  or
$1/\left(z_i-e^\epsilon w_j\right)$ for $M(z,e^{-\epsilon}w)$ or
$M(z,e^\epsilon w)$,
respectively.  Then, the right-hand side of Eq.~\eqnMM\ corresponds to the
regulated Feynman diagrams of a fermion gas where each particle is joined to
exactly two particles of opposite charge via the regulated, fermionic
propagator $i\sqrt{zw}/\left(z-e^{-\epsilon}w\right)$ or $i\sqrt{zw}/\left(
z-e^\epsilon w\right)$; and
there is a factor of minus one for every closed loop.  Because all the
interactions are quadratic, we can completely solve this fermionized theory
(see Ref.~\PD).

Even when we look at correlation functions containing
$\dot x(t)$'s, it is still possible to use partial fractions to express all
the diagrams in the perturbation series in terms of diagrams for a gas of
fermions with quadratic interactions.  However, we will not make use of it
here because the treatment of the regulator in this derivation ruins several
properties of the original $\dot x(t)$ correlation functions,
especially the form
of the contact terms.  We do point out, though, that this  theory in
which the fermionic propagator is regulated should give the same
properties for the long-distance behavior as those
of the original theory in which the
bosonic propagator was regulated.  However, because the fermionized theory
does not treat the bosonic propagators in the numerator and denominator of
$Z_{2N}$ on equal footing with regard to regulators, it changes the relative
normalization of different graphs (including setting many to 0), and
it does not appear to satisfy the Ward identities.
(Of course, if we knew which
counterterms to add, it might again satisfy the Ward identities.  However we
do not know how to proceed in that direction, and,
in any case, the theory will no
longer be so much simpler than the bosonic theory.)  The fact that regulating
the already non-divergent numerator can have such a big effect may seem
surprising.  The catch is that we are multiplying the factors of $\epsilon$ in
the numerator by $1/\epsilon$ poles in the denominator, so the two may cancel
to give finite answers.  This can also have an effect on the short-distance
behavior of the correlation functions.  Therefore, if we hope to solve for
contact terms, we must stick with the regulator that will satisfy the Ward
identities and, therefore, we must regulate the numerator of $Z_{2N}$.

In the next section we will use identity \eqnMM, which is the foundation
for the fermionization, as a starting point for deriving the Feynman rules for
the free energy when the bosonic propagators are regulated.
\goodbreak
\bigskip
\noindent{\bf Duality Transformation}
\medskip
\nobreak
We will  now turn our attention to the calculation of the correlation
functions and the duality transformation.  For simplicity we will consider
only correlation functions of the dimension one operators $\dot x(t)$ and
$\dot y(t)$.
However, most of these calculations can also be directly applied to
correlation functions of the dimension-one operators $e^{\pm ix(t)}$ and
$e^{\pm iy(t)}$.  With some additional work, these calculations can
be generalized to include higher-dimensional operators.

With the use of properties of Gaussian propagators, Eq.~\cordef\ for the
connected $m$-point function of order $V^{2n}$ becomes
\eqn\Ctndef{
\eqalign{C_{2n} (r_1,\ldots,r_m) &= \ll \dot x^{\mu_1} (r_1)
\ldots \dot x^{\mu_m} (r_m)
\rr^{\rm con} \cr
&= c_{2n} \sum_{\{{\svec q}_i\} = \pm \hat x,\pm \hat y} \int \prod^{2n}_{i=1}
dt_i Z'_{n_x,n_y} R_{2n} (r_1,\ldots, r_m) \ \ ,\cr}}
where
\eqnn\ctndef
\eqnn\Zpdef
$$\eqalignno{c_{2n} &= {1\over (2n)!} \left({V_R\over 2}\right)^{2n}\ \ ;
&\ctndef\cr
Z'_{n_x,n_y} &= \ll e^{i{\svec q}_1\cdot {\svec x}(t_1)}\ldots e^{i{\svec
q}_{2n} \cdot {\svec x} (t_{2n})}\rr^{\rm con}_0\ \ ;&\Zpdef\cr}$$
and
\eqn\Rmndef{
R_{2n} (r_1,\ldots, r_m) = \prod^m_{j=1} \sum^{2n}_{k=1} {\svec q}_k \cdot
\ll \dot x^{\mu_j} (r_j) {\svec x} (t_k) \rr_0\ \ .}
The subscript $0$ in Eqs.~\Zpdef\ and \Rmndef\ denotes that the correlation
functions are calculated in terms of the regulated free propagator
$G^{\mu\nu}(t_1-t_2)$ given by Eqs.~\Gdiag\ and \Goffd.
In Eq.~\Zpdef, and throughout the remainder of the paper, we will omit the
self-interactions of the operators in the free correlation functions.
Instead, we will include their effects in
the renormalized potential, $V_R = 2 V_0 T \sinh(\epsilon/2)$.
According to Eqs.~\Zdef\ and \Hdef,
$Z'_{n_x,n_y}$ is the connected part of
$Z_{n_x,n_y}$, which we have just
analyzed.  According to Ref.~\CT, when $m>2$,
$C_{2n}(r_1,\ldots, r_m)$ will
equal zero unless at least two points are coincident.  Because delta-functions
are well-defined functions in Fourier space, but rather singular in real
space, we will calculate the Fourier transform of $C_{2n}$ instead.
Taking the Fourier transform of $C_{2n}$ with respect to $r_1,\ldots, r_m$,
we have
\eqn\Cft{\tilde C_{2n} \left( \ell_1,\ldots,\ell_m\right)
= c_{2n} \sum_{{\svec q}_i}
\int \prod^m_{j=1} {dr_j\over T} e^{-i\ell_j r_j {2\pi\over T}} \int
\prod^{2n}_{i=1} dt_i R_{2n} Z'_{n_x,n_y}\ \ .}

The only $r$-dependence in the original expression for the correlation
function was in $R_{2n}$, so we will restrict our attention first to
\eqn\Rft{
\tilde R(\ell_1,\ldots,\ell_m) = \int {dr_1\over T}\ldots {dr_m\over T}
R_{2n}(r_1,\ldots,r_m) \prod^m_{j=1} e^{-i\ell_j r_j 2\pi/T}\ \ .}
Upon substituting Eq.~\Rmndef\ into this equation for the Fourier transform
of $R_{2n}$ and rearranging the order of the terms, we obtain
\eqn\Rcalc{
\tilde R^{\mu_1\ldots\mu_m}\left( \ell_1,\ldots, \ell_m\right) =
\prod^m_{j=1} \sum^{2n}_{k=1} {\svec q}_k\cdot \int {dr_j\over T} e^{-i\ell_j
r_j 2\pi/T} \ll \dot x^{\mu_j} (r_j) {\svec x} (t_k) \rr_0 \ \ ,}
where the $j^{\rm th}$ factor in the product depends only on $r_j$.
Taking a derivative of Eqs.~\Gdiag\ and \Goffd, we can write $\ll \dot
x^\mu(r) x^\nu(t) \rr_0 = {d\over dr} G^{\mu\nu} (r-t)$ as the Fourier series
\eqn\xdotx{
\ll \dot x^\mu(r) x^\nu(t)\rr_0 = {2\pi i\over T} \
{\alpha\over\alpha^2+\beta^2} \sum^\infty_{{m=-\infty}\atop{m\not=0}} {\rm
sign}(m) r^{\mu\nu} (m) e^{im(r-t) 2\pi /T}e^{-\epsilon|m|}\ \ ,}
where
\eqn\rmndef{
r^{\mu\nu} (m) = \delta^{\mu\nu} - {\beta\over\alpha}{\rm sign}(m)
\epsilon^{\mu\nu}\ \ .}
We will find it useful to define ${\svec r}^\mu(m)$ to be
\eqn\rmudef{
{\svec r}^\mu (m) = r^{\mu x}(m) \hat x + r^{\mu y} (m) \hat y\ \ .}
The integral over $r_j$ in the expression for $\tilde R$, given by
\eqn\rjint{
\int {dr_j\over T} e^{-i\ell_j r_j 2\pi/T}
\ll\dot x^{\mu_j} (r_j) {\svec x}(t_k)
\rr_0\ \ ,}
picks out the $\ell^{\rm th}_j$ term in the Fourier series of $\ll\dot
x^{\mu_j}(r_j) {\svec x}(t_k)\rr_0$.  Therefore, according to Eq.~\xdotx,
after we perform the integrals over $r_j$ in Eq.~\Rcalc, $\tilde R$ becomes
\eqn\fullR{
\tilde R^{\mu_1\ldots \mu_m} \left( \ell_1,\ldots,\ell_m\right) =
a_m
\prod^m_{j=1}\sum^{2n}_{k=1} {\rm sign} (\ell_j) {\svec q}_k\cdot {\svec
r}^{\mu_j} (\ell_j) e^{-i\ell_j t_k(2\pi/T)}e^{-\epsilon|\ell_j|}\ \ ,}
where $a_m$ is given by
\eqn\amdef{a_m= \left( {2\pi i\over T}\right)^m
\left( {\alpha\over\alpha^2+\beta^2}\right)^m\ \ .}

When there is no magnetic field, the connected part of $Z_{n_x,n_y}$ has
either all $e^{\pm ix(t)}$'s or all $e^{\pm iy(t)}$'s, because without a
magnetic field
the $x$-particles and $y$-particles do not interact.  This means that when
$\beta=0$, all the ${\svec q}_k = \pm \hat x$ or they are all $\pm\hat y$.  For
example, take all the ${\svec q}_k = q_k\hat x$ with $q_k=\pm1$.  Then $\tilde
R^{\mu_1\ldots \mu_m}(\ell_1,\ldots, \ell_m)=0$ unless all the $\mu_i=x$.  In
that case $\tilde R$ is given by
\eqn\Rzerom{
\tilde R \left(\ell_1,\ldots,\ell_m;0\right) = a_m \prod^m_{j=1}
\sum^{2n}_{k=1} q_k {\rm sign} (\ell_j) e^{-i\ell_j t_k(2\pi/T)}
e^{-\epsilon|\ell_j|}\ \ ;}
and $Z'_{n_x,n_y}$ is equal to $Z'_{2n}$, the connected part of $Z_{2n}$.
(In Eq.~\Rzerom\ we have added the last argument in $\tilde R$ to denote the
value of $\beta/\alpha$ at which it is evaluated.)  As
an example, consider the case when $2n=2$.  Upon rearrangement of the
terms, $\tilde R\left( \ell_1,\ldots, \ell_m\right)$ is
\eqn\Rtnt{
\eqalign{\tilde R\left(\ell_1,\ldots, \ell_m;0\right) &= a_m
\prod^m_{j=1} {\rm sign}(\ell_j) e^{-\epsilon|\ell_j|} \cr
&\times\sum^m_{M=0} \sum_{\sigma_M}
\prod_{j\in\sigma_M}q_1e^{-i\ell_jt_1} \times \prod_{k\notin\sigma_M} q_2
e^{-i\ell_k t_2}\ \ ,\cr}}
where $\sigma_M$ is summed over all subsets of $\{1,2,\ldots,m\}$ with $M$
elements.

For the general case, we have
\eqn\Rgc{\eqalign{
\tilde R \left( \ell_1,\ldots, \ell_m;0\right) &= a_m \prod^m_{j=1}
\left({\rm sign}(\ell_j) e^{-\epsilon|\ell_j|} \right) \cr
&\times \sum_{(\sigma_1,\ldots, \sigma_{2n})} \prod_{j_1\in \sigma_1} q_1
e^{-i\ell_{j_1}t_1} \prod_{j_2\in\sigma_2} q_2 e^{-i\ell_{j_2} t_2} \ldots
\prod_{j_{2n}\in\sigma_{2n}} q_{2n} e^{-i\ell_{j_{2n}} t_{2n}}\ \
,\cr}}
where the sum is over all partitions,
$(\sigma_1,\ldots,\sigma_{2n})$, of $\{ 1,2,\ldots, m\}$ into $2n$ ordered
sets.  If we substitute this back into the expression for $\tilde C_{2n}
\left( \ell_1,\ldots,\ell_m;0\right)$, we obtain
\eqn\Czmag{
\tilde C_{2n} \left( \ell_1,\ldots, \ell_m;0\right) = a_m c_{2n}
\prod^m_{j=1}\left( {\rm sign} (\ell_j) e^{-\epsilon|\ell_j|}\right)
\sum_{{ \{ q_i\}=\pm1}\atop{\Sigma q_i = 0}}
\sum_{(\sigma_1,\ldots,\sigma_{2n})} I\left( Z'_{2n}\right) \prod^{2n}_{k=1}
q^{|\sigma_k|}_k \ \ ,}
where $|\sigma_k|=$ number of elements in $\sigma_k$ and
\eqn\Idef{I\left( Z'_{2n}\right) =
\int \prod^{2n}_{i=1} dt_i\, Z'_{2n} \prod^{2n}_{k=1} \exp\left\{-i\left(
\sum\limits_{j\in\sigma_k} \ell_j\right) t_k 2\pi/T\right\}\ \ .}
$I(Z'_{2n})$ is equal to the Fourier transform of
$Z'_{2n} = \ll e^{iq_1x(t_1)}\ldots e^{iq_{2n} x(t_{2n})}\rr^{\rm con}_0$
with respect to the Fourier-space
variables, ${\svec k} = (k_1,\ldots,k_{2n})$, given by
\eqn\kdef{{\svec k} = \left(\sum_{j\in\sigma_1} \ell_j, \sum_{j\in \sigma_2}
\ell_j,\ldots,\sum_{j\in\sigma_{2n}}\ell_j\right)\ \ .}
Therefore, $\tilde
C_{2n}(\ell_1,\ldots, \ell_m)$ is just a weighted sum over Fourier
coefficients of $Z'_{2n}$.   In the
following sections we will address the calculation of the Fourier series of
$Z'_{2n}$, and we will prove that it always equals a constant plus a finite,
piecewise-linear, homogeneous function of the Fourier variables.

With a non-zero magnetic field, the equations are more complicated.  If we
keep the magnetic field in Eq.~\fullR\ and repeat our calculations, we find
that now
\eqn\Cmag{
\eqalign{\tilde C^{\mu_1\ldots\mu_m}_{2n} \left(\ell_1,\ldots,
\ell_m;{\beta\over\alpha}\right) &= a_m c_{2n} \prod^m_{j=1} \left[{\rm
sign}(\ell_j) e^{-\epsilon|\ell_j|}\right] \cr
&\times \sum_{\{q_i\}=\pm\hat x\pm\hat y}
\sum_{(\sigma_1,\ldots ,\sigma_{2n})} I (Z'_{n_x,n_y})
\prod^{2n}_{k=1} \prod_{i\in\sigma_k}
{\svec r}^{\mu_i} (\ell_i) \cdot {\svec q}_k\ \ ,\cr}}
where $a_m$, $c_{2n}$ and $I$ are the same as before, except that in this
equation $I$ depends on $Z'_{n_x,n_y}$ instead of $Z'_{2n}$.  When
$\beta/\alpha\in \IZ$ and we do not regulate the magnetic field propagator in
$Z'_{n_x,n_y}$, we can make some simplifications.  As we discussed earlier, in
that case $Z_{n_x,n_y}$ looks the same as when $\beta=0$;  the $x$-
and $y$-particles do not interact; and the connected part of $Z_{n_x,n_y}$ has
only all $x$-particles or all $y$-particles.  Consequently, if $n_x+n_y=2n$ we
have
\eqn\ZnxZny{ Z'_{n_x,n_y}
= \cases{ 0 & if $n_x\not=2n$ and $n_y\not=2n$\cr\noalign{\vskip 0.2cm}
Z'_{n_x} & if $n_x=2n$ \cr\noalign{\vskip 0.2cm}
Z'_{n_y} & if $n_y = 2n$\ \ . \cr}}
It follows that when we sum over the ${\svec q}$'s to get $Z'_{n_x,n_y}$, the
only values for ${\svec q}$ that do not give zero occur when all the ${\svec
q}$'s are $\pm\hat x$ or all are $\pm\hat y$.  We will define ${\svec q}_i =
q_i \hat e_\nu$, where $\hat e_\nu = \hat e_x \equiv \hat x$
when all the $q$'s are
$\pm\hat x$ and $\hat e_\nu = \hat y$ when all the
${\svec q}$'s are $\pm\hat y$.  Then
the sum over ${\svec q_i}$ in $\tilde C$ reduces to
\eqn\qsum{\eqalign{
&\sum_{\{q_i\}=\pm1}\sum_{\hat e_\nu=\hat x,\hat y} \sum_{(\sigma_1,\ldots,
\sigma_{2n})} \prod^{2n}_{k=1}\left[ \prod_{i\in\sigma_k} {\svec r}^{\mu_i}
(\ell_i)\cdot {\svec q}_k\right]
I \left( Z'_{n_x,n_y}\right) \cr
= &\sum_{\{q_i\}=\pm 1} \sum_{(\sigma_1,\ldots,
\sigma_{2n})} \left\{ \prod^{2n}_{k=1} \left[
\prod_{i\in\sigma_k} r^{\mu_ix}(\ell_i) q_k\right] I
\left(Z'_{n_x}\right)
+ \prod^{2n}_{k=1} \left[\prod_{i\in\sigma_k} r^{\mu_iy} (\ell_i)
q_k \right] I \left(Z'_{n_y}\right)\right\}\ \ .\cr}}
Because $r^{\mu_ix}$ and $r^{\mu_iy}$ are
independent of the variables $\sigma$, $q$ and $k$, and
since $Z'_{n_x} = Z'_{n_y} = Z'_{2n}$, we can write the sum over $q_i$ as
\eqn\qsumfin{
\left[ \prod^m_{j=1} r^{\mu_jx} (\ell_j) + \prod^m_{j=1} r^{\mu_jy}
(\ell_j)\right] \sum_{\{ q_i\} = \pm 1} \sum_{(\sigma_1,\ldots, \sigma_{2n})}
I\left(Z'_{2n}\right)\prod^{2n}_{k=1} q^{|\sigma_k|}_k\ \ .}
Upon substituting this back into Eq.~\Cmag\ for $\tilde C$ with
$\beta/\alpha\not=0$ and comparing it to Eq.~\Czmag\ for $\tilde C$ with no
magnetic field, we find
\eqn\dual{
\tilde C^{\mu_1\ldots\mu_m}_{2n} \left( \ell_1, \ell_2, \ldots,
\ell_m;{\beta\over\alpha}\right) = \left[ \prod^m_{j=1} r^{\mu_jx}(\ell_j) +
\prod^m_{j=1} r^{\mu_jy} (\ell_j)\right]\tilde C_{2n} \left(
\ell_1,\ldots, \ell_m;0\right)\ \ .}

Therefore, we have found a transformation which relates the correlation
functions, $\tilde C^{\mu_1\ldots\mu_m}_{2n}
\left(\ell_1,\ldots,\ell_m;{\beta\over\alpha}\right)$, at magnetic field
$\beta/\alpha\in\IZ$ to the correlation functions, $\tilde C_{2n}(\ell_1,
\ldots, \ell_m;0)$, at zero magnetic field.   This is exactly what is
predicted by the duality transformation of Ref.~\PD.
(In Ref.~\PD, the
duality transformation was only calculated explicitly for the two-point
function, but the general transformation can be derived similarly.)
Because we can write
all the correlation functions of $\beta/\alpha\in\IZ$ in terms of those at
$\beta=0$, we have a way of expressing functions which are finite at large
separation of their variables in terms of functions which are purely contact
terms.  This is the motivation of our claim that the contact terms must have
some physical meaning and are not just relics of the regulator.  However, in
deriving Eq.~\dual, we did not regulate the off-diagonal propagator when it
appears in $Z_{n_x,n_y}$, but we did regulate it in $R$.  Furthermore, we
ignored what happens when two particles of differing species are coincident in
$Z_{n_x, n_y}$.  Because the contact terms can be affected by these
different treatments of the regulator, we do
not know if this derivation really gives the correct formula we would obtain
if we were to consistently use the full bosonic regulator that satisfies the
Ward identities.  This is the second main question we will be pursuing in the
following sections; whether the duality transformation remains true even when
the magnetic field is regulated.  In this paper we will not actually show that
Eq.~\dual\ is still true.  However, from Eq.~\dual, we see that if we fix
$\beta/\alpha$, all the $m$-point functions can be calculated in terms of one
another by multiplying and dividing by objects of the form
$\prod^m_{j=1} r^{\mu_jx}(\ell_j) + \prod^m_{j=1} r^{\mu_jy}
(\ell_j)$.  We will find that even when we regulate the magnetic field,
this relation remains true.  In other words, we will prove that for any
$\beta/\alpha\in\IZ$ and $m$, there is a function $F\left(\ell_1,\ldots,
\ell_m;\beta/\alpha\right)$ such that
\eqn\wdual{
\tilde C^{\mu_1\ldots\mu_m}  \left( \ell_1,\ldots, \ell_m;{\beta\over
\alpha}\right) = \left[ \prod^m_{j=1} r^{\mu_jx} (\ell_j) + \prod^m_{j=1}
r^{\mu_jy} (\ell_j) \right] F\left( \ell_1,\ldots, \ell_m;
{\beta\over\alpha}\right)\ \ .}
As is the case when $\beta=0$, we will prove that $F$ is finite, homogeneous
and
piecewise linear.

When there is no magnetic field, $\tilde C(\ell_1,\ldots,\ell_m;0)$ has
several symmetries.  First, it is symmetric under interchange of the
$\ell_i$'s and $\ell_j$'s.  Because $Z_{2n}'$ is symmetric under taking all
the $q_i$'s to $-q_i$, $\tilde C(\ell_1,\ldots,\ell_m;0)$ equals zero
whenever $m$ is odd.  Furthermore, $Z_{2n}'$ is invariant under
$t_i \to -t_i$, so $\tilde C(\ell_1,\ldots,\ell_m;0)$ is symmetric under
taking all the $\ell_j$'s to $-\ell_j$.  Lastly, $\tilde C$ comes from
``anti-symmetrizing" over all partitions, $(\sigma_1,\ldots,\sigma_{2n})$, of
$m$ elements into $2n$ ordered sets.  In other words, it contains the sum
\eqn\Ssum{S(\vec \ell) = \sum_{(\sigma_1,\ldots,\sigma_{2n})} f(\vec k)
\prod_{j=1}^{2n} q_j^{|\sigma_j|}\ \ ,}
where $|\sigma_j|$ equals the number of elements in $\sigma_j$; the
$q_j=\pm 1$ with $\sum_{j=1}^{2n} q_j=0$; and $\vec k$ is given by Eq.~\kdef.
Such a sum has the property that if any of the $\ell_j$ are zero,
the sum is zero.  Therefore, $\tilde C(\ell_1,\ldots,\ell_m;0)$ equals
$\prod_{j=1}^m {\rm sign} (\ell_j)e^{-\epsilon |\ell_j|}$ times a function
that vanishes whenever any one of the $\ell_j$ equals zero.  Our results
from the following sections will imply that
$F(\ell_1,\ldots,\ell_m; \beta/\alpha)$ also has these properties.

In a future paper, we will show how these symmetries, plus the piecewise
linearity and homogeneity of $F$ and the Ward identities, are sufficient for
solving for many of the correlation functions
to all orders in perturbation theory.  We conjecture that they are actually
sufficient for calculating all the $\dot x$ and $\dot y$
correlation functions to all
orders in perturbation theory, but so far  we are unable to prove or
disprove this conjecture.  It is straightforward to show, however, that in any
case where this information is sufficient for exactly solving for the
correlation functions, the duality transformation in Eq.~\dual\
will also be true.
\newsec{FEYNMAN RULES}
We have seen that, for special values of the magnetic field, the unregulated,
critical DHM simplifies
greatly.  In particular, there is an identity that can be used to fermionize
the partition function.  However, because the treatment of the regulator in
the fermionized theory affects the value of the contact terms, in this section
we will analyze the graphs for the partition function when the bosonic
propagators are regulated.  Even when we take into account the complications
due to the regulator, the fermionization identity still plays a crucial role
in simplifying the graphs.  Using this identity, we will characterize
and derive Feynman rules for the connected graphs when $\beta/\alpha$ is an
integer.  These are
the graphs for the free energy. They are also useful in calculating
correlation functions because they are precisely
the graphs for $Z'_{n_x,n_y}$.

Most of the results in this section might also be applicable for
any $\alpha$ and $\beta$ on
the critical circle $\alpha=\alpha^2+\beta^2$.  The only difference is that
when $\beta/\alpha$ is not an integer, at any order in $V_0$ we obtain an
infinite sum of graphs.  Any particular graph has the same form as when
$\beta/\alpha\in \IZ$.  However, for small $\epsilon$ the sums are purely
formal
and they either might not converge or might not give correct results when used
in further calculations.

Lastly, we remark that the fermionization identity, Eq.~\fermid, can be
generalized for graphs with non-zero total charge.  That means we can derive
similar Feynman rules for graphs in the charged sector.  These graphs are
necessary for calculating the Ward identities and the open string boundary
state.  In this paper, though, we will restrict our attention only to the
zero-charge sector.
\goodbreak
\bigskip
\noindent{\bf Graphs with $\pmb{\beta=0}$}
\medskip
\nobreak
Before we turn to graphs  with non-zero magnetic field, we will begin by
considering the one-dimensional neutral graphs with $\beta=0$.
For simplicity, we will ignore some overall factors of $e^\epsilon$ and
$-1$ which do not affect the form of our results.  Then
the integrand for the ${\cal O}(V_0^{2n})$ term in
the partition function is given by
\eqn\Ztnreg{Z_{2n} = (-1)^n {\prod\limits_{i<j} \left( z^2_i+
z^2_j-(e^{-\epsilon}+e^\epsilon)z_iz_j\right)
\left( w^2_i + w^2_j -(e^{-\epsilon} + e^\epsilon)w_iw_j\right)
\over \prod\limits_{i,j} \left( z_i - e^{-\epsilon} w_j\right)
\left( z_i - e^\epsilon w_j\right)} \prod_{i=1}^n z_i w_i \ \ ,}
where $z_j = e^{2\pi i t_j/T}$; $w_j = e^{2\pi i s_j/T}$;
the positive charges are
at the $t_j$'s and the negative charges are at the $s_j$'s.  As we saw in the
previous section, when we set $\epsilon$ to 0 in the numerator,
$Z_{2n}$ can be written as
\eqn\Ztnferm{Z_{2n}
 = (-1)^n \det \left( {1\over z_i - e^{-\epsilon}w_j}\right) \det \left(
{1\over z_i - e^\epsilon w_j}\right) \prod_{i=1}^n z_i w_i\ \ .}
This is equivalent to the regulated, fermionized partition function.
However, as we explained in Section~3, we must also regulate the
numerator in Eq.~\Ztnreg.

We can express the regulated propagator in the numerator in terms of the
unregulated propagator as
\eqn\regnum{
\left( z^2_i + z^2_j - (e^{-\epsilon} + e^\epsilon)z_i z_j\right)
=\left( z_i -
z_j\right)^2 + \left( 1 - \cosh\epsilon\right) 2z_i z_j\ \ .}
With this identity, we can write $Z_{2n}$ as
\eqn\ZAB{Z_{2n} = A_{2n} (z,w) B (z,w) \ \ ,}
where
\eqn\Adef{
A_{2n} (z,w) = (-1)^n {\prod\limits_{i<j} \left( z_i - z_j\right)^2
\prod\limits_{i<j} \left(w_i - w_j\right)^2 \over \prod\limits_{i,j} \left(
z_i - e^{-\epsilon}w_j\right) \left( z_i - e^\epsilon w_j\right)} \prod_i z_i
w_i\ \ ;}
and
\eqn\Bdef{
\eqalign{B(z,w)
&= \sum^{n(n-1)/2}_{M,N=0} \left\{ 2(1-\cosh\epsilon)\right\}^{M+N} \cr
&\times \sum_{\pi,\kappa} \sum_{\sigma,\lambda}\prod^M_{\ell=1} {z_{\pi(\ell)}
z_{\kappa(\ell)} \over \left( z_{\pi(\ell)} - z_{\kappa(\ell)}\right)^2}
\prod^N_{m=1} {w_{\sigma(m)} w_{\lambda(m)}\over \left( w_{\sigma(m)} -
w_{\lambda(m)}\right)^2}\ \ .\cr}}
Here, the sum over $\pi,\kappa$ runs over all possible sets of $M$ pairs of
integers,
\break
$\left\{ \left( \pi(1), \kappa(1)\right),\ldots, \left( \pi(M),
\kappa(M)\right)\right\}$, with $1\le \pi(i)< \kappa(i)\le n$.  The sum over
$\sigma$ and $\lambda$ is defined similarly.

$A$ is the integrand for the fermionized partition function given by
Eq.~\Ztnferm.  If we represent
each $w_i$ and $z_i$ by a vertex and let each $i\sqrt{z_i w_j}/\left(
z_i - e^{-\epsilon} w_j\right)$ or $i\sqrt{z_i w_j}/\left(w_j -
e^{-\epsilon}z_i\right)$
correspond to a directed edge joining $z_i$  and $w_j$,
then it just the sum over all
possible ways to make directed loops out of the $2n$ vertices by connecting
each vertex to two vertices of the opposite charge, with a factor of minus one
for each closed loop.  For example, $A_4$ is given in Figure 1.
The expression for $B$ can be interpreted as telling us to join any $M$ pairs
of $z$ charges with the propagator $2(1-\cosh\epsilon)z_i z_j/(z_i-z_j)^2$ and
similarly for any $N$ pairs of $w$ vertices. This propagator counts as two
edges because there are two factors of $z_i-z_j$ in the denominator.
As an example, the graphs for $B_4$ are given in Figure 2.
One can verify that all the disconnected graphs in the product $AB$ are
exactly the graphs that are subtracted off when we take the logarithm of
$Z_{2n}$ to evaluate
the free energy in the zero-charge sector.  Consequently, the connected graphs
in $AB$ are the graphs for $Z'_{2n}$ and the free energy.

We note that the graphs for $Z_{2n}$ no longer look properly regulated because
$B$ has factors of $(z_i-z_j)$ and $(w_i-w_j)$ in the denominator.  However,
because we started with a properly regulated expression for $Z_{2n}$, we know
that this cannot be the case.  In fact, we can use simple properties of
partial fractions to add the connected graphs due to a particular term in
$B$, which we have just enumerated,
to obtain properly regulated connected graphs
with the same number of $(z_i - e^{-\epsilon}w_j)$ propagators in the
denominator and no propagators in the numerator.

To summarize, we have found a way to enumerate all the graphs for the free
energy.  They are all the connected graphs that are made of directed loops
which
alternate in $+$-vertices and $-$-vertices, times factors of $2(1-\cosh
\epsilon) z_i z_j/(z_i-z_j)^2$ joining pairs of like-charged vertices.

For the remaining sections of this paper, it is important to note
that each graph for the free energy
due to $Z_{2n}$ has the following properties.
\medskip
\item{1)}It is connected and 1PI.
\medskip
\item{2)}If it has $V$ vertices and $E$ edges and $-f$ factors of $\epsilon$ in
the numerator (where $1-\cosh\epsilon\approx - {1\over 2} \epsilon^2$ yields
two factors of $\epsilon)$, then
\eqn\EVf{E - V + f = 0 \ \ .}
\item{3)}The graphs are homogeneous in the $z$'s and $w$'s with degree $0$.
\medskip
\item{4)}There are exactly half as many of each $z_i$ in the numerator as
there are in the denominator.  More specifically, if $p_{ij}$ is the number
of $z_i$'s in the numerator due to propagators joining vertices $z_i$
and $z_j$, and $a(i,j)$ is the number of edges joining vertices $z_i$ and
$z_j$, then
\eqn\pha{p_{ij} = {1\over 2} a(i,j)\ \ .}
\noindent These four properties follow directly from our description of the
graphs for the free energy.
\goodbreak
\bigskip
\noindent{\bf Graphs with a Magnetic Field}
\medskip
\nobreak
When we add the magnetic field, the integrand for the partition function
becomes
\eqn\ZZZC{Z_{2n,2m} = Z_{2n}(z) Z_{2m}(w) C_{2n,2m}(z,w)\ \ ,}
where $Z_{2n}(z)$ is the partition function for a one-dimensional particle in
the absence of the magnetic field, and $C_{2n,2m}$ is due to the magnetic
field.  If we define
\eqn\qidef{q_i = \cases{+1 & for $1\le i\le n$\cr\noalign{\vskip 0.2cm}
-1 & for $n+1\le i\le 2n$\ \ ,\cr}}
and
\eqn\pidef{p_i = \cases{ +1 & for $1\le i \le m$ \cr\noalign{\vskip 0.2cm}
-1 & for $m+1\le i \le 2m$\ \ ,\cr}}
then according to Eqs.~\Znxdef, \Znydef\ and \Cnxnydef, we can write
\eqnn\Ztnq
\eqnn\Ztmp
\eqnn\Ctntmdef
$$\eqalignno{Z_{2n} (z) &= (-1)^n \prod^{2n}_{{i,j=1}\atop{i<j}} \left( z_i -
e^{-\epsilon} z_j\right)^{q_i q_j} \left( z_i - e^\epsilon z_j\right)^{q_i
q_j}\ \ ;&\Ztnq \cr
Z_{2m}(w) &= (-1)^m \prod^{2m}_{{i,j=1}\atop{i<j}} \left( w_i - e^{-\epsilon}
w_j\right)^{p_ip_j} \left( w_i - e^\epsilon w_j\right)^{p_ip_j}\ \ ;&\Ztmp
\cr\noalign{\hbox{and}}
C_{2n,2m}(z,w) &= \prod_{i,j} \left( {z_i - e^\epsilon w_j\over \omega_j -
e^\epsilon z_i}\ {z_i\over w_j} \right)^{-q_i p_j
\beta/\alpha}\ \ .&\Ctntmdef\cr}$$
Because $Z_{2n}$ and $Z_{2m}$ are exactly the same expressions we already
considered for charge-0 graphs with no magnetic field, we already know how to
enumerate the graphs they produce.  Therefore, we need only to analyze
$C_{2n,2m}$, the part that explicitly depends on the magnetic field.
Because $\sum q_i = \sum p_i = 0$, we can write $C_{2n,2m}$ as
\eqn\Ctntmfac{
C_{2n,2m} = \prod_{{i,j}\atop{-q_i p_j>0}} \left( e^{-\epsilon} {z_i -
e^\epsilon w_j\over z_i - e^{-\epsilon}w_j}\right)^{\beta/ \alpha}
\prod_{{i,j}\atop{-q_i p_j<0}} \left( e^{-\epsilon} {w_j - e^\epsilon z_i\over
w_j - e^{-\epsilon} z_i}\right)^{\beta/\alpha}
\ \ .}
Any particular factor in $C_{2n,2m}$ has the form
\eqn\Cfac{
e^{-\epsilon} {z_i - e^\epsilon w_j\over z_i - e^{-\epsilon} w_j} = 1 + \left(
e^{-\epsilon} - 1 \right)+{\left( e^{-2\epsilon} - 1\right) w_j\over z_i -
e^{-\epsilon}w_j}\ \ ,}
(or \Cfac\ with $z$ and $w$ interchanged). We can interpret this as saying
that we can join $z_i$ and $w_j$ with any of the propagators $\left(
e^{-\epsilon}-1\right)^\ell \left( {(e^{-2\epsilon} - 1)w_j\over z_i -
e^{-\epsilon} w_j}\right)^k$, for $0\le k$, $0\le\ell$, and $0\le
k+\ell\le\beta
/ \alpha$.  As long as $k+\ell\not=0$, we then say that $z_i$ and
$w_j$ are ``connected'' (in the looser sense that these graphs must be
evaluated
for the free energy); and for every factor of $\left(
e^{-2\epsilon}-1\right)w_j\big/ \left( z_i - e^{-\epsilon} w_j\right)$
we say the graph has an additional edge joining vertices $z_i$ and $w_j$.

If the $z_i$'s are $x$-vertices and the $w_j$'s are $y$-vertices, then we
can summarize the Feynman rules for the free energy when
$\alpha/(\alpha^2+\beta^2)=1$ and $\beta/\alpha\in \IZ$ as follows:
\medskip
\item{1.}All the $x$-vertices (and $y$-vertices) are
joined in directed loops with
alternating charges by the propagator
\eqn\xxprop{{i\sqrt{z_iz_j}\over z_i - e^{-\epsilon} z_j} \qquad \qquad
({\rm or} \qquad
{i\sqrt{w_iw_j}\over w_i - e^{-\epsilon}w_j} )\ \ ,}
times minus one for each closed loop.
\medskip
\item{2.}Arbitrary numbers of pairs of like-charged $x$-vertices are
joined with the propagator
\eqn\lcprop{{2(1-\cosh\epsilon)z_iz_j\over (z_i - z_j)^2}\ \ ,}
and similarly for the $y$-vertices.
\medskip
\item{3.}Arbitrary numbers of pairs of like-charged $x$- and $y$-vertices
are joined with
\eqn\lcxyprop{
\left( {(e^{-2\epsilon} - 1) w_j\over z_i - e^{-\epsilon} w_j}\right)^k
(e^{-\epsilon}-1)^\ell\ \ ;}
and arbitrary numbers of pairs of oppositely charged $x$- and $y$-vertices
are joined with
\eqn\ocxyprop{
\left( {\left( e^{-2\epsilon}-1\right) z_i \over w_j -
e^{-\epsilon}z_i}\right)^k \left( e^{-\epsilon}-1\right)^\ell \ \ .}
for $0\le\ell+ k\le\beta/\alpha$.

For the following sections, it is useful to note that each graph,
$G$, for the free
energy is made up of several subgraphs, $G_1,\ldots, G_\gamma$. Each subgraph,
$G_i$, is neutral and has either all $x$-vertices or all $y$-vertices.  Each
$G_i$ is identical to a connected graph for $Z_{2n_i}$, where $2n_i$
equals the number of vertices in $G_i$, and $Z_{2n_i}$ is from
the partition function for the one-dimensional system with
$\beta=0$, analyzed in the first part of this section.  The graphs, $G_i$,
are joined to at least one other graph, $G_j$, by the products of
propagators in \lcxyprop\ and \ocxyprop, where
$z_i\in G_i$ and $w_j\in G_j$ or {\it vice versa}.  An example of a
magnetic-field graph is given in Figure 3.  In the figure, the $G_i$
for $1 \le i \le 6$ are the zero-field subgraphs; and the edges,
$e_1, \ldots, e_7$, are from the magnetic-field propagators.
Each magnetic-field graph, $G$, has
the following properties:
\medskip
\item{1)}Each subgraph, $G_\ell$, is connected and 1PI with respect to
each edge corresponding to
${i\sqrt{z_i z_j}\over z_i - e^{-\epsilon} z_j}$
and each of the two edges corresponding to
${2(1-\cosh \epsilon) z_i z_j\over \left( z_i - z_j\right)^2}$,
for $z_i,z_j\in G_\ell$. (Here, $z_i$ and $z_j$ can stand for either
$x$-vertices or $y$-vertices.)
However, the entire graph may no longer be either connected or 1PI with
respect to all of its edges.
\medskip
\item{2)}If the full graph has $V$ vertices, $E$ edges, $-f$ factors of
$\epsilon$ in the numerator and $g$ connected components, then
\eqn\EVfg{E - V + f + g \le 1\ \ .}
\item{3)}The graphs are homogeneous of degree 0.
\medskip
\item{4)}In each subgraph, $G_\ell$, each variable, $z_i$, appears
twice as many times in the denominator as in the numerator.
However, this is no longer true
in the full graph.  Instead, if $p_{ij}$ is the number of $z_i$'s in the
numerator due to propagators joining vertex $z_i$ to the vertex $z_j$;
if $p_{ij}$ is defined similarly; and if $a(i,j)$ is the number of edges
between the vertices, then
\eqn\ppa{p_{ij} + p_{ji} = a(i,j)\ \ .}
%
\newsec{DERIVATION OF THE RECURSION RELATIONS}
In this section we will consider integrals of the form
\eqn\INdef{
I(N) = \oint\prod^N_{j=1} {dz_j\over 2\pi i z_j} F(N,{\svec z})\ \ ,}
where
\eqn\FNzdef{F(N,{\svec z}) = q \left( e^{-\epsilon}\right)
\prod_{i,j,k} {1\over z_i - e^{-n(i,j,k)\epsilon}z_j}
\prod_i z^{p_i+k_i}_i\ \ .}
In this expression, $n\in \IZ$, $p_i+k_i\in \IZ$ and $q(e^{-\epsilon})$
is a rational
function in $e^{-\epsilon}$ which goes as $\epsilon^{-f}$ as $\epsilon$ goes
to 0, for some $f\in\IZ$.  This type of integral includes those in the
perturbation series for the free energy and the Fourier transform of
$I(t_1,\ldots, t_{N})= \ll \prod_j e^{i{\svec q}_j\cdot {\svec
x}(t_j)}\rr_0^{\rm con}$, for any charge when $\beta/\alpha$ is an integer.

The integrand corresponds to a graph with vertices $z_1,\ldots, z_{N}$; edges
$1/(z_i - e^{-n(i,j,k)\epsilon}z_j)$; and factors $z_j^{p_j}$
and $z_j^{k_j}$ associated with each vertex.
We will say that any such graph, described by Eq.~\FNzdef, is of
``standard form".  When we perform an integration,
for each residue that is evaluated  we will obtain a new graph with
$N-1$ vertices. In this section, we will
prove that these new graphs have the same form as the old ones,
except that their
connectedness and 1 particle irreducibility may change.  We will begin by
defining a few other properties of the graph, which also will
either remain the same when
a residue is evaluated, or change in a well-defined way.

We will let $V=$ the number of vertices in the graph, and $E=$ the number of
edges. Also, we will let $g=$ the number of connected components of the graph.
Then we can define the degree of divergence, $d$, of any graph to be
\eqn\ddef{d(V) = E - V + f + g \ \ .}
This is, in fact, the superficial degree of divergence of the graph.  In what
follows, we will prove that at each integration, the degree of divergence
remains constant or decreases so that $d(V)\ge d(V-1)$.  Therefore, the graphs
diverge no faster than their superficial degree of divergence.
We will also find
the conditions for when the degree of divergence remains constant and when it
decreases.

We also note that these graphs are homogeneous with degree $h$ given by
\eqn\hdef{h = \sum^V_{i=1} p_i + \sum^V_{i=1} k_i -E\ \ .}
We will prove that after each integration, $h$ remains constant.

Finally, we will define $a(i,j)=a(j,i)$ to be the number of edges joining the
vertex $z_i$ with the vertex $z_j$, and we define $p_{ij}$ by any non-negative
half-integers or integers such that
\eqn\defpij{p_i = \sum^V_{j=1} p_{ij}\ \ .}
Then we will be interested in graphs which satisfy the following conditions:
\medskip
\item{1)}$p_{ij} + p_{ji} = a(i,j)$
\medskip
\item{1a)}$\sum_{i=1}^V k_i=K$
\medskip
\item{2)}If $a(i,j)\not=0$, then $p_{ij}$ and $p_{ji}\not=0$.
\medskip
\noindent Note that we can always choose $p_i$ and $k_i$ so that condition (1)
is true, but that condition (2) is not always satisfied by graphs with
standard form.

In this section, we will prove that any graph with standard form
which satisfies (1) and (2)
will still have standard form and satisfy (1) and (2) after being integrated.
According to our definitions of $h$, $a(i,j)$ and $p_{ij}$, this means that we
will find, in addition, that $\sum^V_{i=1} k_i = K(V)$ will remain fixed also.

We now turn our attention to performing the integral $I(N)$.  Each integral in
$z_j$ is a contour integral over the unit circle and can be integrated by
evaluating the residues with respect to $z_j$.  The only possible poles of
$z_j$ inside the unit circle occur when $z_j = e^{-b\epsilon}z_k$ for some
$b>0$ and some $z_k$, or when $z_j=0$.
If there are no such poles, the integral is zero.  Otherwise, for every such
pole, when we evaluate the residue, we will obtain a new graph.
(In some cases we will have expressions like $z_i - z_j$ in the denominator,
so it will appear that we have a pole on the contour of integration.  However,
in all such cases, there are always other graphs we can add to this graph so
that the singularity cancels.  Thus, as long as we integrate all the
variables in each graph in the same order, we can just ignore such poles.)
To distinguish between the old and new graphs, let $q(e^{-\epsilon};V)$,
$p_j(V)$, $p_{ij}(V)$, $k_j(V)$ and $a(i,j;V)$ be the values of
$q(e^{-\epsilon})$, $p_j$, $p_{ij}$, $k_j$ and $a(i,j)$ for the graph with
$V$ vertices.  We will drop the argument, $V$, whenever it is clear which
graph we are referring to.
\goodbreak
\bigskip
\noindent{\bf Pole at $\pmb{z_{N}=e^{-a\epsilon}z_k}$ with
Multiplicity $\pmb{= 1}$}
\medskip
\nobreak
We will begin with a graph with $N$ vertices and perform the integral over
$z_N$.
Suppose there is a pole at $z_N = e^{-a\epsilon}z_k$, and, for simplicity,
suppose it has multiplicity equal to 1.  The only part of $F(N;{\svec z})$
that is affected are those factors which contain a $z_N$ or a $z_k$, so we will
split $F$ into two parts, one that does not change when we evaluate the
residue, and one that does. We will define
\eqn\FoFt{F(N;{\svec z}) = F_1({\svec z}) F_2({\svec z})\ \ ,}
where
\eqnn\Fonedef
\eqnn\Ftwodef
$$\eqalignno{
F_1({\svec z}) &= \prod^{N-1}_{{j=1}\atop{j\not=k}} \left[
z^{p_j+k_j}_j \prod^{j-1}_{{\ell=1}\atop{\ell\not=k}} \prod^{a(j,\ell)}_{i=1}
{1\over z_j - e^{-b_i\epsilon} z_\ell}\right] \ \ ;
&\Fonedef \cr\noalign{\hbox{and}}
F_2 ({\svec z}) &= q\left(e^{-\epsilon}\right)
z^{p_N+k_N}_N z^{p_k+k_k}_k \prod^{a(N,k)}_{i=1}
\left[{1\over z_N - e^{-b_i\epsilon} z_k} \right] \cr
&\times
\prod^{N-1}_{{\ell=1}\atop{\ell\not=k}}
\left[ \prod^{a(N,\ell)}_{i=1} \left({1\over z_N - e^{-b_i\epsilon}z_\ell}
\right)
\prod^{a(k,\ell)}_{i=1} \left({1\over z_k - e^{-b_i\epsilon} z_\ell}\right)
\right]\ \ .&\Ftwodef \cr}$$
Here, for every value of $j$ and $\ell$ we let $b_i$ be an integer equal to
$\pm n(j,\ell,i)$; and we have ignored some factors of $-1$ and
$e^{-\epsilon}$,
which will not affect the degree of divergence or any other property we are
considering.

Only $F_2({\svec z})$ is affected when we evaluate the residue, and the
result is
\eqn\Rtwo{\eqalign{
R_2 =&\hbox{Res} \left[ {1\over 2\pi i z_N} F_2\right]\bigg|_{z_N =
e^{-a\epsilon} z_k} \cr
=&q\left( e^{-\epsilon}\right)
z^{p_k + k_k + p_N + k_N-1}_k e^{-a\epsilon (p_N + k_N)}
\left[ \prod^{a(N,k)-1}_{i=1} {1\over z_k\left( e^{-a\epsilon} -
e^{-b_i\epsilon}\right)}\right] \cr
&\times
\left\{ \prod^{N-1}_{{\ell=1}\atop{\ell\not=k}} \left[
\prod^{a(N,\ell)}_{i=1} {1\over z_k e^{-a\epsilon} - z_\ell e^{-b_i\epsilon}}
\prod^{a(k,\ell)}_{i=1} {1\over z_k - z_\ell e^{-b_i\epsilon}}\right]
\right\}\ \ .\cr}}
The new total power of $z_k$ in the numerator is
\eqn\newzk{p_k + p_N + k_k + k_N - a(N,k)\ \ .}
In the new graph, we can define all of the new $k_j$'s, $p_j$'s and
$a(i,j)$'s, except for $k_k$, $p_k$ and $a(k,\ell)$ for $1\le\ell\le N$,
to be the same as in the old graph.  Then the new $k_k$ is
\eqn\newkko{k_k (N-1) = k_k(N) + k_N(N)\ \ ;}
the new $p_k$ is
\eqn\newpko{p_k(N-1) = p_k(N) + p_N(N) - a(N,k;N)\ \ ;}
and the new $a(k,\ell)$ is
\eqn\newako{a(k,\ell;N-1) = a(N,\ell;N)+a(k,\ell;N)\ \ .}
Then, ignoring an overall
factor of $\pm e^{-A\epsilon}$ for some $A$ independent of the $p$'s
and $k$'s, we find that the total residue is
\eqn\Resko{
\eqalign{R =&\hbox{Res} \left[ {1\over 2\pi i z_N} F\right]\bigg|_{z_N =
e^{-a\epsilon}z_k}  \cr
=& F_1({\svec z})\times q\left( e^{-\epsilon}; N-1\right)
z^{p_k(N-1) + k_k (N-1)}_k
\prod^{a(k,\ell;N-1)}_{\ell\not=k} {1\over z_k - e^{-\tilde b_i
\epsilon} z_\ell}\ \ ,\cr}}
where $\tilde b_i$ equals either $b_i-a$ or $b_i$ and is still an
integer; and the
new $q\left( e^{-\epsilon};N-1\right)$ is given in terms of the old
$q(e^{-\epsilon};N)$ as
\eqn\newqko{q\left( e^{-\epsilon};N-1\right)
= \left[ \prod^{a(N,k)-1}_{i=1} {1\over
e^{-a\epsilon} - e^{-b_i\epsilon}}\right] e^{-a\epsilon p_N} e^{-a\epsilon k_N}
q\left( e^{-\epsilon};N\right)\ \ .}
If $q(e^{-\epsilon};N)$ goes as $\epsilon^{-f(N)}$ as $\epsilon\to0$, then in
the same limit, $q\left( e^{-\epsilon};N-1\right)$ goes as $\epsilon^{-f(N-1)}$
where $f(N-1)=f(N) + a(N,k)-1$.

In terms of the graph, what we have just done is to remove the edge
corresponding to $1/(z_N - e^{-a\epsilon}z_k)$; remove the vertex $z_N$;
and reconnect to $z_k$ all the edges that originally were
attached to $z_N$.  Therefore, a 1PI
graph will remain 1PI; a connected graph will remain connected or
become 1PI; and a disconnected graph with $g$ connected components will
remain disconnected with $g$ connected components.
Furthermore, in performing the integral, we have replaced all factors of $z_N$
in the numerator with $z_k$.  In addition, we have removed all the edges
that connect $z_N$ to $z_k$.  For each such edge,
we get rid of one factor of $z_k$ in the numerator, and for all but one of
these edges we get a factor of $1/(e^{-a\epsilon}-e^{-b_i\epsilon})
\approx 1/\epsilon$.  An example is given in Figure 4.  In this figure,
we evaluate the residue at the $z_1 = e^{-a\epsilon}z_2$ pole
due to the edge, $e_1$.  The edges, $e_1$ and $e_2$, correspond to
$1/(z_1-e^{-a\epsilon}z_2)$ and $1/(z_1-e^{-b\epsilon}z_2)$, respectively.

The description in the previous paragraph implies that the change in
the number of edges
is $\Delta E = - a(N,k)$; the number of vertices decreases by $1$; the number
of factors of $\epsilon$ in the denominator is increased by $\Delta f =
a(N,k)-1$; the change in $g$ is zero; and the change in factors
of the $z$'s in the numerator is $-a(N,k)$.  From this, we find that
\eqn\newdko{\eqalign{d(N-1) =&d(N) - a(N,k) + 1 + a(N,k) - 1 + 0 \cr
=&d(N)\ \ ,\cr}}
so the degree of divergence remains constant.  Also,
\eqn\newhko{\eqalign{h(N-1) =&h(N) - a(N,k) +a(N,k) \cr =&h(N)\ \ ,\cr}}
so $h$ also remains constant.

Next, we will show that if $p_{ij} + p_{ji} = a(i,j)$ for all $i,j=1,2,\ldots,
N$, and if $\sum_{i=1}^N k_i=K$, then after we evaluate the residue,
these continue to
be true.  From the definition of $k_j(N-1)$ in Eq.~\newkko\ and the preceding
sentence, we conclude that
\eqn\newKko{\sum^{N-1}_{i=1} k_i (N-1)
= \sum^N_{i=1}k_i(N) = K\ \ ,}
so this sum does indeed remain fixed.

For $j,\ell \ne N$, we will define the new $p_{j\ell}$ as follows:
\eqn\newpjlko{p_{j\ell}(N-1) = \cases{p_{j\ell}(N) &for $j,\ell \ne k$;
\cr\noalign{\vskip 0.2cm}
p_{k\ell}(N) + p_{N\ell}(N) &for $j=k$, $\ell\ne k$;
\cr\noalign{\vskip 0.2cm}
p_{jk}(N) + p_{jN}(N) &for $j\ne k$, $\ell = k$.\cr}}
If the $p_{j\ell}(N)$'s are non-negative integers, then so are the
$p_{j\ell}(N-1)$'s.  For Eq.~\newpjlko\ to be a valid definition, we must also
check that
\eqn\pjldefko{
p_j(N-1) = \sum^{N-1}_{{\ell=1}\atop{\ell\not=j}} p_{j\ell}(N-1)\ \ .}
As long as $j\ne k$, it is automatically true.  When $j=k$, using
definitions \newpko\ and \defpij, we have
\eqn\pkcalc{
p_k (N-1) = \sum^N_{{\ell=1}\atop{\ell\not=k}} \left[ p_{k\ell}(N)\right] +
\sum^{N-1}_{\ell=1} \left[p_{N\ell} (N)\right] - a (N,k;N)\ \ .}
We are assuming the original graph satisfies condition (1), so
that
\eqn\ppacalc{
p_{kN}(N) + p_{Nk} (N) = a(N,k;N)\ \ .}
Substituting this into \pkcalc\ and using the definition of
$p_{k \ell} (N-1)$, we conclude that
\eqn\pkldefko{
p_k(N-1) = \sum^{N-1}_{{\ell=1}\atop{\ell\not=k}} p_{k\ell}(N-1)\ \ ,}
and the definition of $p_{j\ell}(N-1)$ is a valid one.

To verify condition (1) in the new graph, we must check whether
\eqn\appko{
a\left( j,\ell;N-1\right) = p_{j\ell} (N-1) + p_{\ell j} (N-1)\ \ .}
For $j \ne k$ and $\ell \ne k$, this follows immediately from the
definitions of the $a$'s and $p$'s.  When $j=k$, we have
\eqn\appcalc{\eqalign{p_{k\ell}(N-1)+p_{\ell k}(N-1)
&= p_{k\ell} (N) + p_{N\ell} (N) + p_{\ell k} (N) + p_{\ell N}(N)
\cr
&= a(k,\ell;N) + a(N,\ell;N) \ \ .\cr}}
This last line equals $a(k,\ell;N-1)$, so we are done.

Finally, we want to show that if condition (2) was true before we
performed the integral, then it remains true afterwards.  Suppose
$a(j,\ell;N-1)$ is not zero.  If neither $j$ nor $\ell$ equals $k$, then
condition (2) is clearly satisfied.  Thus, we need only to consider the
case when $\ell = k$.  Then at least one of $a(j,k;N)$ and
$a(j,N;N)$ must be non-zero.  It follows by condition (2) for the
original graph
that at least one of $p_{jk}(N)$ and $p_{jN}(N) \ge 1/2$.  Therefore,
$p_{jk}(N-1) = p_{jk}(N) + p_{jN}(N) \ge 1/2$.  Similarly, we have
$p_{kj}(N-1) \ge 1/2$.  Thus, condition (2) is satisfied.
\goodbreak
\bigskip
\noindent{\bf Pole at $\pmb{z_{N}=e^{-a\epsilon}z_k}$ with
Multiplicity $\pmb{> 1}$}
\medskip
\nobreak
We now consider the case when the multiplicity is greater than 1.  If we are
evaluating a residue from a pole at $z_N = e^{-a\epsilon}z_k$ of multiplicity
$m>1$, the only change from when $m=1$ is that we must get rid of all the edges
corresponding to $1/\left(z_N-e^{-a\epsilon}z_k\right)^m$ and then take the
$m-1$ derivatives, $1/(m-1)! \left( \partial^{m-1}/\partial z^{m-1}_N\right)$,
before we set $z_N=e^{-a\epsilon}z_k$.  The derivatives can either act on
$z_N^{p_N+k_N}$ or on some of the
$1/\left(z_N-e^{-b_i\epsilon}z_\ell\right)$'s.
Let
\eqn\msdef{m-1=m(p) + m(K) + \sum_{\ell\not=N}m(\ell)\ \ ,}
where $m(p)$, $m(K)$ and $m(\ell)$ are  non-negative integers;
$m(p)$ and $m(K)$
equal the number of derivatives acting on $z_N^{p_N}$ and $z_N^{k_N}$,
respectively; and $m(\ell)$ equals the number of derivatives acting on
$1/(z_N
-e^{-b_i\epsilon}z_\ell)$ for all $b_i$.  For every choice of $m(p)$,
$m(K)$ and $m(\ell)$, we will obtain a new graph.
The new graphs differ from the graph
for multiplicity $=1$ only in the following ways.  When $m(p)$ derivatives act
on $z^{p_N}_N$, we just decrease $p_N$ by $m(p)$,
and similarly for $m(K)$ and $k_N$; when $m(\ell)$ derivatives act
on $1/(z_N-e^{-b_i}z_k)$ with $\ell = k$, then,
ignoring factors of $\pm e^{-A\epsilon}$,  this
just increases the number of factors of $1/\left[z_k\left( e^{-a\epsilon} -
e^{-b_i\epsilon}\right)\right]$ in the answer by $m(k)$;
and when $m(\ell)$ derivatives act
on $1/\left(z_N-e^{-b_i\epsilon}z_\ell\right)$ for $\ell\not=k$,
we just increase the new
number of edges, $a(k,\ell;N-1)$, joining $z_k$ with $z_\ell$ by $m(k)$.
Finally, the graph will be multiplied by
${p_N\choose m(p)}$ and ${k_N\choose m(K)}$.  (All the
other factorials due to the definition of the residue and taking derivatives
cancel.)  In the special case where only one derivative acts on
$z_N^{p_N+k_N}$, we have $m(p)=1$ and $m(K)=0$, or {\it vice versa}, so
these factors reduce to $p_N+k_N$.  In Figure 5, we give an example where
a residue at a $z_1=e^{-a\epsilon}z_2$ pole of multiplicity $2$ is
evaluated.  In this diagram, the edges, $e_1$ and $e_1'$, both correspond to
$1/(z_1-e^{-a\epsilon}z_2)$.

Apart from the changes listed above, we can repeat the $m=1$ calculation
exactly as before.  We find that
\eqnn\dEkm
\eqnn\dVkm
\eqnn\dfkm
\eqnn\dgkm
\eqnn\dpkkm
$$\eqalignno{
\Delta E &= - a(N,k) + \sum^{N-1}_{{j=1}\atop{j\not= k}}m(j)\ \ ;
&\dEkm\cr
\Delta V &= - 1\ \ ; &\dVkm\cr
\Delta f &= a(N,k) - m +m(k)\ \ ; &\dfkm\cr
\Delta g &= 0 \ \ ; &\dgkm\cr
\noalign{\hbox{and}}
\Delta p_k + \Delta k_k &= p_N - m(p) - a (N,k) + m - m(k) - 1 + k_N - m(K)
\ \ . &\dpkkm\cr}$$
{}From this, we conclude that
\eqn\dcalc{
d(N-1) = d(N) - a(N,k) + \sum^{N-1}_{{j=1}\atop{j\not=k}} m(j) + 1 + a(N,k)
- m+m(k) + 0\ \ .}
With the definition of $m(j)$ in Eq.~\msdef, this becomes
\eqn\newdkm{d(N-1) = d(N) - m(p)-m(K) \ge d(N)\ \ ,}
and we have equality only when no derivatives act on $z^{p_N+k_N}_N$.  Thus,
the degree of divergence either remains constant or decreases.  Similarly, we
can evaluate $h(N-1)$, with the result
\eqn\newhkm{h(N-1) = h(N)\ \ .}

Finally, we will verify that conditions (1), (1a) and (2) remain true if they
were true before the integration. To do so, we will define
\eqn\newpkm{p_\ell(N-1) = \cases{
p_\ell(N) & $\ell\not=k$;\cr\noalign{\vskip 0.2cm}
p_k(N) + p_N(N) - m(p) - m(k) - m(K) + m-1-a(N,k) & $\ell=k$; \cr}}
and
\eqn\newkkm{k_\ell(N-1) = \cases{
k_\ell(N) & for $\ell\not=k$; \cr\noalign{\vskip 0.2cm}
k_k(N) + k_N(N) & for $\ell=k$ \cr}}
as before.  It is straightforward to check that
with these definitions $p_k(N-1) + k_k(N-1) = \Delta p_k + \Delta
k_k +  p_k (N) +k_k(N)$ as it should, and also that $p_\ell$ is a
non-negative integer for all $\ell$.  Once again
$\sum^{N-1}_{\ell=1} k_\ell (N-1)= \sum^N_{\ell=1}
k_\ell (N) = K$, so it remains constant.

When we count the number of edges joining $z_j$ and $z_\ell$, we find that
the new $a(j,\ell)$ are given by
\eqn\newakm{a(j,\ell;N-1) = \cases{
a(j,\ell;N) & for $j,\ell\not=N,k$; \cr\noalign{\vskip 0.2cm}
a(k,\ell;N) + a(N,\ell;N) + m(\ell) & for $j=k$.\cr}}
We will define the new $p_{j\ell}$'s to be
\eqn\newpjlkm{p_{j\ell} (N-1) = \cases{ p_{j\ell}(N) & for $j,\ell\not=N,k$;
\cr\noalign{\vskip 0.2cm}
p_{k\ell}(N) + p_{N\ell}(N) + m(\ell) & for $j=k$; \cr\noalign{\vskip 0.2cm}
p_{jk}(N) + p_{jN} (N) & for $\ell=k$. \cr}}
This is a valid definition for the $p_{j\ell}$, first of all because
they are still non-negative integers or half-integers.  Secondly, for
$j\not=k$ or $N$, we clearly still have
\eqn\pjldefkm{
\sum^{N-1}_{{\ell=1}\atop{\ell\not=j}} p_{j\ell} (N-1) = p_j (N-1)\ \ .}
Using the definitions of $p_{k\ell}(N-1)$, $p_k(N-1)$, and $m(\ell)$
and recalling that $a(N,k;N)$ satisfies condition (1), we can verify that
even when $j=k$, Eq.~\pjldefkm\ is still satisfied.
It immediately follows from the definitions of the new $a(j,\ell)$ and
$p_{j\ell}$ that when $j$ and $\ell\not=N$ or $k$, if
$a(j,\ell)=p_{j\ell}+p_{\ell j}$ in the old graph, the equality
is still true in the new graph.  We need only to check whether
\eqn\appkm{a(k,\ell;N-1) = p_{k\ell}(N-1) + p_{\ell k} (N-1)\ \ .}
The right-hand side is
\eqn\appcalckm{
\eqalign{p_{k\ell}(N-1)+p_{\ell k}(N-1) &= p_{k\ell}(N) + p_{N\ell}(N) +
m(\ell) + p_{\ell k} (N) + p_{\ell N} (N) \cr
&= a(k,\ell;N) + a(N,\ell;N) + m(\ell)\ \ . \cr}}
This last expression is exactly $a(k,\ell;N-1)$, so we have shown that if
condition (1) is true, then after evaluating the residue it remains true.

Finally, we must verify that if condition (2) is true before the integration,
it is still true after we evaluate the residue.  Since the only part of the
graph that changed is the part in some way associated with the vertices
$z_N$ and $z_k$, we need only to see whether when $a(k,\ell;N-1)>0$,
we also have
$p_{k\ell}(N-1)\ge {1\over 2}$ and $p_{\ell k}(N-1)\ge {1\over 2}$.
This time, if $a(k,\ell;N-1)$ is $>0$, then
$a(k,\ell;N)+a(N,\ell;N)+m(\ell)>0$.  All three are non-negative integers, so
this condition implies either $a(k,\ell;N)$, $a(N,\ell;N)$ or $m(\ell)$ is
greater than zero.  Because $m(\ell)$ is the number of derivatives acting on
$1/\left(z_N-e^{-b_i\epsilon}z_\ell\right)$ and $a(N,\ell;N)$
is the number of such factors in the integrand before
the derivatives are taken, $m(\ell)$ can only be non-zero when $a(N,\ell;N)$ is
non-zero.  Thus, at least one of $a(k,\ell;N)$ and
$a(N,\ell;N)$ is positive.  Since we are assuming the original graph
satisfied condition (2), it follows that at least one of $p_{k\ell}(N)$ or
$p_{N\ell}(N)$ is non-zero.  This implies that $p_{k\ell}(N-1) = p_{k\ell}(N)
+ p_{N\ell}(N) + m(\ell)$ is greater than zero.  It is also either an integer
or half-integer, so we conclude that $p_{k\ell} (N-1)\ge 1/2$.
Similarly $p_{\ell k}(N-1)\ge 1/2$, and condition (2) is still satisfied.

To summarize our results so far, we have found that if we evaluate a $z_N =
e^{-a\epsilon}z_k$ residue of the function, $F(N; {\svec z})$,
which is defined in Eq.~\FNzdef\ and which also
satisfies condition (1), then we obtain the following recursion relations
for the new graph:
\medskip
\item{R.1)}It still has the same form as $F(N;{\svec z})$, but with one less
integration variable (or vertex).
\item{R.2)}If the original graph is connected or 1PI, it remains connected
or 1PI; and if the original graph has $g$ connected components,
it continues to have $g$ connected components.
\medskip
\item{R.3)}It still satisfies condition (1).
\medskip
\item{R.4)}Its degree of divergence remains constant or decreases.
More specifically, it decreases
only when a derivative acts on a $z^{p_N+k_N}_N$, and in that case, it
decreases exactly by the number of derivatives acting on $z_N$.
\medskip
\item{R.5)}It is still homogeneous of degree $h$.
\medskip
\item{R.6)}$\Sigma k_i$ remains constant.
\medskip
\item{R.7)}The new $k_i$ are partial sums of the old $k_j$. Namely,
\eqn\newkSi{k_i(N-1) = \sum\limits_{j\in S_i} k_i\ \ ,}
with
\eqnn\SinotN
\eqnn\USi
\eqnn\intSiSl
$$\eqalignno{
{\rm i)}& \qquad S_i\subset \{1,2,\ldots N\}\ \ ; &\SinotN\cr
{\rm ii)}& \qquad \bigcup\limits_{i=1}^{N-1} S_i = \{1,2,\ldots, N\}\ \ ;
&\USi\cr
\noalign{\hbox{and}}
{\rm iii)}& \qquad S_i \cap S_\ell = \emptyset \qquad {\rm for} \qquad
i\ne\ell\ \ . &\intSiSl\cr}$$
\medskip
\item{R.8)}If the old graph satisfies condition (2), then the new graph
corresponding to the residue will also satisfy condition (2).
\medskip
\item{R.9)}When no derivatives act on $z^{p_N+k_N}_N$, the new rational
function,
$q\left( e^{-\epsilon};N-1\right)$, is related to the old one
in the following way:
\eqn\qePq{q\left( e^{-\epsilon}; N-1\right) = e^{-a\epsilon k_N} P\left(
e^{-\epsilon}\right) q\left( e^{-\epsilon};N\right)\ \ ,}
where $P(e^{-\epsilon})$ is a rational function in $e^{-\epsilon}$ that is
completely independent of $k_N$.
\medskip
\item{R.10)}When one derivative acts on $z^{p_N+k_N}_N$, the new rational
function has the form
\eqn\qkpePq{
q\left( e^{-\epsilon};N-1\right) = \left( k_N + p_N\right) e^{-a\epsilon
k_N} P\left( e^{-\epsilon}\right) q \left( e^{-\epsilon};N\right)\ \ ,}
where again $P(e^{-\epsilon})$ is a rational function that is completely
independent of $k_N$.

In the following section, we will apply these properties of integrals of $F$
to the calculation of the free energy and correlation functions.  However,
first we must also consider what happens when we have $z=0$ poles.
\goodbreak
\bigskip
\noindent{\bf Pole at $\pmb{z_{N}=0}$}
\medskip
\nobreak
When we evaluate a $z_N=0$ pole, the only part of $F(N;{\svec z})$ that is
affected are those factors which contain $z_N$, so we again can break $F$
into two parts.  This time they are
\eqn\FoFtzz{F(N;{\svec z}) = F_1({\svec z}) F_2({\svec z})\ \ ,}
where
\eqnn\Fodefzz
\eqnn\Ftdefzz
$$\eqalignno{
F_1\left({\svec z}\right) &= q\left(e^{-\epsilon}\right)
\prod\limits^{N-1}_{j=1} \left[ z^{p_j + k_j}_j
\prod\limits^{j-1}_{\ell=1} \prod\limits^{a(j,\ell)}_{i=1}
{1\over z_j - e^{-b_i\epsilon} z_\ell}\right] \ \ , &\Fodefzz
\cr\noalign{\hbox{and}}
F_2({\svec z}) &= z^{p_N + k_N}_N \prod\limits^{N-1}_{\ell=1}
\prod\limits^{a(N,\ell)}_{i=1} {1\over z_N - e^{-b_i\epsilon}z_\ell}
\ \ .&\Ftdefzz\cr}$$
Only $F_2({\svec z})$ will be affected when we evaluate the residue.  Because
we are assuming there is a pole at $z_N=0$, we must have the condition $p_N +
k_N \le 0$.  We will call $p_N + k_N = -m$.  Then the residue of $F_2 ({\svec
z})$ is given by
\eqn\ResFtzz{
\hbox{Res}\; \left[{1\over 2\pi i z_N} F_2({\svec z})\right] \bigg|_{z_N=0}
  = {1\over m!} \
{\partial^m\over \partial z^m_N} \prod\limits^{N-1}_{\ell=1}
\prod\limits^{a(N,\ell)}_{i=1} {1\over z_N - e^{-b_i\epsilon}
z_\ell}\bigg|_{z_N=0}\ \ .}
If we let $m(\ell)$ equal the number of times a derivative acts on
$1/\left(z_N - e^{-b_i\epsilon}z_\ell\right)$
for any $b_i$, then, for a given choice of $m(\ell)$,
after we perform the derivatives and set $z_N = 0$, we find
\eqn\ResFtprop{
\hbox{Res}\; \left[{1\over 2\pi i z_N} F_2({\svec z})\right] \bigg|_{z_N=0}
\propto \prod^{N-1}_{\ell=1} z^{-m(\ell) -
a(N,\ell)}_\ell\ \ ,}
where the constant of proportionality is a polynomial in $e^{\pm\epsilon}$
which
depends on the $m(\ell)$.  This constant of proportionality is
finite as $\epsilon\to 0$, so we do not introduce any new factors of
$1/\epsilon$.  As a result, $\Delta f = 0$.

In terms of the graph, what we have just done is to remove the vertex
corresponding to $z_N$; remove all factors of $z_N$ in the numerator; remove
all edges connected to $z_N$; and include the factors $z^{-m(\ell) -
a(N,\ell)}_\ell$ in the numerator.  In Figure 6, we give an example where a
residue at a $z_1=0$ pole is evaluated and $m(2)$ is taken to equal $m$.
Each graph obtained by evaluating a residue at $z_N=0$
still has the standard form described by Eq.~\FNzdef,
but with one less vertex.  However, because we just removed all the edges
connected to $z_N$, a 1PI or connected graph may no longer be 1PI or
connected.  Let $e$ be the number of edges we remove, and let $c$ be the
number of connected components in the final graph.  Suppose
we start with a 1PI graph. Then, when we remove a vertex
and all edges adjoining it, we must, by definition, remove at least two edges.
In addition, in order to end up with $c$ connected components, we must remove
at least $2c$ edges.  Thus, if the original graph is 1PI, we have
\eqn\gEopi{1 + \Delta g \le -{1\over 2} \Delta E\ \ ,}
where the change in the number of connected components is given by
$\Delta g = c-1$; and the change in the number of edges is given by
$\Delta E = - e$.  Similarly, if the original graph is connected, but
not 1PI, we must remove at least one edge for each connected component
in the final graph, so this time we have
\eqn\gEc{1 + \Delta g \le - \Delta E\ \ .}
Lastly, if the graph is disconnected, the evaluation of the $z=0$ residue
affects only one connected component, so again we have Eq.~\gEopi\
if the component is 1PI and Eq.~\gEc\ if the component is only connected.

We are now in a position to calculate the degree of divergence of the new
graph according to Eq.~\ddef, with the result
\eqn\dopi{d(N-1) \le d(N) - \Delta g-1 \le d(N)-1 }
for a 1PI graph; and
\eqn\dc{d(N-1)\le d(N)}
for a connected graph.  Thus, the degree of divergence always remains
constant or decreases.  More specifically, when the graph
is 1PI, it decreases by at least 1; and it decreases by exactly one only
when two edges are removed and the graph remains connected.  For a connected
graph, $d$ remains constant only when the final number of connected
components equals the number of edges that are removed.  Otherwise, it
decreases.  We also note that the final
graph is still homogeneous with degree $h(N-1) = h(N)$.  To show this, we
observe that
\eqnn\dEzz
\eqnn\dpkzz
$$\eqalignno{\Delta E &= - \sum\limits^{N-1}_{\ell=1} a(\ell,N)
\ \ ;&\dEzz \cr
\Delta (p_\ell + k_\ell) &= - m(\ell) - a(\ell,N) \quad\hbox{for}\ \ \ell\not=N
\ \ ;&\dpkzz\cr}$$
and we removed $p_N + k_N =-\sum\limits^{N-1}_{\ell=1} m(\ell)$ factors of
$z_N$ from the numerator.  Substituting these expressions back into
Eq.~\hdef\ for $h$, we find that for the total graph, $\Delta h = 0$.
However, if the graph becomes disconnected, each connected component
could have a different degree of homogeneity.

Finally, for $i,j\ne N$, let us define
\eqnn\newpzz
\eqnn\newpijzz
\eqnn\newazz
\eqnn\newkzz
$$\eqalignno{p_i(N-1)=& p_i(N)- p_{iN}(N)\ \ ;&\newpzz\cr
p_{i j}(N-1) =& p_{i j}(N) \ \ ;&\newpijzz\cr
a\left( j,i;N-1\right) =& a (j,i;N) \ \ ;&\newazz\cr
\noalign{\hbox{and}}
k_i (N-1) =& k_i(N)+ p_{i N}(N) - a (N,i;N) - m(i)\ \ .
&\newkzz\cr}$$
It follows from these definitions that if conditions (1) and (2) are
satisfied before we evaluate the residue, they
are still true afterwards.  Furthermore, if condition (1) is true, then it is
straightforward to show that
\eqn\ksumzz{
\sum\limits^{N-1}_{i=1} k_i(N-1) = \sum\limits^N_{i=1} k_i(N)
\ \ ,}
so the sum over the $k$'s remain constant.

We conclude that results (R.1), (R.3), (R.5), (R.6) and (R.8)
for the $z_N = e^{-\epsilon}z_j$
pole are still valid for the $z_N=0$ pole.  Also, the degree of divergence
still
decreases or remains constant upon integration.

We will now focus on the 1PI graphs with degree of divergence equal to one
because all the graphs for the free energy with zero magnetic field have these
two properties.  Then, because $d(N)$ is monotonically decreasing,
in order to obtain non-vanishing results as $\epsilon\to 0$, we need only
to consider residues where $d$ remains constant or is decreased exactly by 1.
Thus, according to Eq.~\dopi, when we evaluate the residue at
$z_N=0$ of a 1PI graph, the only situation that gives an
answer that does not vanish as $\epsilon$ goes to zero is when $z_N$ is
connected to exactly two edges.  In that case the resulting graph is
connected and possibly 1PI, and it has degree of divergence equal to 0.
We will now directly evaluate these residues in more detail.
As before, we need only consider the integral
of $F_2({\svec z})$, where $F_2({\svec z})$ is the part of $F(N;{\svec z})$
depending on $z_N$.  In this case the integral of $F_2({\svec z})$ is given by
\eqn\Ftzint{
\oint {dz_N\over 2\pi i z_N} F_2({\svec z}) = \oint {dz_N\over 2\pi i z_N}
z^{p_N}_N {1\over z_N - e^{-a\epsilon} z_\ell}\ {1\over z_N -e^{-b\epsilon}
z_j} z^{k_N}_N\ \ .}
The multiplicity, $m+1$, of the pole at $z=0$ is equal to $-k_N-p_N+1$.
We will define
\eqn\resRes{{\rm res} = { \hbox{Res}}\, \left[{1\over 2 \pi i z_N}
F_2({\svec z})\right]\bigg|_{z_N=0}\ \ .}
Calculating this residue, we find
\eqn\resFtzz{\eqalign{{\rm res} &= \theta(m+1) {1\over
m!} \sum^m_{n=0} {m\choose n} {\partial^n\over \partial z^n_N} \left( {1\over
z_N - e^{-a\epsilon}z_\ell}\right) {\partial^{m-n}\over \partial z^{m-n}_N}
\left( {1\over z_N - e^{-b\epsilon}z_j}\right)\Bigg|_{z_N=0} \cr
&= \theta(m+1) \sum^m_{n=0} e^{(n+1)a\epsilon}e^{(m-n+1)b\epsilon}
z^{-(n+1)}_\ell z^{-(m-n+1)}_j\ \ .\cr}}

First we will consider the case where $\ell$ and $j$ are equal. Then the
residue reduces to
\eqn\reszsq{\hbox{res} = \sum^m_{n=0} e^{(n+1)a\epsilon} e^{(m-n+1)b\epsilon}
z^{-m-2}_\ell \theta(m+1)\ \ .}
Using the definition of $m$ and taking the limit as $\epsilon$ goes to 0, we
find
\eqn\limres{\lim_{\epsilon\to0}(\hbox{res})  = \theta\left( - k_N - p_N
+1\right) \left[ - k_N - p_N + 1 + {\cal O}(\epsilon)\right] z^{k_N + p_N
-2}_\ell\ \ .}
For $i,j\ne N$, we can define the new $p$'s, $a$'s and $k$'s as in
Eqs.~\newpzz, \newpijzz, \newazz\ and \newkzz.
Then the $k_i$ are given by
\eqn\newki{
k_i (N-1) = \cases{k_j(N) &for $i\ne \ell$ \cr\noalign{\vskip 0.2cm}
k_\ell(N) + k_N + p_N + p_{\ell N} - 2 &for $i=\ell$;\cr}}
and, when condition (1) is satisfied, the expression for $k_\ell(N-1)$
reduces to $k_\ell (N-1) = k_N + k_\ell(N)$.  We also note that,
in effect, we have just removed a tadpole from a 1PI graph, so the remaining
graph is still 1PI.  We conclude that the full residue is given by
\eqn\Resjeql{
\hbox{Res}\; \left[{1\over2\pi i z_N}F(N;{\svec z})\right] \bigg|_{z_N=0}
= \theta \left( - k_N - p_N +
1\right) \left[ - k_N - p_N + 1 + {\cal O}(\epsilon) \right] \tilde
F(N-1;{\svec z})\ \ ,}
where $\tilde F (N-1;{\svec z})$ is a 1PI graph with $N-1$ vertices,
which has degree of divergence equal to 0 and satisfies condition (1),
and for which
$\sum\limits^{N-1}_{j=1}k_j(N-1) = \sum\limits^N_{j=1} k_j(N)$.

Next, we consider the case when $z_\ell\not=z_j$.  In the discussion of the
general case of the $z_N=0$ pole, we have seen that each term in the sum in
the last line of Eq.~\resFtzz\ corresponds to a graph with $N-1$ vertices
with degree of divergence equal to 0.  In all subsequent integrations, we have
shown that $d$ decreases or remains the same, so the final answer for each
graph will be finite or zero as $\epsilon\to 0$.  To show that the finite part
is piecewise linear and homogeneous in ${\svec k}$, we will rewrite all the
graphs in Eq.~\resFtzz\ as a sum of two graphs which will be 1PI, have the
standard form given by Eq.~\FNzdef\ and satisfy condition (1).

We begin with the identity
\eqn\xyid{{x^{m+1} - y^{m+1}\over x-y} = \sum^m_{n=0} x^n y^{(m-n)}\ \ .}
We can use this identity to evaluate the sum
\eqn\Sxydef{S = \sum^m_{n=0} x^{-(n+1)} y^{-(m-n+1)}\ \ ,}
with the result
\eqn\Sxyfin{S = {y^{-m-1}\over x-y} - {x^{-m-1}\over x-y}\ \ .}
The last line in Eq.~\resFtzz\ has the same form as $S$, with the
identifications $x= z_\ell\, e^{-a\epsilon}$, $y = z_j e^{-b\epsilon}$, and
$m = - k_N-p_N$.  It follows that
\eqn\restg{\eqalign{\hbox{res}\; &= \theta \left( - k_N -
p_N+1\right) {e^{-b\left( k_N + p_N -1\right)\epsilon} z^{k_N + p_N-1}_j \over
e^{-a\epsilon} z_\ell - e^{-b\epsilon}z_j} \cr
&- \theta \left( - k_N - p_N+1\right) {e^{-a\left( k_N - p_N-1\right)\epsilon}
z^{k_N + p_N-1}_\ell \over
e^{-a\epsilon} z_\ell - e^{-b\epsilon}z_j} \ \ .\cr}}
Without loss of generality, consider the graph which corresponds to the
full residue due to the first term in the preceding equation.  It is
\eqn\Fjdef{
\tilde F(j) = \theta \left( - k_N - p_N + 1\right) {e^{-b\left( k_N + p_N
-1\right)\epsilon} z^{k_N+p_N-1}_j \over
e^{-a\epsilon}z_\ell - e^{-b\epsilon}z_j} F_1({\svec z})\ \ .}
This graph is obtained from $F(N;{\svec z})$ by removing the vertex $z_N$
and the two edges that join it to $z_\ell$ and $z_j$, respectively, and then
reconnecting $z_\ell$ and $z_j$ with one edge.  Because these were the only
edges coming from $z_N$, it follows that if $F(N;{\svec z})$ was originally
1PI, then the new graph is 1PI; and if it was originally connected, the
new graph is connected.

For the purpose of calculating the degree of divergence, we observe that in
obtaining the new graph, $\tilde F(j)$, the number of vertices and the total
number of edges each decrease by one, while the number of factors of
$\epsilon$ and the number of connected components remain constant.  Then,
according to Eq.~\ddef, the degree of divergence for $\tilde F(j)$ is the
same as for $F(N;{\svec z})$.  Therefore, if the original graph had
$d(N)=1$, then the new graph still has degree of divergence equal to one.
However, because the total residue of $F(N;{\svec z})$ must be finite, we know
that the divergent parts of the two graphs obtained from Eq.~\restg\ must
cancel.

Finally, we will show that if $F$
satisfies conditions (1), (1a) and (2), then $\tilde F(j)$ does also.
First, we can define the new $p$'s, $a$'s and $k$'s
for $\tilde F(j)$ as follows:
\eqnn\pikopi
\eqnn\aopi
\eqnn\kopi
$$\eqalignno{p_{ik} (N-1)&= \cases{p_{ik}(N)
&for  $\{i,k\}\not=\{j,\ell\}$ \cr \noalign{\vskip 0.2cm}
p_{\ell j} (N) + p_{\ell N} &for $i=\ell$, $k=j$
\cr\noalign{\vskip 0.2cm}
p_{j\ell} (N) + p_{jN}+p_N-1 &for $i=j$, $k=\ell$;\cr}&\pikopi \cr
\noalign{\vskip 0.3cm}
a(i,k;N-1)&= \cases{a(i,k;N) &for $\{i,k\}\not= \{j,\ell\}$ \cr
\noalign{\vskip 0.2cm}
a(j,\ell;N) + 1 &for $\{i,k\}=\{j,\ell\}$\ \ ;\cr}&\aopi \cr
\noalign{\hbox{and}}
k_i (N-1)&= \cases{k_i(N) &for $i\not=j$\cr \noalign{\vskip 0.2cm}
k_j(N) + k_N &for $i=j$.\cr}&\kopi \cr}$$
Then straightforward calculations show that the full residue is
\eqn\resFopi{\eqalign{
\hbox{Res}\;\left[{1\over 2\pi i z_N} F(N;{\svec z})\right]\bigg|_{z_N=0}
&= \theta\left( - k_N - p_N +1\right)
e^{-b(k_N + p_N -1)\epsilon} \tilde F \left( {\svec z}, {\svec k} + k_N
 \hat e_j\right) \cr
&- \theta \left( - k_N - p_N +1\right) e^{-a\left( k_N + p_N-1\right)\epsilon}
\tilde F\left( {\svec z},{\svec k} + k_N  \hat e_\ell
\right)\ \ ,\cr}}
where $\tilde F({\svec z},{\svec k})$ is a 1PI graph of standard form
with the same degree of divergence as $F(N;{\svec z})$.
Given the definition for the new $k_i$'s, it is clear that the sum over the
$k$'s remains constant.  The original graph had
exactly one edge connecting $z_N$ to $z_j$ and
$z_\ell$, respectively, so $a(N,j;N) = a(N,\ell;N) = 1$ and
$p_N = p_{N\ell} + p_{N j}$.  Therefore, if $F$ satisfied condition (1), then
the expression for $p_{j\ell}(N-1)$ reduces to
\eqn\pjlopi{p_{j\ell}(N-1)=p_{j\ell}(N)+p_{N\ell};}
and $\tilde F(j)$ also satisfies condition (1).  If $F$ also satisfied
condition (2), then, to determine whether $\tilde F(j)$ satisfies condition
(2) we need only to check whether $p_{j\ell}(N-1)$ and $p_{\ell j}(N-1)$ are
positive.  Because the
$p_{ik}(N)$ are integers or half-integers, conditions (1) and (2) imply
that $p_{N\ell}(N) = p_{\ell N}(N) = {1\over 2}$.
It follows that condition (2) is true in the new graph.

For connected graphs, we will also encounter $z_N=0$ poles where $z_N$ has
only one vertex, $z_\ell$, connected to it.  The calculation of the residue is
similar to those we have just done, and the result is
\eqn\zzoe{
\hbox{Res}\;\left[{1\over 2\pi i z_N} F(N; {\svec z})\right]\bigg|_{z_N=0}
= - \theta \left( - k_N - p_N + 1\right)
e^{-a \left( k_N + p_N - 1\right)\epsilon} \tilde F({\svec z})
\ \ ,}
where $\tilde F({\svec z})$ is a connected graph with the same degree of
divergence as $F(N;{\svec z})$.  When $F(N; {\svec z})$ satisfies condition
(1), then $\tilde F({\svec z})$ is obtained from $F(N;{\svec z})$ by removing
the vertex $z_N$, the edge joining $z_N$ to $z_\ell$, and the factors of
$z_\ell$ in the numerator due to this edge; and the new ${\svec k}$ is given by
\eqn\kjcon{k_j (N-1) = \cases{k_j(N) &for $j\not=\ell$
\cr\noalign{\vskip 0.2cm}
k_\ell (N-1) + k_N &for $j=\ell$.\cr}}

All of these calculations are valid until we reach the final integration in a
connected component.  The final integration has the form
\eqn\finint{q\left(e^{-\epsilon};{\svec k}\right)=
\oint {dz_1\over 2\pi i z_1} z^{p_1(1) + k_1(1)}_1 \times q\left(
e^{-\epsilon}; {\svec k};1\right)\ \ ,}
where $q(e^{-\epsilon}; {\svec k};1)$ is some piecewise-rational
function of $e^{-\epsilon}$
which depends on the original ${\svec k}$.  The degree of divergence for the
graph in the integrand is
given by $d(1)=f$, where $q\left(e^{-\epsilon};{\svec
k};1\right) \sim \epsilon^{-f}$.   This integral equals
$q(e^{-\epsilon};{\svec k};1)$ if $p_1+ k_1 = 0$, and it equals 0 otherwise.
We have shown that if the graph always remains connected, then the degree of
homogeneity remains constant. Consequently, $p_1+k_1=0$ only if $h=0$
originally.  In addition, if the graph satisfies condition (1), we must have
$p_1(1)=0$.  This implies that the integral is zero, unless $k_1(1)=0$.
If condition (1a) is also satisfied, then, to obtain a non-vanishing result,
we must have $\sum_{i=1}^N k_i(N)=0$.  In
that case the integral is given by
\eqn\qfin{q\left( e^\epsilon;{\svec k}\right) \sim \epsilon^{-f}\ \ .}
The actual degree of divergence of this is $f$.  Therefore, when the
final residue is non-zero, the actual degree of divergence is always equal to
the degree of divergence of the graph with one remaining vertex, which
in turn is always less than or equal to the superficial degree of
divergence of the original graph.

In summary, when a $z_N=0$ pole is evaluated, we obtain the following
recursion relations:
\medskip
\item{R.11)}If the graph (or connected component containing
$z_N)$ is 1PI, then
the degree of divergence decreases by at least 1. If the graph  (or
connected component containing $z_N$) is only connected, then the degree of
divergence either remains constant or decreases.
\medskip
\item{R.12)}The resulting graph or graphs have the same form \FNzdef\ as the
original graph, but with one less vertex. Also,
if the original graph satisfies conditions (1) or (2), then so do the
new graphs.  However, the sum over $k$'s does not necessarily remain constant.
\medskip
\item{R.13)}If the graph is 1PI with $d=1$, the only non-vanishing
graphs that we can obtain
after a $z_N=0$ residue is evaluated are the following:
\medskip
\item{a)}A 1PI graph of the form given by Eq.~\FNzdef\ with $N-1$ vertices
and degree of divergence$ =0$.
It still satisfies conditions (1) or (2) if the original graph
does.  The sum over the $k$'s remains constant, and the new $k$'s are given
by Eq.~\newki\
for an $\ell\in (1,2,\ldots,N-1)$.  The new $q$ is given by
\eqn\newqog{
q\left( e^{-\epsilon};N-1\right) = \theta\left( - k_N - p_N + 1\right)
\left[ - k_N - p_N + 1 + {\cal O}(\epsilon)\right]
q\left( e^{-\epsilon};N\right)\ \ .}
\item{b)}The difference of two 1PI graphs of the standard form \FNzdef\
with $N-1$ vertices.  Each graph has
degree of divergence $=1$ and still satisfies condition (1) or (2) if
the original graph did.  The new $k$'s are given by Eq.~\kopi\
for some $j \in (1,2,\ldots,N-1)$, and each of the two graphs has a different
value of $j$.  The sum over the $k$'s remains constant.
The new $q$ for each graph has the form
\eqn\qopi{
q(e^{-\epsilon};N-1) = \theta\left( - k_N - p_N +1\right) e^{-b(k_N + p_N-1)
\epsilon} q (e^{-\epsilon};N)\ \ ,}
where all that can differ between the two graphs is the value of $b$.  When
the graph satisfies condition (2), this reduces to
\eqn\qopit{
q\left( e^{-\epsilon};N-1\right)= \theta (-k_N) e^{-\epsilon b k_N}
q\left(e^{-\epsilon};N\right)\ \ .}
\item{R.14)}If conditions (1) and (1a) are satisfied, then the final
integral equals zero unless $\sum_{i=1}^N k_i(N)=0$.
\newsec{FREE ENERGY AND FREE CORRELATION FUNCTIONS OF
$\pmb{e^{i{\svec q} \cdot {\svec x}(t)}}$}
In this section, we will calculate the free energy and the
Fourier transform of the free correlation functions of
$e^{i{\svec q} \cdot {\svec x}(t)}$.

We will begin with the free energy.
According to the results of Section 4, all the graphs for the free energy
can be described by Eq.~\FNzdef\ with $k=0$.  They all initially
have degree of divergence equal to 1 and are homogeneous of degree 0.
The results from the previous section show that
the degree of divergence decreases monotonically as the
residues are evaluated.  This implies that any graph for the free
energy diverges at most as
$\epsilon^{-1}$ as $\epsilon\to 0$.  As has
been remarked before in Ref.~\CT, the graphs have no logarithmic
divergences in
$\epsilon$.  Therefore, at any order in $V_0$, the free energy for the charge-0
sector diverges at most as $\epsilon^{-1}$ with no logarithm subdivergences.
\goodbreak
\bigskip
\noindent{\bf Correlation Functions with no Magnetic Field}
\medskip
\nobreak
Next, we will consider the Fourier transform of $I({\svec t})$,
the connected part of
\break
$\ll e^{i {\svec q}_1 \cdot {\svec x}(t_1)}\ldots
e^{i {\svec q}_{2n} \cdot {\svec x}(t_{2n})}\rr_0$.
When there is no magnetic field,
these graphs are the ones described in the first part of Section~4.
They have the form given by Eq.~\FNzdef\ with $k=0$ and they are
1PI. They have degree of divergence equal to 1, and are homogeneous
of degree 0. They also satisfy conditions (1) and (2), which are defined at
the beginning of Section~5.  We will be taking the Fourier transform of
$I({\svec t})$, so we will evaluate
\eqn\Itkdef{
\tilde I({\svec k}) = \oint \prod\limits^{2n}_{j=1} {dz_j\over 2\pi i z_j}
I({\svec t}) \prod\limits^{2n}_{j=1} z^{k_j}_j\ \ ,}
where $z_j = e^{2\pi i t_j/T}$.  Therefore, in Eq.~\FNzdef, the initial
$k_i$ will equal the Fourier variables, $k_j$, for $j=1$ to $2n$.

When we perform each integration by evaluating residues, the degree
of divergence will either remain constant or decrease.  According to the
results of the previous section, for a 1PI graph it only remains constant
when a residue at a $z_j = z_k
e^{-a\epsilon}$ pole is evaluated and no derivatives act on $z_j$.
Since we start out with degree of divergence equal to 1, the only way to get a
divergent answer is if we never evaluate a residue at a $z=0$
pole and never take any
derivatives of any $z_j$.  In that case, the degree of divergence is still 1.
By successively applying results (R.6), (R.7) and (R.9) at each
integration, (and (R.14) for the final integration), we
conclude that the final answer has the form
\eqn\qnzpe{
q\left( e^{-\epsilon}\right)= \prod_j
\exp\left\{-a_j \epsilon \sum\limits_{i\in S_j} k_i\right\}
P\left(e^{-\epsilon}\right)
\delta_{\Sigma_{i=1}^{2n}k_i,0}\ \ ,}
where $S_j$ are arbitrary subsets of $\{1,2,3,\ldots, 2n\}$, and
$P(e^{-\epsilon})$ is a rational function of $e^{-\epsilon}$ which is
completely independent of ${\svec k}$ and has degree of divergence equal to 1.
Therefore, as $\epsilon\to 0$,
\eqn\qnzp{
q\left( e^{-\epsilon}\right) = {c_{-1}\over\epsilon} + c_0 - c_{-1}
\sum_{j} \left( a_j \sum_{i\in S_j} k_i\right) + {\cal O}(\epsilon)
\ \ ,}
where $c_1$, $c_0$, and $a_j$ depend only on the form of the initial graph,
and not on the $k_i$'s.  (In Eq.~\qnzp, and in the rest of the paper, we
will omit the momentum-conserving delta functions.)

Result (R.4) says that when we evaluate a residue at
$z_N = e^{-a\epsilon}z_k$ and $m$ derivatives act on
$z_N$, the degree of divergence is reduced by $m$.  Therefore, to get a finite
graph, we must have only one derivative acting on $z_N$. Because the
resulting graph is still 1PI, after that, in order to obtain
non-vanishing results, we can evaluate only $z_i =
e^{-a\epsilon} z_j$ residues with no derivatives acting on $z_i$.

By applying results (R.6), (R.7), (R.9) and (R.10), we find that
the final answer is
\eqn\znzped{
q\left( e^{-\epsilon}\right) = \left( \sum\limits_{i\in S_N} k_i +
p_N\right) \prod\limits_{j} e^{-a_j\epsilon\sum\limits_{i\in S_j} k_i}
P(e^{-\epsilon})\ \ ,}
where $S_j$ and $P(e^{-\epsilon})$ have the same form as before, but now
$P(e^{-\epsilon})$ has degree of divergence equal to 0.  As $\epsilon\to 0$,
the residue becomes
\eqn\qnzpd{
q\left(e^{-\epsilon}\right) = c_0 \left( \sum\limits_{i\in S_N} k_i +
p_N\right)\ \ .}
Again, $c_0$, $S_N$ and $p_N$ are completely independent of ${\svec k}$.

Next, suppose we evaluate a $z_N=0$ residue where $z_N$ is directly
joined to only one other vertex, $z_\ell$, by two edges.  In that case,
result (R.13a) says that the degree of divergence is reduced by one,
so the graph becomes finite. Also, it remains 1PI.
Consequently, in the remaining integrals we can
no longer evaluate any $z=0$ residues or take any derivatives of any $z_i$.
Applying results (R.6), (R.7), (R.9) and (R.13a),
we find that the final integral is
\eqn\qzzee{
q\left( e^{-\epsilon}\right) = \theta\left( - \sum_{i\in S_N} k_i - p_N +
1\right) \left[ - \sum_{i\in S_N} k_i - p_N + 1 + {\cal O}(\epsilon)\right]
\prod_j e^{-a_j \epsilon\sum\limits_{i\in S_j} k_i} P(e^{-\epsilon})
\ \ ,}
where $S_j$ and $P(e^{-\epsilon})$ are defined as before, with
$P(e^{-\epsilon})\sim c_0 + {\cal O}(\epsilon)$.  In this equation and
the following equations, the value of $k_N$ when the $z_N=0$ residue is
evaluated is given by $k_N = \sum_{i\in S_N}k_i$.

Our initial graph satisfied condition (2), so at the $N^{\rm th}$ integration,
it must also satisfy condition (2).  Also, because $z_N$ was connected to
$z_\ell$ by exactly two edges, $a(N,\ell;N)=2$.  By the definition
of the $p$'s, we know that
$p_N=p_{N\ell}$ and $p_{\ell N}$ are non-negative, and $p_N\in\IZ$.
Condition (2) further implies that $p_N$ and
$p_{\ell N}$ are non-zero.  Therefore, because $a(N,\ell;N) = p_N + p_{\ell
N}$, we conclude that $p_N=1$.
Using this result, we find that as $\epsilon$ goes to zero,
$q(e^{-\epsilon})$ becomes
\eqn\qzze{
q(e^{-\epsilon}) = - c_0 \theta\left( - \sum\limits_{i\in S_N}
k_i\right) \sum\limits_{i\in S_N} k_i\ \ .}

So far, all the residues we have evaluated have the form
\eqn\indhyp{
R({\svec k}) = {c_{-1}\over \epsilon}+ c_0 + {\svec a} \cdot {\svec k} +
{\svec b}_r ({\svec k}) \cdot {\svec k} + {\cal O}(\epsilon)\ \ ,}
where
\eqn\brind{{\svec b}_r({\svec k}) = \sum_{\cal S} {\svec b}_{\cal S}
\prod_{S\in {\cal S}}
\theta \left( - \sum_{j\in S} k_j\right)\ \ .}
In this equation, ${\cal S}$ is summed over all sets of subsets of
$\{1,2,\ldots, 2n\}$, and the $S$'s are subsets of $\{1,2,\ldots,2n\}$.
If two graphs are the same except for differing values of ${\svec k}$, then
$c_{-1}$, $c_0$, ${\svec a}$ and ${\svec b}_r$ are the same for both
graphs.  In addition, ${\svec b}_{\cal S}$ has the form
\eqn\bsehat{{\svec b}_{\cal S} = \sum_{S\in {\cal S}} c(S)
\sum_{j\in S}\hat e_j\ \ ,}
for $c(S) \in \IR$.  This implies that ${\svec b} \cdot {\svec k}$
depends only on the $k_i$ through the sums $k_S = \sum_{j\in S} k_j$, where
$k_S$ was the momentum at some vertex, $z_{i_S}$, when a $z_{i_S}=0$ pole
was evaluated.

We will assume that all residues of all graphs with $N-1$ vertices have this
form, and induct on $N$.  In fact, the only type of pole that gives
non-vanishing residues as $\epsilon \to 0$
that we have yet to consider is a
$z_N=0$ pole when $z_N$ is directly
connected to exactly two vertices, $z_j$ and $z_\ell$, by one edge each.  In
that case, according to result (R.13b) the residue is
\eqn\RFF{
R({\svec k}) = \theta (-k_N) \left[ e^{-\epsilon bk_N} F\left( {\svec z},
{\svec k}+k_N \hat e_j\right) - e^{-\epsilon ak_N}
F\left( {\svec z}, {\svec k} + k_N \hat e_\ell\right)\right]
\ \ ,}
where $F({\svec z}, {\svec k})$ is a 1PI graph with $N-1$ vertices that
has the form given by Eq.~\FNzdef, satisfies conditions (1) and (2),
and has degree of divergence
equal to 1.  Therefore, by the induction hypothesis, after we integrate over
$F$, we obtain
\eqn\indF{
F\left( {\svec z}, {\svec k}\right) = {c_{-1}\over \epsilon} + c_0 + {\svec
a} \cdot {\svec k} + {\svec b}_r \cdot {\svec k} + {\cal O}(\epsilon)
\ \ ,}
for some $c_{-1}$, $c_0$, ${\svec a}$ and ${\svec b}_r$.  Substituting this
into Eq.~\RFF\ and simplifying, we find that the
leading terms of $R({\svec k})$ are
\eqn\newRk{R({\svec k}) = \theta(-k_N) \left[ k_N c_{-1}(a-b)
+ k_N{\svec a} \cdot (\hat e_j - \hat e_\ell)
+ {\svec b}_r({\svec k}_j) \cdot {\svec k}_j
- {\svec b}_r({\svec k}_\ell) \cdot {\svec k}_\ell\right]\ \ ,}
where we have defined ${\svec k}_j = {\svec k} + k_N \hat e_j$ and ${\svec
k}_\ell = {\svec k} + k_N \hat e_\ell$.  This has the form $R({\svec k})=
{\svec b}_r\cdot {\svec k}$ for some new value of ${\svec b}_r$,
so by induction, all
non-vanishing residues have the form given by Eq.~\indhyp.  Note also that
as $k_N\to 0$, the expression in brackets goes to zero.
This is sufficient for showing that the Fourier coefficients of the
correlation functions are continuous functions of ${\svec k}$
if we view them as functions on $\IR^{2n}$. This result is useful for solving
for the correlation functions, but we will not show it in detail here.

The full integral is the sum over all of these residues.
Therefore, we conclude that the Fourier transform of the free, connected
correlation functions of $e^{\pm ix(t)}$ are all piecewise linear in
${\svec k}$ plus a constant.  They have the form
\eqn\fullint{\tilde I({\svec k}) = {c_{-1}\over\epsilon} + c_0 +
{\svec a} \cdot {\svec k} + {\svec b}_r \cdot {\svec k} + {\cal O}(\epsilon)
\ \ ,}
for $c_{-1}$ and $c_0\in \IR$, ${\svec a}\in \IR^{2n}$ and
for ${\svec b}_r$ as defined in Eq.~\brind.
We stress that this form is independent of the number of $e^{\pm ix(t)}$'s in
the correlation function, as long as there are an equal number of $+x$'s and
$-x$'s.  This is one of the main results of this paper.
\goodbreak
\bigskip
\noindent{\bf Correlation Functions with a Magnetic Field}
\medskip
\nobreak
Next, we wish to evaluate the Fourier transform of $I({\svec t})$ in
the presence of a
non-zero magnetic field when $\beta/\alpha\in\IZ$.  Almost all of the preceding
proof still applies, except that now the graphs are not always 1PI and do not
always satisfy condition (2).  In fact, according to the second part of
Section~4, any graph, $G(2n)$, can now
have several subgraphs,
$G_1,\ldots, G_M$, each of which are 1PI, have degree of divergence $= 1$,
and satisfy conditions (1) and (2).  Furthermore,  each subgraph has
only $x$-vertices or only $y$-vertices and its total
charge is 0.  Thus any one of these graphs is identical to the ones we just
considered when there is no magnetic field.
Each of the subgraphs, $G_i$, is ``connected'' to at least one other
subgraph, $G_j$, either by some factors of
$(e^{-\epsilon}-1)\sim -\epsilon$, or by at least one edge and factor of $z$
corresponding to
\eqn\maglink{
\left( e^{-2\epsilon}-1\right) z_k/\left(z_k - e^{-\epsilon}z_\ell\right)
\ , \quad\hbox{ or }\quad \left(
e^{-2\epsilon}-1\right) z_\ell/\left(z_\ell - e^{-\epsilon}z_k\right)\ \ ,}
where $z_k\in G_i$ and $z_\ell\in G_j$.  The full graph still has total
degree of divergence less than or equal to 1.

For convenience, we will define the following notation.  Let $S_j$ denote
the set of integers, $i\in\{1,2,\ldots,2n\}$, such that if $i\in S_j$, then
$z_i$ is in the $j^{\rm th}$ subgraph, $G_j$.  Also, we will define
${\svec K} = (K_1,\ldots,K_M)$ by $K_j = \sum_{i\in S_j} k_i$, so $K_j$ is
the total momentum in the subgraph $G_j$.  We will call the edges in \maglink\
``magnetic-field edges".  Lastly, any vertex that is not connected to a
magnetic-field edge will be called an ``internal vertex" and any vertex that
is connected to a magnetic-field edge will be called an ``external vertex".

We will begin by evaluating the Fourier transform of the 1PI magnetic
field graphs.  Most of the preceding proofs for zero-magnetic field graphs
still apply, with the result that the Fourier transforms are
piecewise-linear plus piecewise-constant functions.  However, these
functions are no longer homogeneous in the momenta.  Also, to prove the
duality transformation, we would like the correlation functions of
$\dot x(t)$ and $\dot y(t)$ to depend on the momenta in only one subgraph.
In the next section, we will show that for the weaker version of the duality
transform to be true, it is sufficient that each term in the Fourier
transform of the magnetic field graphs depends only on ${\svec K}$ and on
the $k_j$ which are in a particular subgraph.

We will define our order of integration as follows.  First integrate over
all the internal vertices of $G_M$ and then over all the external vertices
in $G_M$.  Then repeat this process with
$G_{M-1}, G_{M-2}, \ldots , G_1$, where a
point is considered internal or external depending on whether it was
internal or external in the original graph.  We will call the original graph
$G$ and any graph obtained after the $j^{\rm th}$ integration we will call
$G(2n-j)$.  We will also assume that the vertices are labeled so that the
order of integration is $z_{2n}, z_{2n-1}, \ldots, z_1$.
In the calculation for zero magnetic field, all integrations that
give non-vanishing results consist of collapsing an edge and vertex, and
possibly adding extra copies of an edge.  By collapsing an edge and vertex,
we mean that we first remove a vertex, $z_j$, and all edges connecting $z_j$
to another vertex, $z_i$.  Then, all the remaining edges that had been
connected to $z_j$ are re-connected to $z_i$.  (According to the definition
of the magnetic-field edges and the 1PI subgraphs and
the redefinition of the $p_{jk}$'s in \newpjlko\ and \newpjlkm,
after we evaluate a residue at a $z_N= e^{-a\epsilon} z_j$
pole, where $z_N \in G_1$ and $z_j \in G_2$ are joined
by  magnetic-field edges, the resulting subgraph due to $G_1$ and $G_2$
again looks like the usual 1PI subgraph with no magnetic-field edges.)
If no $z=0$ residue is
evaluated until the final integration, we will obtain an answer of the form
\eqn\Rkmnzp{R({\svec k})
= {c_{-1}\over \epsilon}+ c_0 + {\svec a}\cdot {\svec k}\ \ ,}
as before.  In addition to being a linear function plus a constant function,
$R({\svec k})$ can be divided into a sum of functions, each depending only
on the $k_i \in G_j$ for a particular $j$.

When we evaluate a $z=0$ residue, if the vertex we are integrating is not
connected to any magnetic-field edges, then the factors of $z$'s in the
numerator work the same as for the zero magnetic field graphs.
Consequently, these integrations only yield homogeneous results.  It remains
to be shown that such residues depend only on ${\svec K}$ and on $k_i\in
G_j$ for a particular $G_j$.

First, consider the case where residues at $z_j=0$ poles are evaluated only
for $z_j$ which are internal vertices with respect to the original graph.
Suppose we have integrated over all the vertices in $G_{m+1}, G_{m+2}, \ldots,
G_{M}$, and we are about to integrate over an internal vertex of $G_m$.
Then the integrand can be factored into two parts, ${\cal G}_a$ and
$I({\cal G}_b)$.  ${\cal G}_a$ is equal to the subgraphs $G_1,
G_2,\ldots, G_m$ and the edges that connect them.  $I({\cal G}_b)$ is due
to integrating over the subgraphs $G_{m+1},\ldots,G_M$ and depends only on
the external vertices in $G_1, G_2, \ldots,G_m$.  If, for some vertex,
$z_i \in G_J$ for $m+1 \le J \le M$, we have evaluated a $z_i = 0$ residue,
then $I({\cal G}_b) {\cal G}_a$ has degree of divergence equal to zero.  By
repeated applications of Eqs.~\resFtzz\ and \zzoe, we find that
$I({\cal G}_b) = \sum_i f_i({\cal G}_b)$, where each
$f_i({\cal G}_b){\cal G}_a$ is a graph with degree of divergence less
than or equal to $0$.  Because ${\cal G}_a$ is the union of
1PI graphs, when we evaluate a $z_j=0$ residue for $z_j$ an internal
vertex of ${\cal G}_a$, the degree of divergence is reduced by one.  We
conclude that in order to obtain a non-vanishing result, we cannot evaluate a
series of $z_j=0$ residues when the $z_j$'s are internal vertices from more
than one subgraph.
Except for this additional condition, as long as no residues at $z_j=0$,
for $z_j$ an external vertex, are evaluated until the final integration, the
whole calculation for zero magnetic field still applies.  We obtain
\eqn\Rkivzz{
R({\svec k}) = {c_{-1} \over \epsilon}+c_0+{\svec a}\cdot{\svec k} +
{\svec b}_r \cdot {\svec k}\ \ ,}
with
\eqn\bridef{(b_r)_i = \cases{\sum\limits_{\cal S} b_{i{\cal S}}
\prod\limits_{S\in {\cal S}}
\theta\left(-\sum\limits_{j\in S} k_j\right) &for $i\in S_m$\cr
\noalign{\vskip 0.2cm}
0 &for $i\notin S_m$,}}
where $S_m$ corresponds to a particular subgraph, $G_m$; ${\cal S}$ is
summed over all sets of subsets of $S_m$; $S$ runs through all the subsets
of $S_m$ contained in ${\cal S}$; and $b_{i {\cal S}} \in \IR$.

Even when $z_j \in G_m$ is originally an external vertex, it is possible at
the $(2n-j)^{\rm th}$ integration for it no longer to be connected to any
magnetic-field edges.  In order for the residue at $z_j=0$ to be
non-vanishing, the counting in Eqs.~\gEopi\ and \dopi\ implies that it must
must be connected to exactly two edges. Because neither of these are
magnetic-field edges, this can only happen if all but one of the subgraphs,
$G_m,\ldots,G_J$,  which at some stage in the integration
are connected to $z_j$, have collapsed to a point.  For convenience, we will
take $G_m$ to be the subgraph that has not collapsed to a point and assume
that the subscripts of $G_m,\ldots, G_J$ are consecutive integers.  The rules
for the change in ${\svec k}$ imply that the only way the $k_i$, for
$i \in G_{m+1},\ldots,G_J$, appear in the graph, $G(j+1)$, is through the
${\svec K}_i$ for $m+1 \le i \le J$.
In addition, they appear only in the momentum at
the $z_j^{\rm th}$ vertex, which has the form
\eqn\kjKk{k_j(j+1)
= \sum_{i=1}^J K_i(2n) + \sum_{\ell \in S} k_\ell(2n)\ \ ,}
where $S \subset S_m$.  On the left-hand side of this equation, the momentum
is evaluated just before the integration over $z_j$, and on the right-hand
side, the momenta are evaluated with respect to the original graph.
Apart from this change in the momentum at $z_j$, the evaluation of the
$z_j=0$ residue is the same as if $z_j$ were an internal vertex with respect
to the original graph.  Therefore, the residues still have the same form as
in Eq.~\Rkivzz, except that now ${\svec b}_r$ has the form
\eqn\brivzz{
{\svec b}_r = \sum_{\cal S} \sum_{\cal S'} {\svec b}_{\cal S,S'}
\prod_{S\in{\cal S}} \prod_{S'\in{\cal S'}}
\theta\left(-\sum_{j\in S} k_j - \sum_{j\in S'} K_j\right)\ \ ,}
where ${\cal S}$ and ${\cal S'}$ are summed over all sets of subsets of
$S_m$ and $\{1,2,\ldots,M\}$, respectively; and $S$ and $S'$ are subsets of
$S_m$ and $\{1,2,\ldots,M\}$, respectively.  Also, ${\svec b}_{\cal S,S'}$
is chosen so that ${\svec b}_r \cdot {\svec k}$ depends only on the $k_j$
with $j \in S_m$ and on ${\svec K}$.

Most of the remainder of this section is devoted to evaluating the residue
of a $z_N=0$ pole when $z_N$ is connected to a magnetic-field edge.  We will
do this case carefully, because this is where the inhomogeneity in
${\svec k}$ arises.
According to the parenthetical comment before Eq.~\Rkmnzp, after we have
already performed several integrations, the 1PI subgraph containing $z_N$
could actually be due to several subgraphs of the original graph, $G$.  For
simplicity, we will assume that it just came from one subgraph, $G_m$, in the
original graph, but everything works the same if we take the general case
instead.

Because we are starting with a 1PI graph with degree of divergence $=1$,
Eqs.~\gEopi\ and \dopi\ imply that the only way to get a non-vanishing answer
is if $z_N$ is connected to exactly two edges.  $G_m$ is also 1PI, even
after some of its vertices have been integrated, which means that if $z_N$
is connected to an edge that is not a magnetic-field edge, it must be
connected to at least two such edges.  However, because we chose $z_N$ to
be connected to at least one magnetic-field edge, it follows that $z_N$ is
connected to exactly two magnetic-field edges.  This situation only arises
when $z_N$ is the final integration variable in $G_m$,
and the two edges that are connected to $z_N$ have the form
\eqn\tezN{
{\left( e^{-2\epsilon}-1\right) z^{p_{Nj}}_N z^{p_{jN}}_j \over z_N -
e^{-a\epsilon}z_j} \ {\left( e^{-2\epsilon}-1\right) z^{p_{N\ell}}_N
z^{p_{\ell N}}_\ell \over z_N - e^{-b\epsilon}z_\ell}\ \ ,}
for $z_j, z_\ell \notin G_m$.  In this equation,
$p_{Nj}$, $p_{jN}$, $p_{N\ell}$, $p_{\ell N}$ are non-negative
integers with $p_{Nj} + p_{jN} = p_{N\ell}+p_{\ell N}=1$.
In addition, because the graph is 1PI,
the two edges must go to the same connected component.  We will call this
graph ``$G(N)$" and these edges ``$e_1$" and ``$e_2$".

In order to calculate the correlation functions recursively, we will find it
useful to divide the original graph, $G(2n)$, into two disjoint parts, one
that contains $z_N$ and $G_m$ and one that contains $z_j$ and $z_\ell$.  To
do this, we recall that all the integrations so far that do not give
vanishing results as $\epsilon\to 0$ just consist of collapsing an edge and
a vertex and possibly also making extra copies of already existing edges.
It follows that the two edges, $e_1$ and $e_2$, which are
attached to $z_N$ in the graph $G(N)$, originally came from either one or
two edges, $e_1'$ and $e_2'$, in the initial graph, $G(2n)$.  Suppose there
is a path in $G(2n)$ from $z_N$ to either $z_j$ or $z_\ell$ which does not
pass through $e_1'$ or $e_2'$.  Then, because all we are doing is collapsing
edges, in $G(N)$ there will still be a path joining $z_N$ to $z_j$ or
$z_\ell$ without passing through $e_1$ or $e_2$.  However, there is no such
path in $G(N)$.  Thus, we conclude that when we remove $e_1'$ and $e_2'$
(and the factors of $z$'s and $\epsilon$'s associated with these edges), the
original graph must have two separate components, ${\cal G}_a$ and ${\cal
G}_b$, with $z_N \in {\cal G}_a$ and $z_j, z_\ell\in{\cal G}_b$.  Similar
reasoning shows that $G_m \subset {\cal G}_a$.
${\cal G}_a$ and ${\cal G}_b$ look like connected graphs for non-zero magnetic
field with fewer than $2n$ vertices; they still have the correct form for
such graphs with $d=-1$, except that $\sum_{i \in {\cal G}_a} k_i$ and
$\sum_{i \in {\cal G}_b} k_i$ do not necessarily equal zero.

In the graph, $G(N)$, the vertex, $z_N$,
must be the only remaining vertex from ${\cal G}_a$.  After
having integrated over all the vertices in ${\cal G}_a$ except for $z_N$,
Eq.~\finint\ and the following discussion imply that we will be left with
\eqn\Gaint{{\cal G}_a \to I({\cal G}_a) z_N^{p_N(1) + k_N(1)}\ \ ,}
where $I({\cal G}_a)$ equals the integral of ${\cal G}_a$ when
$p_N(1)+k_N(1)=0$.  Because ${\cal G}_a$ satisfies condition (1), we have
$p_N(1) = 0$.  Also, our redefinitions of ${\svec k}$ at each integration
imply that $k_N(1) = \sum_{i \in {\cal G}_a} k_i$.  We note that
$I({\cal G}_a)$ is a function only of those $k_i$ which are in
${\cal G}_a$.

The full $z_N$ integral can then be written as
\eqn\magzz{
I(G;N) = I ({\cal G}_a) \times \oint {dz_N\over 2\pi i z_N} \  z^{p_{Nj} +
p_{N\ell} + k_N}_N {\left( e^{-2\epsilon}-1\right)^2\over z_N - e^{-\epsilon
a} z_j} \  {z^{p_{jN}}_j z^{p_{\ell N}}_\ell \over z_N - e^{-\epsilon b}
z_\ell} \times {\cal G}_b\ \ .}
In this equation, $I(G;N)$ is defined to be a sum of residues of the graph,
$G$, after the $N^{\rm th}$ integration, and $I(G)$ is from the integral
over all the variables in $G$.
Now we can apply our general results for integrating the $z=0$ poles.  If
$z_j = z_\ell$  we get from Eq.~\reszsq\
\eqn\magzsq{
\eqalign{I(G;N) &= I({\cal G}_a) \times \theta\left( m + 1 \right)
z^{p_{\ell N}  - m - 2}_\ell
\times {\cal G}_b\cr
&\times \left( e^{-2\epsilon}-1\right)^2
\sum\limits^{m}_{n=0} e^{(n+1)a\epsilon}
e^{(m-n+1)b\epsilon\ \ ,} \cr}}
where $m=-p_{N\ell}-k_N$.
Next, we use the condition that $p_{N\ell} + p_{\ell N}=2$ to
obtain, as $\epsilon\to0$,
\eqn\magzeq{
I(G;N) = 4 \epsilon^2 \theta\left(- p_{N\ell} - k_N + 1\right)
\left[ - p_{N\ell} - k_N+1\right] I({\cal G}_a) \times
\left[ {\cal G}_b\cdot z^{k_N}_\ell\right]\ \ .}
The recursion rules for ${\svec k}$ and total momentum conservation imply
that $k_N$ in Eq.~\magzeq\ is given by
\eqn\kNGaGb{k_N = \sum\limits_{i\in{\cal G}_a} k_i(2n)
= - \sum\limits_{i\in {\cal G}_b} k_i(2n)\ \ .}
In the graph, $z^{k_N}_\ell {\cal G}_b$, we can define the
new $k_\ell$ to be $k_\ell+k_N$.
Then ${\cal G}_b z^{k_N}_\ell$ has total momentum equal to zero and
is exactly a 1PI magnetic-field graph with fewer vertices than the
original graph, $G(2n)$.  Because all magnetic-field
graphs have degree of divergence less than or equal to 1,
the only way we can obtain a finite answer is if
$I({\cal G}_a)\sim 1/\epsilon$ and the integral over ${\cal
G}_bz^{k_N}_\ell$ is also $\sim 1/\epsilon$.  If both graphs were 1PI,
the proof that the divergent part of the graphs is
always independent of ${\svec k}$ would apply.  However, for connected,
one-particle reducible magnetic-field graphs, we will find that this
is not the case.  Instead, we will assume that the divergent part of a
magnetic field graph is given by ${\cal E}_{-1}({\svec K})$, a
piecewise-constant function of the total momentum in each subgraph,
${\svec K}$.  For 1PI graphs, ${\cal E}_{-1}({\svec K})$ is simply
$c_{-1}/\epsilon$.  We conclude that
\eqn\magzzl{I(G) = 4\epsilon c_{-1} {\cal E}_{-1}({\svec K})
\theta\left( - p_{N\ell} - k_N+1\right)
\left[ -p_{N\ell} - k_N +1\right]
\quad\hbox{with}\quad
p_{N\ell} = 0,1,\ \hbox{or}\ 2\ \ .}
Note that this is no longer homogeneous.  However, it depends only on
${\svec K}$ and the total
sum $k_N = \sum\limits_{i\in{\cal G}_a}k_i$, where ${\cal G}_a$ is the union of
some $G_i$'s and all the edges joining these $G_i$'s.
This will be sufficient to show that the correlation functions of
$\dot x(t)$ and $\dot y(t)$ are homogeneous in ${\svec k}$.

We return to Eq.~\magzz\ and now consider the case when $z_\ell\not=z_j$.
Applying Eqs.~\Ftzint\ and \restg\ to Eq.~\magzz, we find
\eqn\magtg{
I(G;N) = \left( e^{-2\epsilon}-1\right) I({\cal G}_a)
\theta \left( - p_N - k_N + 1\right)
\times \left[ e^{-b(k_N + p_N-1)\epsilon} {\cal H}_j
- e^{-a(k_N+p_N-1)\epsilon} {\cal H}_\ell\right]
\ \ ,}
where
\eqnn\Hjdef
\eqnn\Hldef
$$\eqalignno{{\cal H}_j &= z_j^{k_N}
{z_j^{p_{N\ell}} z_\ell^{p_{\ell N}} (e^{-2\epsilon}-1) \over
e^{-a \epsilon} z_\ell - e^{-b \epsilon} z_j} {\cal G}_b\ \ ,&\Hjdef\cr
\noalign{\hbox{and}}
{\cal H}_\ell &= z_\ell^{k_N}
{z_\ell^{p_{N j}} z_j^{p_{j N}} (e^{-2\epsilon}-1) \over
e^{-a \epsilon} z_\ell - e^{-b \epsilon} z_j} {\cal G}_b\ \ .&\Hldef\cr}$$
To obtain this equation, we have used the relations $p_N=p_{N j} +
p_{N\ell}$ and $p_{N \ell} + p_{\ell N} = p_{N j}+ p_{j N} = 1.$  As in
Eq.~\kopi, we can define the new $k_i$'s in ${\cal H}_j$ and ${\cal H}_\ell$
to be the same as the old, except that $k_j(N-1)=k_j(N) + k_N$ in
${\cal H}_j$ and $k_\ell(N-1) = k_\ell(N) + k_N$ in ${\cal H}_\ell$.  Then
${\cal H}_j$ and ${\cal H}_\ell$ are both 1PI magnetic-field graphs with
degree of divergence less than or equal to one.
Because $k_N = \sum_{i \in {\cal G}_a} k_i
= \sum_{G_j \in {\cal G}_a} K_j$, the graphs ${\cal H}_j$ and
${\cal H}_\ell$ depend only on ${\svec K}$ and the original
$k_i \in {\cal G}_b$.

{}From the discussion following Eqs.~\restg\ and \Fjdef, we know that the
integral of the expression in square brackets in Eq.~\magtg\ goes as
$\epsilon^0$.  We also know that $I({\cal G}_a)$ diverges at most as
$\epsilon^{-1}$, and its divergent part is given by
${\cal E}_{-1}({\svec K})$.
Therefore, to get a non-vanishing result for $I(G;N)$ as $\epsilon \to 0$, in
Eq.~\magtg\ we must take only the divergent part of $I({\cal G}_a)$, with the
result,
\eqn\orint{
I(G;N) = -2\epsilon {\cal E}_{-1}({\svec K}) \,
\theta\left(- p_N - k_N + 1\right)
\left[ e^{-b(k_N + p_N - 1)\epsilon} {\cal H}_j -
e^{-a(k_N+p_N-1)\epsilon}{\cal H}_\ell\right]\ \ .}

We will now use induction on the number of vertices to calculate $I(G)$.
Because ${\cal H}_j$ and ${\cal H}_\ell$ are 1PI magnetic-field graphs
with fewer than $2n$ vertices, we will assume we know how to integrate them.
In fact, as long as we do not evaluate any more residues at $z_i=0$ poles
where $z_i$ is connected to two different vertices by magnetic-field edges,
we do already know how to integrate ${\cal H}_j$ and ${\cal H}_\ell$,
with the results given by Eqs.~\Rkmnzp, \Rkivzz\ and \magzzl.
Let ${\cal E}({\svec K})$ be a finite, piecewise-linear function of
${\svec K}$ plus a piecewise-constant function of ${\svec K}$ that
diverges at most as $\epsilon^{-1}$.
Then, for our induction hypothesis, we will assume that the
integrals of ${\cal H}_j$ and ${\cal H}_\ell$ have the form
\eqn\genopi{
I({\cal H}) = {c_{-1}\over \epsilon} + c_0 + {\svec a}\cdot {\svec k}
+ \sum_{i=1}^M
{\svec b}_i \cdot {\svec k}_i + {\cal E}({\svec K})\ \ ,}
where $c_0$, ${\svec b}_i$ and ${\cal E}({\svec K})$ may have different
values for ${\cal H}_j$ and ${\cal H}_\ell$, but $c_{-1}$ must be the same
for both.  In addition, inspection of Eqs.~\Hjdef\ and \Hldef\ for ${\cal
H}_j$ and ${\cal H}_\ell$ reveal they are the same if we perform a simple
shift of ${\svec k}$ in ${\cal H}_j$.  In Eq.~\genopi, $c_{-1}$, $c_0$ and the
components of ${\svec a}$ are real numbers.  The vector ${\svec b}_i$ is
given by
\eqn\biopi{{\svec b}_i= \sum_{{\cal S}_i}\sum_{\cal S'}
{\svec T}_{{\cal S}_i,{\cal S'}}({\svec K}) \prod_{S \in {\cal S}_i}
\prod_{S' \in {\cal S'}}
\theta\left(-\sum_{j \in S} k_j - \sum_{j \in S'} K_j\right)\ \ .}
In the sums, ${\cal S}_i$ and ${\cal S'}$ run over all sets of subsets of
$S_i$ and $\{1,2,\ldots,M\}$, respectively.
${\svec T}_{{\cal S}_i,{\cal S'}}({\svec K})$ is a piecewise-constant function
of ${\svec K}$ which is finite as $\epsilon \to 0$.  For convenience, we will
drop the subscript, ${\cal S'}$, and just write
${\svec T}_{{\cal S}_i}({\svec K})$.  The components of
${\svec T}_{{\cal S}_i}({\svec K})$ are chosen so that
${\svec b}_i \cdot {\svec k}$ depends only on ${\svec K}$ and
on the $k_j \in S_i$.

Note that Eq.~\Rkmnzp, Eq.~\Rkivzz\ with ${\svec b}_r$ given by \bridef\ and
\brivzz, and Eq.~\magzzl\ are special cases of
Eq.~\genopi\ for the integral of ${\cal H}$.  Thus, it only remains to check
whether the integral over the graph continues to have the form given by
Eqs.~\genopi\ and \biopi\ when we evaluate a $z_N=0$ residue for $z_N$
connected to two different vertices by magnetic-field edges.
Upon substituting Eq.~\genopi\ for the integral of
${\cal H}_j$ and ${\cal H}_\ell$ back into Eq.~\orint\ for such $z_N=0$
residues, we find that
\eqn\IGopi{
I(G)=\sum_{i=1}^M {\svec b}_i \cdot {\svec k} + {\cal E}({\svec K})
\ \ ,}
for some new ${\svec b}_i$ of the form given by Eq.\biopi\ and some new,
finite,
piecewise-constant plus piecewise-linear function, ${\cal E}({\svec K})$.
This is again a special case of Eq.~\genopi.  Therefore, by induction, we
conclude that the integral of any 1PI magnetic-field graph has the form
given by Eqs.~\genopi\ and \biopi.

When $G(2n)$ is connected but not 1PI, the new situations we must consider
arise when $z_N \in G_m$ is connected to $g$ different connected components
(other than $G_m$) by one magnetic-field edge, respectively.  Keeping this
in mind, we will divide our graph into subgraphs as follows.  First, we
have the subgraphs, $G_1,\ldots, G_M$, which were defined above and which
look like the graphs for zero magnetic field.  Next, we identify all the
edges, $e_1,\ldots, e_L$, such that, when any one of them is removed, the
original graph is divided into two disconnected pieces.  Then, we define the
${\cal G}_1,\ldots,{\cal G}_J$ to be the 1PI graphs that we obtain from
the original graph by removing all the edges $e_1,\ldots, e_L$.
Any ${\cal G}_j$ is then the union of some $G_i$'s and the magnetic-field
edges which connect them.  The edges, $e_1, \ldots, e_L$, are
also magnetic-field edges.  Now consider the graph
where each ${\cal G}_j$ is a vertex and the magnetic-field edges joining
different ${\cal G}_j$'s are the edges.  Because this graph becomes
disconnected whenever any one edge is removed, it must have a vertex,
${\cal G}_1$, which has only one edge, $e_1$, leading to it.

We will begin by integrating over all the vertices in ${\cal G}_1$, except
for the one vertex, $z_N$, that is connected to the magnetic-field edge,
$e_1$.  These vertices have the exact same form as the vertices in 1PI
graphs, so we already know how to integrate them.  Using similar
considerations that led to Eq.~\magzz, after all these integrations we
obtain
\eqn\conint{
I(G;N) = I({\cal G}_1) \times \oint {dz_N\over 2\pi i z_N} z^{p_{Nj} + k_N}_N
{\left( e^{-2\epsilon}-1\right)\over z_N - e^{-\epsilon a} z_j} z^{p_{jN}}_j
{\cal G}_b\ \ ,}
where $I({\cal G}_1)$ is the integral of the graph ${\cal G}_1$, and
${\cal G}_b$ is the subgraph of $G(2n)$ obtained by removing ${\cal G}_1$
and $e_1$.

If $a>0$, then the $z_N = e^{-\epsilon a} z_j$ pole lies inside the unit
circle and yields the residue
\eqn\conRa{
R_a= I({\cal G}_1)(e^{-2\epsilon}-1)e^{-\epsilon a (p_{Nj}+ k_N)}
z_j^{k_N} {\cal G}_b\ \ .}
To obtain this equation, we used the fact that $p_{Nj}+p_{jN}=1$.  For the
residue of the $z_N=0$ pole, according to Eq.~\zzoe\ we have
\eqn\conRz{
R_0=-(e^{-2\epsilon}-1)I({\cal G}_1) \theta\left(-p_{Nj}-k_N+1\right)
e^{-\epsilon a(p_{Nj}+k_N-1)} z_j^{k_N} {\cal G}_b\ \ .}
Then the full integral is
\eqn\IRaRz{I(G;N)=\theta(a) R_a + R_0\ \ .}
We can define ${\cal H} = z_j^{k_N}{\cal G}_b$ to be a new magnetic-field
graph with fewer vertices and with the new $k_j$ given by
$k_j(N-1)=k_j(N)+k_N$.  Because $k_N = \sum_{i\in {\cal G}_1} k_i
=-\sum_{i \in {\cal G}_b} k_i$ for the $k_i$ in the original graph, the
total momentum in ${\cal H}$ equals zero.  In addition, because $k_N$
can be written as $k_N=\sum_{G_j \in {\cal G}_1}K_j$,
the graph, ${\cal H}$, depends only on ${\svec K}$
and the original momenta in ${\cal G}_b$.

${\cal G}_1$ is a 1PI graph like the ones we just integrated, so its
integral is given by Eq.~\genopi.  If ${\cal H}$ does not contain any more of
the edges, $e_2,\ldots,e_L$, then it is 1PI and its integral is given by
Eq.~\genopi\ also.  Substituting these values of ${\cal G}_1$ and ${\cal H}$
into the integral \IRaRz\ of the full connected graph, we find that $I(G)$ is
still given by Eqs.~\genopi\ and \biopi.
Simliarly, when ${\cal H}$ is not 1PI, if we
assume that its integral is still of the form \genopi\ and \biopi,
then, when we substitute this
hypothesis for ${\cal H}$ into the full integral, we find that $I(G)$
is again given by Eqs.~\genopi\ and \biopi. Therefore, by induction, the
integral of any connected magnetic field graph must have this form.

Finally, if the integral for the magnetic field corresponds to a graph with
$g$ connected components, each component will have an integral equal to
$I_j$, for $1\le j\le g$, of the form \genopi\ and \biopi.
Such a graph must also have at least $g-1$
extra factors of $\epsilon$, due to the magnetic field propagators, so the
integral over the full graph is
\eqn\IGdis{
I(G)=\prod^g_{j=1} I_j \times \left( e^{-2\epsilon}-1\right)^\ell
\quad\hbox{for}\ \ \ell\ge g-1\ \ .}
When $\ell\ge g+1$, this expression goes to 0 as $\epsilon\to 0$.  Otherwise,
substituting in \genopi\ for each $I_j$, we find that as $\epsilon\to0$,
the integral of the magnetic field graph still has the form given by
Eq.~\genopi.

We conclude that the free, connected correlation functions of
$e^{i{\svec q} \cdot {\svec x}(t)}$'s with $\beta /\alpha \in \IZ$ have the
form
\eqn\finalI{
\tilde I({\svec k}) = {c_{-1}\over \epsilon} + c_0 +
{\svec a}\cdot {\svec k} + \sum_{i=1}^M
{\svec b}_i \cdot {\svec k}_i + {\cal E}({\svec K})\ \ .}
In this equation, $c_{-1}$, $c_0$ and ${\svec a}$ are constants; the
${\svec b}_i$ are the finite, piecewise-constant functions of ${\svec k}$
defined in Eq.~\biopi; and ${\cal E}({\svec K})$ is the sum of a finite,
piecewise-linear function and a piecewise-constant function that diverges at
most as $\epsilon^{-1}$.
\newsec{THE COORDINATE CORRELATION FUNCTIONS}
In this section, we will return to the calculation of the connected correlation
functions of $\dot x(t)$ and $\dot y(t)$.  We have already seen that they are
completely determined if we know the Fourier transform of the free, connected
correlation functions
$\ll \prod\limits^{2n}_{j=1} e^{i{\svec q}_j \cdot {\svec X}(t_j)}
\rr_0^{\rm con}$.
We have found that these functions are piecewise-constant plus
piecewise-linear functions in the Fourier-space variables.
We will now combine these two results to show that the Fourier transform of
the correlation functions of $\dot x(t)$ and $\dot y(t)$ are finite in
$\epsilon$ and piecewise-linear and homogeneous in the Fourier variables.

We will first concentrate on the case where the magnetic field is zero.
According to Eqs.~\Czmag\ and \Idef, the ${\cal O}(V^{2N})$ part of the
$\dot x(t)$ $m$-point function is
\eqn\Czmdef{
\tilde C_{2N} \left( \ell_1,\ldots, \ell_m;0\right)=a_m c_{2N} \prod^m_{j=1}
\left( \hbox{sign} (\ell_j) e^{-\epsilon|\ell_j|}\right) F_{2N} \left(
\ell_1,\ldots, \ell_m\right)\ \ ,}
where
\eqn\Fdef{
F_{2N} \left( \ell_1,\ldots, \ell_m\right)= \sum_{{\{q_i\}=\pm 1}\atop{\Sigma
q_i=0}} \sum_{\left( \sigma_1,\ldots, \sigma_{2N}\right)}
\tilde I \left( k_1,\ldots, k_{2N} \right)
\prod^{2N}_{j=1} q^{|\sigma_j|}_j \ \ .}
In the sum, $(\sigma_1,\ldots,\sigma_{2N})$ ranges over all partitions of
$m$ objects into $2N$ sets, and $|\sigma_j|$ is the number of elements in
the $j^{\rm th}$ set, $\sigma_j$.  $\tilde I$ is given by
\eqn\Itilde{
\tilde I \left( k_1,\ldots, k_{2N}\right) = \int \prod^{2N}_{j=1} dt_j\,
e^{2\pi i t_j k_j/T} I \left( t_1,\ldots,t_{2N}\right)\ \ ,}
where $I({\svec t})$ is the free, connected correlation function,
$\ll \prod\limits^{2N}_{j=1} e^{i q_j x(t_j)}\rr_0^{\rm con}$.  The
Fourier-space variables, $k_j$, are defined to be
\eqn\kjdef{k_j = \sum_{i\in \sigma_j} \ell_i\ \ .}

We calculated $\tilde I({\svec k})$ in the previous section and found that
\eqn\Icbk{
\tilde I ({\svec k}) = c + {\svec b}({\svec k})\cdot {\svec k}\ \ ,}
where $c = (c_{-1}/\epsilon) + c_0 + {\cal O}(\epsilon)$ is independent of
${\svec k}$.  The function ${\svec b}({\svec k})$ is finite in $\epsilon$
and piecewise constant in ${\svec k}$. It
depends only on the signs of the partial sums of ${\svec k}$,
where the partial sums are defined by
$\sum\limits_{j\in S}k_j$ with $S\subset \{1,2,\ldots,2N\}$.  Consequently,
${\svec b}({\svec k})\cdot {\svec k}$ is homogeneous in ${\svec k}$, with
degree one.  (The comments after Eq.~\newRk\ imply
that if we let ${\svec k}$ take on values in $\IR^{2N}$, the function
${\svec b}({\svec k})\cdot{\svec k}$ is continuous.)  If we substitute this
expression for $\tilde I$ back into Eq.~\Fdef\ for $F$, we find that
\eqn\FFzFo{F_{2N} \left( \ell_1,\ldots, \ell_m\right) =
F^{(0)}_{2N} \left( \ell_1,\ldots, \ell_m\right)
+ F^{(1)}_{2N} \left( \ell_1,\ldots, \ell_m\right)\ \ ,}
where
\eqn\Fzero{
F^{(0)}_{2N} = c\sum_{{\{q_i\}=\pm1}\atop{\Sigma q_i=0}} \sum_{\left(
\sigma_1,\ldots, \sigma_{2N}\right)} \prod^{2N}_{j=1} q^{|\sigma_j|}_j
\ \ ,}
and
\eqn\Fone{F^{(1)}_{2N}
= \sum_{{\{q_i\}=\pm1}\atop{\Sigma q_i=0}}
\sum_{\left( \sigma_1,\ldots,\sigma_{2N}\right)}
{\svec b}({\svec k})\cdot {\svec k}
\prod^{2N}_{j=1} q^{|\sigma_j|}_j \ \ .}
In Eq.~\Fzero, we can rearrange the order of the product over $j$'s
and the sum over $\sigma$'s so that it is again the same as in
Eq.~\Rzerom.  We find that
\eqn\finFz{
F^{(0)}_{2N} = c\sum_{{\{q_i\} =\pm1}\atop{\Sigma q_i=0}} \prod^m_{j=1}
\left[ \sum^{2N}_{k=1} q_k\right] = 0 \ \ .}
Therefore, the divergent ${\svec k}$-independent part of $\tilde I$ drops out
of the calculation, and we are left only with $F_{2N}=F^{(1)}_{2N}$,
which is finite as $\epsilon\to 0$.  Substituting this value for $F_{2N}$
back into the equation for $C_{2N}$ and recalling the definition of
${\svec b}({\svec k})$, we find
that as $\epsilon\to 0$, $\tilde C_{2N}\left( \ell_1,\ldots,\ell_m;0\right)$
is finite and is a piecewise-linear, homogeneous function of the $\ell_i$'s.
This result holds for any value of $2N\in 2\IZ$.  Therefore, to all orders in
perturbation theory, the $m$-point function of $\dot x(t)$,
$\tilde C\left(\ell_1,\ldots,\ell_m;0\right)$, is finite
in $\epsilon$ and is piecewise linear and homogeneous in the $\ell_i$'s.  In
particular, it has the form
\eqn\finCzm{
\tilde C \left( \ell_1,\ldots, \ell_m;0\right) = \prod^m_{j=1} \left( {\rm
sign}(\ell_j)\right) \times {\svec b}({\svec\ell})\cdot {\svec\ell}
\ \ ,}
where the $j^{\rm th}$ component of ${\svec b}$ has the form
\eqn\finbzm{
b^j ({\svec\ell})= a^j + \sum_{\cal S} b^j_{\cal S} \prod_{S\in {\cal S}}
\theta \left(-\sum_{i\in S} \ell_i\right)\ \ .}
Here, ${\svec a}$ and ${\svec b}_{\cal S}$ are constants; ${\cal S}$ is
summed over sets of subsets of $\{1,2,\ldots, m\}$;
and the $S$ are subsets of $\{1,2,\ldots, m\}$.  In addition, when we
let ${\svec\ell}$ take on values in $\IR^m$,
${\svec b}({\svec \ell}) \cdot {\svec \ell}$ is continuous.

According to Eq.~\Cmag, in the presence of a magnetic field, the
${\cal O}(V^{2N})$ contribution to the correlation function is
\eqn\CFmag{
\tilde C^{\mu_1\ldots \mu_m}_{2N} \left( \ell_1,\ldots, \ell_m;
{\beta\over\alpha}\right) = a_m c_{2N} \prod^m_{j=1} \left[{\rm sign}(\ell_j)
e^{-\epsilon|\ell_j|}\right] F^{\mu_1\ldots\mu_m}_{2N} \left( \ell_1,\ldots,
\ell_m;{\beta\over\alpha}\right)\ \ ,}
where
\eqn\magFdef{
F^{\mu_1\ldots \mu_m}_{2N} \left( \ell_1,\ldots,\ell_m\right) =
\sum_{\{q_i\} = \pm\hat x,\pm\hat y} \sum_{(\sigma_1,\ldots,\sigma_{2N})}
\tilde I \left( k_1,\ldots, k_{2N}\right)
\prod^{2N}_{k=1} \prod_{i\in\sigma_k}
\left[ {\svec r}^{\mu_i} (\ell_i) \cdot {\svec q}_k \right]\ \ .}
The tensor, $r^{\mu\nu}(l)$, is defined in Eq.~\rmndef; and the function,
$\tilde I$, is the Fourier transform of $\ll \prod\limits^{2N}_{j=1}
e^{i{\svec q}_j\cdot {\svec X}(t_j)}\rr_0^{\rm con}$, with respect to the
Fourier variables, $k_1,\ldots, k_{2N}$, defined in Eq.~\kjdef.
We have seen that each graph, $G$, for $\tilde I$ is made up of a collection of
subgraphs, $G_1,\ldots, G_\gamma$, and the propagators that join them,
where each subgraph, $G_j$, looks like a zero-magnetic field graph.
Using the notation of the previous sections, we let
$S_j\subset \{1,2,\ldots,2N\}$ be the set of all $i$ such
that the vertex $z_i$ is in $G_j$.  Then, according to the definition of
zero-magnetic field graphs, the charges, ${\svec q}$, satisfy
\eqn\qiej{
{\svec q}_i = q_i \hat e_j \qquad {\rm for} \quad i \in S_j\ \ ,}
where $\hat e_j$ is either $\hat x$ or $\hat y$.  In addition, the sum of
the charges in $G_j$ must be zero, which implies
\eqn\qsumz{\sum_{i \in S_j} q_i = 0\ \ .}
We will define $K_j$ to be the total momentum in each $G_j$, so that
$K_j = \sum_{i\in S_j} k_i$; and we will call ${\svec K} =
(K_1,\ldots,K_\gamma)$.

In the previous section, we found that for each graph, $\tilde I$ has the form
\eqn\Imagfin{
\tilde I_G ({\svec k}) = c + \sum_{i=1}^\gamma {\svec b}(S_i) \cdot {\svec k}
 + {\cal E} ({\svec K})\ \ ,}
where $c = (c_{-1}/\epsilon) + c_0 + {\cal O}(\epsilon)$ is a constant,
independent of ${\svec k}$.
${\cal E}({\svec K})$ is a sum of a piecewise-constant function that
diverges at most as $\epsilon^{-1}$ and a piecewise-linear function that is
finite as $\epsilon \to 0$.  ${\svec b}(S_i)$ is again a piecewise-constant
function of the $k_i$'s, but now it is given by
\eqn\bmagdef{
{\svec b}(S_i)= {\svec a} + \sum_{{\cal S}_i}\sum_{\cal S'}
{\svec T}_{{\cal S}_i}({\svec K}) \prod_{S \in {\cal S}_i}
\prod_{S' \in {\cal S'}}
\theta\left(-\sum_{j \in S} k_j - \sum_{j \in S'} K_j\right)\ \ .}
${\cal S}_i$ and ${\cal S'}$ are summed over the sets of subsets of $S_i$
and $\{1,2,\cdots,\gamma\}$, respectively.  ${\svec a}$ is a constant; and
${\svec T}_{{\cal S}_i}({\svec K})$ is a finite, piecewise-constant function
whose components are chosen so that ${\svec b}(S_i) \cdot {\svec k}$ is a
function only of ${\svec K}$ and those $k_j$ with $j \in S_i$.

Because each graph for $\tilde I$ depends on the total momentum in its
subgraphs, we will find it useful to look at $F_{2N}$ for a particular
graph, $G$, and to rearrange the sums over the partitions in the
equation for $F_{2N}$.  In Eq.~\magFdef, we sum over
all partitions of $m$ elements into $2N$ sets, $\sigma_1,\ldots,
\sigma_{2N}$.  This can be viewed as summing over all possible ways to attach
each of the $m$ momenta, $\ell_1,\ldots, \ell_m$, to exactly one of the
$2N$ vertices in the graph for $\tilde I$.  We will break this sum up
into, first, a sum over all the ways to
attach each of the $m$ momenta to the different subgraphs
$G_1,G_2,\ldots, G_\gamma$, and, second, a sum over all the ways to
connect the $m$ momenta to each of the vertices within the subgraph to
which they are attached.  If $\ell_n$ is connected to the vertex labeled by
$j_n$ which is in the $i_n^{\rm th}$ subgraph,
then we can write $F$ for the graph, $G$, as
\eqn\FSsum{
\eqalign{F^{\mu_1\ldots\mu_m}_G &\left( \ell_1,\ldots, \ell_m;{\beta\over
\alpha}\right) = \sum_{\{{\svec q}_i\}} \sum^\gamma_{i_1=1} \ldots
\sum^\gamma_{i_m=1}\cr
&\times \left[ \sum_{j_1\in S_{i_1}} \ldots \sum_{j_m\in S_{i_m}}
\tilde I_G \left( k_1,\ldots, k_{2N}\right)
\prod^m_{n=1} {\svec q}_{j_n} \cdot {\svec r}^{\mu_n} \left( \ell_n\right)
\right]\ \ ,\cr}}
where
\eqn\kiln{k_i = \sum_{\{n: j_n=i\}} \ell_n\ \ .}
With these definitions, we note that $K_j =
\sum\limits_{\{n: i_n=j\}} \ell_n$, where $K_j$ is the total momentum in the
$j^{\rm th}$ subgraph and the $\ell_n$'s in the sum are attached to
the $j^{\rm th}$ subgraph.  We will define the momenta, $L_j$,
by $L_j = K_j$.  For every term
in the sums over the $i_n$'s, the expression in square brackets in
Eq.~\FSsum\ has a fixed value of $L_j$ for $1\le j\le \gamma$.  When we
substitute expression \Imagfin\ for $\tilde I_G$ into
Eq.~\FSsum\ for $F_G$, we find that
\eqn\FGFcFS{
F_G^{\mu_1\ldots\mu_m} \left({\svec \ell};{\beta\over
\alpha}\right) = \sum_{\{{\svec q}_i\}}
\sum^\gamma_{i_1=1} \ldots \sum^\gamma_{i_m=1}
\Biggl\{ F^{\mu_1\ldots \mu_m}_c \left(
{\svec \ell};{\beta\over\alpha}\right)
+ \sum_{i=1}^\gamma \sum_{{\cal S}_i}
F^{\mu_1\ldots \mu_m}_{{\cal S}_i} \left({\svec \ell};{\beta\over\alpha}
\right)\Biggr\}\ \ ,}
where
\eqnn\Fcdef
\eqnn\FSidef
$$\eqalignno{
F^{\mu_1\ldots \mu_m}_c \left( \ell_1,\ldots,\ell_m;{\beta\over\alpha}\right)
&= \left(c + {\cal E}({\svec L})\right)
\sum_{j_1\in S_{i_1}} \ldots \sum_{j_m\in S_{i_m}}
\prod^m_{n=1} {\svec q}_{j_n} \cdot {\svec r}^{\mu_n} (\ell_n)\ \ ;&\Fcdef\cr
\noalign{\hbox{and}}
F^{\mu_1\ldots \mu_m}_{{\cal S}_i} \left(
\ell_1,\ldots,\ell_m;{\beta\over\alpha}\right) & =
{\svec T}_{{\cal S}_i} ({\svec L}) \cdot \left[ \sum_{j_1\in S_{i_1}}
\ldots \sum_{j_m\in S_{i_m}} {\svec f}_{{\cal S}_i} ({\svec\ell})
\prod^m_{n=1} {\svec
q}_{j_n} \cdot {\svec r}^{\mu_n}(\ell_n)\right]\ \ .&\FSidef\cr}$$
To obtain the last equation, for simplicity we have taken only one term in
the sum over ${\cal S'}$ in Eq.~\bmagdef.
In Eq.~\FSidef, ${\svec f}_{{\cal S}_i}({\svec \ell})$ is a homogeneous,
piecewise-linear function of ${\svec \ell} = (\ell_1,\ldots,\ell_m)$.  It is
given by
\eqn\fSidef{
{\svec f}_{{\cal S}_i}({\svec \ell}) = {\svec \ell}
\prod_{S \in {\cal S}_i} \prod_{S' \in {\cal S'}}
\theta\left(-\sum_{j_n \in S} \ell_n - \sum_{n \in S'} L_n \right)\ \ .}
${\svec T}_{{\cal S}_i}({\svec L})$ is the piecewise-constant function whose
$n^{\rm th}$ component is given by
\eqn\TSiLdef{
\left({\svec T}_{{\cal S}_i}\right)^n({\svec L})=
\left({\svec T}_{{\cal S}_i}\right)^{n'}({\svec K})\ \ ,}
where $\ell_n$ is connected to the vertex labeled by $n'$.  Then, because
${\svec b}(S_i) \cdot {\svec k}$ depends only on ${\svec K}$ and the
$k_j$ with $j \in S_i$, it follows that
${\svec T}_{{\cal S}_i}({\svec L}) \cdot {\svec f}_{{\cal S}_i}({\svec \ell})$
must be a function only of
${\svec L}$ and the $\ell_j$ that are connected to $G_i$.

We can rearrange the order of sums and products in the expression
for $F_c$, with the result
\eqn\qsums{
\left[ \sum_{j_1\in S_{i_1}}\ldots \sum_{j_m\in S_{i_m}} \prod^m_{n=1} {\svec
q}_{j_n} \cdot {\svec r}^{\mu_n} (\ell_n)\right] = \prod^m_{n=1} \left[ {\svec
r}^{\mu_n}(\ell_n)\cdot \sum_{j\in S_{i_n}} {\svec q}_{j_n}\right]
\ \ .}
In each subgraph corresponding to $S_{i_n}$, we have zero total charge, so
the above expression is zero and
\eqn\Fcz{F^{\mu_1\ldots\mu_m}_c \left( \ell_1,\ldots,\ell_m;
{\beta\over\alpha}\right) = 0 \ \ .}
In order to simplify $F_{{\cal S}_i}$, we will take $i=1$ and
consider a term in Eq.~\FSidef\
where the $\ell_j$ for $1 \le j \le p$ are connected to $S_1$ and all the other
$\ell$'s are not connected to $S_1$.  Then we have the expression
\eqn\FScalc{F_{{\cal S}_1}^{\mu_1\ldots \mu_m}
\left({\svec \ell}; {\beta \over \alpha} \right) =
{\svec T}_{{\cal S}_1} ({\svec L}) \cdot
\left[ \sum_{j_1\in S_1}\ldots \sum_{j_p\in S_1}
\sum_{j_{p+1}\in S_{i_{p+1}}} \ldots \sum_{j_m\in S_{i_m}}
{\svec f}_{{\cal S}_1}({\svec \ell})
\prod^m_{n=1}{\svec q}_{j_n} \cdot {\svec r}^{\mu_n}(\ell_n) \right]\ \ .}
Next, we separate the parts that depend on $S_1$ from those that do not to
obtain
\eqn\FEE{F_{{\cal S}_1}^{\mu_1\ldots \mu_m} = E_1 E_0\ \ ,}
where
\eqn\Eone{E_1 = {\svec T}_{{\cal S}_1} ({\svec L}) \cdot
\sum_{j_1\in S_1} \ldots \sum_{j_p\in S_1}
{\svec f}_{{\cal S}_1} ({\svec\ell})
\prod^p_{n=1} \left({\svec q}_{j_n}\cdot {\svec r}^{\mu_n}
(\ell_n)\right)\ \ ,}
and
\eqn\Ezero{E_0 =
\sum_{j_{p+1}\in S_{i_{p+1}}} \ldots \sum_{j_m\in S_{i_m}}
\prod^m_{n=p+1} \left({\svec q}_{j_n} \cdot {\svec r}^{\mu_n} (\ell_n)
\right)\ \ .}
As in the calculation for $F_c$, we can rearrange the order of
the summations and products to find
\eqn\Ezez{ E_0 =
\prod^m_{n=p+1} {\svec r}^{\mu_n} (\ell_n) \cdot \left( \sum_{j\in
S_{i_n}} {\svec q}_j \right) = 0 \ \ .}
Therefore, $F_{{\cal S}_i}$ is zero unless all the $\ell_n$'s are attached
to the same subgraph, $S_i$.  In that case, $F_{{\cal S}_i}=E_i$;
$L_j =0$ for $1\le j\le\gamma$, $j\not=i$; and $L_i =
\sum\limits^m_{n=1} \ell_n = 0$.
The last equation is true by overall momentum conservation, because the
original expressions for the correlation functions are translationally
invariant.  We conclude that ${\svec T}_{{\cal S}_i}({\svec L}) =
{\svec T}_{{\cal S}_i}({\svec 0})$ is just a
constant. Using these definitions and Eq.~\qiej\ for ${\svec q}$
(where $\hat e_i = \hat e_\nu$), $F_{{\cal S}_i}$ becomes
\eqn\FSfin{
F_{{\cal S}_i} = \left[ \prod^m_{n=1} r^{\mu_n\nu} (\ell_n)\right]
\sum_{j_1\in S_i} \ldots \sum_{j_n\in S_i}
{\svec T}_{{\cal S}_i}({\svec 0}) \cdot
{\svec f}_{{\cal S}_i}({\svec \ell})
\prod^m_{n=1} q_{j_n}\ \ .}

We can use the definitions of ${\svec T}_{{\cal S}_i}$ and
${\svec f}_{{\cal S}_i}$ to find
\eqn\Tfbl{
\sum_{{\cal S}_i} {\svec T}_{{\cal S}_i}({\svec 0}) \cdot
{\svec f}_{{\cal S}_i}({\svec \ell})
= {\svec b}_i({\svec \ell}; G) \cdot {\svec \ell}\ \ ,}
where, because all the $\ell_i$ are joined to the same subgraph, and because
${\svec L} = {\svec 0}$, ${\svec b}_i({\svec \ell}; G)$ has the form
\eqn\bifin{{\svec b}_i\left({\svec \ell}; G\right) = {\svec a} +
\sum_{\cal S} {\svec T}_{\cal S} \prod_{S \in {\cal S}}
\theta\left(-\sum_{i \in S} \ell_i\right)\ \ .}
In this equation, ${\svec a}$ and ${\svec T}_{\cal S}$ are constants,
independent of ${\svec \ell}$ and $\epsilon$.  However, they do depend on
our choice of the graph, $G$; the subgraph, $S_i$; and the $j_n$'s, which
determine the vertices to which the $\ell_n$'s are attached.  ${\cal S}$ now
runs through all sets of subsets of $\{1,2,\ldots,m\}$, and $S$ runs through
all the sets contained in ${\cal S}$.  Upon substituting all of this back
into the equation for $F_{2N}$, we obtain
\eqn\FtNblG{F^{\mu_1\ldots\mu_m}_{2N} \left(
\ell_1,\ldots,\ell_m;{\beta\over\alpha}\right) = \sum_{\{{\svec q}_i\}} \sum_G
\sum^{\gamma_G}_{i=1} \left[ \prod^m_{n=1} r^{\mu_n\nu_i}\right]\sum_{j_1\in
S_i} \ldots\sum_{j_m\in S_i} {\svec b}_i ({\svec\ell};G) \cdot
{\svec\ell}\prod^m_{n=1} q_{j_n}\ \ .}
Substituting this into the equation for the correlation functions and taking
the limit as $\epsilon \to 0$, we get
\eqn\CtNfin{\tilde C^{\mu_1\ldots \mu_m}_{2N}\left(
\ell_1,\ldots,\ell_m;{\beta\over\alpha}\right) = a_m c_{2N} \prod^m_{j=1}
\left[{\rm sign} (\ell_j) \right]  F^{\mu_1\ldots\mu_m}_{2N}
\left({\svec \ell};{\beta\over\alpha}\right)\ \ .}

This calculation is valid for any value of $2N$, which implies the result is
true to all orders in $V_0$.  We conclude that the correlation functions of
$\dot x$ and $\dot y$ are finite (or vanishing) as $\epsilon \to 0$, and they
are homogeneous, piecewise-linear functions of the Fourier-space variables.
This is the main result of this paper, but it remains for us to show that
the weak version of the duality transformation is true.

In Eq.~\FtNblG, $G$ is summed over the graphs for $\tilde I_{2N}$ with
charges $\{{\svec q}_i\}$.  Recall that
$\tilde I$ is the Fourier transform of the connected part of $Z_{n_x,n_y}$,
defined in Eq.~\Znxny.  In this equation, the ${\svec q}_i$
appear only in the exponents $\zeta_{ij}$ and $\eta _{ij}$.  Namely, we have
\eqn\xinu{
\zeta_{ij} = - {\svec q}_i\cdot {\svec q}_j\qquad\hbox{and}\qquad \eta_{ij}
= - {\beta\over\alpha} \hat z\cdot \left( {\svec q}_i\times {\svec q}_j\right)
\ \ .}
We note that both of these are invariant under ${\svec q} \to \hat z\times
{\svec q}$.  This implies that for every graph, $G$, with charges
${\svec q}_i$, there is exactly one other graph, ${\bar G}$, which is identical
to $G$, except that it has charges $\hat z\times {\svec q}_i$.  The value of
the Fourier transform of these graphs, which appears in $F_{2N}$ in the form
${\svec b}_i({\svec \ell}; G) \cdot {\svec \ell}$,
must be identical.  In Eq.~\FtNblG, the only other
dependence of $F_{2N}$ on the charges ${\svec q}_i$ appears in the
expression $\prod\limits^m_{n=1}
r^{\mu_n\nu_i}$, which depends on the species, $\nu_i$, of the charges in the
subgraph $S_i$; and in $\prod^{m}_{n=1} q_{j_n}$, which depends on the sign of
the charges in the subgraph $S_i$.  If we start with
$\prod\limits^m_{n=1}\left( r^{\mu_nx}(\ell_n) q_{j_n}\right)$ and take
${\svec q}_{j_n}=\hat x q_{j_n}\to \hat z\times {\svec q}_{j_n}$, then we
obtain $\prod\limits^m_{n=1} (r^{\mu_ny}(\ell_n) q_{j_n})$.  Thus, we can
change our sum over all $\{{\svec q}_j\}$, $G$ and $i$ to be a sum only over
those $i$ such that the vertices in $S_i$ are all $x$-vertices. We obtain the
other half of the graphs by taking the new charges to be $\hat z\times {\svec
q}_{j_n}$.  Putting all this together, we find
\eqn\FtNpen{\eqalign{
F^{\mu_1\ldots\mu_m}_{2N} \left( {\svec\ell};{\beta\over\alpha}\right)&=
\sum_{\{{\svec q}_j\}} \sum_G \sum^{\gamma_G}_{{i=1}\atop{S_i\,
{\rm all}\, x'{\rm s}} }
\left[ \prod^m_{n=1} r^{\mu_nx}(\ell_n) + \prod^m_{n=1}
r^{\mu_ny}(\ell_n)\right] \cr
&\times \sum_{j_1\in S_i}\ldots \sum_{j_m\in S_i} {\svec b}_i \left(
{\svec\ell}; G\right)\cdot {\svec\ell}\prod^m_{n=1} q_{j_n}\ \ .\cr}}
Substituting this back into the expression for
$\tilde C^{\mu_1\ldots\mu_m}_{2N} \left( {\svec\ell};{\beta/\alpha}\right)$
we have
\eqn\CtNpen{
\tilde C^{\mu_1\ldots\mu_m}_{2N} \left( {\svec\ell};{\beta\over\alpha}\right) =
\left[ \prod^m_{n=1} r^{\mu_nx}(\ell_n) + \prod^m_{n=1}
r^{\mu_ny}(\ell_n) \right] F_{2N} \left(
{\svec\ell};{\beta\over\alpha}\right)\ \ ,}
where, as $\epsilon\to0$,
\eqn\FtNfin{F_{2N} \left( {\svec\ell}; {\beta\over\alpha}\right)
= a_m c_{2N} \left(\prod^m_{j=1}{\rm sign} (\ell_j) \right)
\times {\svec b}_{2N} ({\svec\ell}) \cdot
{\svec\ell}\ \ .}
Once again,
${\svec b}_{2N} ({\svec\ell})\cdot{\svec\ell}$ is a piecewise-linear,
homogeneous function in ${\svec\ell}$ which has the form
\eqn\btNl{{\svec b}_{2N}({\svec\ell}) =
{\svec a} + \sum_{\cal S} {\svec b}_{\cal S} \prod_{S\in {\cal S}}
\theta\left(-\sum_{i\in S} \ell_i\right)\ \ ,}
where ${\cal S}$ is summed over sets of subsets of $\{1,2,\ldots,m\}$ and
the $S$ are subsets of $\{1,2,\ldots, m\}$.  If
$F_{2N}\left({\svec \ell};\beta/\alpha\right)$ were independent of
$\beta/\alpha$, Eq.~\CtNpen\ would be presicely of the form predicted by the
naive derivation of the duality transformation.  Instead,
we have proved a weaker version of the duality transformation.

{}From the preceding discussion, it is
straightforward to show that ${\svec b}_{2N}({\svec\ell})\cdot{\svec\ell}$ is
symmetric under interchanges of $\ell_i$ and $\ell_j$ and when
${\svec\ell} \to - {\svec\ell}$.  Also, because it comes from
``antisymmetrizing'' over all ways to partition $m$ objects into $M$ sets,
for some $M$, it is zero whenever any of the $\ell_i$ are $0$.  With some
additional work, one can show that $F_{2N}({\svec \ell})$ is a continuous
function of ${\svec \ell}$ for ${\svec \ell} \in \IR^m$.  Lastly, we recall
that we have dropped the momentum conserving $\delta$-function from all the
equations.

Before concluding this section, we will calculate, as a concrete example,
the correlation functions at order $V_0^2$.  According to Eq.~\Znxdef,
without a magnetic field, $Z_2$ is given by
\eqn\Ztwo{Z_2 = \ll e^{ix(t_1)} e^{-ix(t_2)}\rr_0
=-{e^{-\epsilon}z_1 z_2 \over (z_1-e^\epsilon z_2)(z_1-e^{-\epsilon}z_2)}
\ \ ,}
where $z_1 = e^{i t_1 2\pi/T}$ and $z_2 = e^{it_2 2 \pi /T}$.  The Fourier
series of $Z_2$ is given by
\eqn\Ztwofs{Z_2 = {1 \over e^{-2\epsilon} -1} \sum_{m=-\infty}^\infty
\left({z_1 \over z_2} \right)^m e^{-|m|\epsilon}\ \ ,}
so the Fourier coefficient of $z_1^{k_1} z_2^{k_2}$ is
\eqn\Itwo{
\tilde I(k_1,k_2) = {1 \over e^{-2\epsilon}-1} e^{|k_1|\epsilon}
\delta_{k_1+k_2,0}\ \ .}
In the limit as $\epsilon \to 0$, this becomes
\eqn\Itwolimd{
\tilde I(k_1,k_2)= \left[-{1\over 2\epsilon}-{1\over 2} +
{1\over 2}|k_1|\right] \delta_{k_1+k_2,0}\ \ .}
We have seen that the constant part of $\tilde I(k_1, k_2)$ does not
contribute to the correlation functions. Substituting ${1\over 2}|k_1|$
for $\tilde I (k_1,k_2)$ into Eq.\Fdef, and summing over the $q_i$, we find
\eqn\Ftwoex{F_2(\ell_1,\ldots,\ell_m)
= \delta_{\Sigma_{i=1}^m \ell_i, 0}
\sum_{M=0}^m \sum_{\sigma_M}(-1)^M
\bigg|\sum_{i\in\sigma_M}\ell_i\bigg| \ \ ,}
where $\sigma_M$ is summed over all subsets of $\{1,2,\ldots,m\}$ with
$M$ elements.  Substituting this back into Eq.~\Czmdef\ for the correlation
function, we obtain
\eqn\Ctwo{\tilde C_2(\ell_1,\ldots,\ell_m;0)
= {1\over 2}\left({V_r\over 2}\right)^2
\left({2\pi i \over T}\right)^m
\prod_{j=1}^m {\rm sign}(\ell_j)
\sum_{M=0}^m \sum_{\sigma_M}(-1)^M
\bigg|\sum_{i\in\sigma_M}\ell_i\bigg|
\delta_{\Sigma_{i=1}^m\ell_i,0}\ \ ,}
where we have used Eqs.~\ctndef\ and \amdef\ for $c_2$ and $a_m$.

Even when the magnetic field is non-zero, the Fourier transform of the
graphs is still given by Eq.~\Itwo, so the correlation function is
\eqn\Ctwom{
\tilde C_2^{\mu_1\ldots\mu_m}\left({\svec \ell};{\beta \over \alpha}\right)
=\left[ \prod^m_{n=1} r^{\mu_nx} (\ell_n)
+ \prod^m_{n=1} r^{\mu_ny} (\ell_n) \right]
\tilde C_2\left({\svec \ell};0\right) \ \ .}
Therefore, to this order in $V_r$, the strong version of the duality
transformation is satisfied.

For the general case, we conclude that, to all orders in $V_0$, the
Fourier transform of the
correlation functions of $\dot x(t)$ and $\dot y(t)$ have the form
\eqn\Cfin{
\tilde C^{\mu_1\ldots \mu_m}\left({\svec\ell};{\beta\over\alpha}\right) =
\left[ \prod^m_{n=1} r^{\mu_nx} (\ell_n) + \prod^m_{n=1} r^{\mu_ny} (\ell_n)
\right] F\left( {\svec\ell};{\beta\over\alpha}\right)
\delta_{\Sigma_{i=1}^m \ell_i,0} \ \ ,}
where $F\left( {\svec\ell};\beta/\alpha\right)$ has the following
properties:
\item{1)}It is a homogeneous function of ${\svec\ell}$.
\smallskip
\item{2)}It is a piecewise-linear function of ${\svec\ell}$.
\smallskip
\item{3)}It is symmetric under interchanges of $\ell_i$ and $\ell_j$.
\smallskip
\item{4)}It is symmetric under ${\svec\ell} \to -{\svec\ell}$.
\smallskip
\item{5)}It vanishes when any of the $\ell_i$ are zero.
\smallskip
\item{6)}It is continuous for ${\svec\ell} \in \IR^m$.
\smallskip
\item{7)} $F=0$ is if $m$ is odd.
\newsec{CONCLUSIONS}
In summary, for integer $\beta/\alpha$, we have proved that the correlation
functions of $\dot x(t)$ and $\dot y(t)$ are homogeneous, piecewise-linear
functions of the Fourier-space variables, and at any particular
value of $\beta/\alpha$ the relation between the different $m$-point functions
is exactly what the duality transformation predicts.  Also, these
correlation functions are finite as $\epsilon\to 0$ and the free energy
diverges as $1/\epsilon$.
All of these results are valid to all orders in perturbation
theory and were obtained using the regulators and renormalization scheme which
also satisfy the reparameterization invariance Ward identities.  To
${\cal O}(V_0^2)$ we have calculated all the coordinate correlation
functions, thereby finding an explicit form for all the functions shown to
be contact terms in Ref.~\CT.  Our methods
of analyzing the Feynman graphs and evaluating the integrals are fairly general
and can be applied to the correlation functions of $e^{i{\svec q}\cdot {\svec
x}(t)}$ and $\dot{\svec x}(t)$ to show that they are also piecewise-polynomial.
With some additional work, they may also be useful for evaluating the
correlation functions of the other dimensionful operators and correlation
functions in the charged sector.

In this paper, we have focused on the coordinate correlation functions for
several reasons.  These are the functions needed to calculate the boundary
state and thereby find solutions to open string theory in a non-trivial
background.  Also, the duality
transformation relates coordinate correlation functions that are
contact terms to
those that are finite and well-defined functions at large times.  A careful
calculation of the properties arising from the fermionization and duality
symmetries of this model can aid us in determining whether such symmetries and
large-time behavior are sufficient for fully fixing the physical contact
terms of the theory, independent of our choice of regularization.

There is an interesting connection between our results and the
Duistermaat--Heckman theorem \DH.  This theorem is the root of exact
solvability  in many systems.  It can be interpreted as saying that the
stationary phase approximation for calculating certain intergrals is
exact.  Taking $\beta=0$ and $T\to\infty$, the
Fourier transform of the free correlation functions of $e^{\pm ix(t)}$ can be
expressed in a way such that, ignoring issues of regularization, one can
formally apply the Duistermaat--Heckman theorem.  This formal application
immediately implies that these correlation functions are piecewise linear.
What we have done in this paper is to prove piecewise
linearity, with the regulator in place, both for these free correlation
functions and for the correlation functions of $\dot x(t)$ and $\dot y(t)$.

The result that the Fourier coefficients are homogeneous functions
of the Fourier-space variables also has an interesting interpretation in real
space;  it says that the correlation functions obey a symmetry under
reparametrizations of time which are not one--to--one.  This extends the
symmetries of the Ward identities, which tell how the correlation functions
transform under one--to--one reparametrizations of time.

Perhaps now that we have uncovered the properties of the
correlation functions of the dissipative Hofstadter model, one can find a more
elegant way of proving them.  Also, the connection between the dissipative
Hofstadter model and the Duistermaat--Heckman theorem is tantalizing, and
suggests
that the dissipative Hofstadter model may be exactly solvable. The properties
we have derived in this paper actually do determine the four-point function
exactly, up to normalization. In a future paper, we will show how these
properties, combined with the Ward identities, are sufficient for determining
the four-point function and many other $m$-point functions exactly, and that
in all cases in which they are determined, the strong statement of the duality
transformation is also satisfied.  However, much still remains to be done in
determining whether these properties, based on fermionization and duality
symmetry, combined with the Ward identities are sufficient for determining all
of the coordinate correlation functions and for fully solving this theory.
\goodbreak
\bigskip
\noindent{\bf Acknowledgements \quad}
I would like to thank S.~Axelrod, J.~Cohn and M.~Crescimanno for helpful
discussions.
\medskip
\nobreak
\par
\vfill
\listrefs
\listfigs
\end